\documentclass[iop]{emulateapj-rtx4}

\usepackage{natbib}
\usepackage{graphics,graphicx}
\usepackage{float}
\usepackage{natbib}
\usepackage{amsfonts,amsmath,amssymb}
\usepackage{color}
\usepackage{verbatimbox}
\usepackage{hyperref}
\hypersetup{backref=true, pagebackref=true, hyperindex=true, breaklinks=true,colorlinks=true,urlcolor=blue, linkcolor=blue,  citecolor=blue,pagecolor=red, bookmarks=true, bookmarksopen=true}

\slugcomment{}

\newcommand{\TW}{TW~Hya}
\defcitealias{2014A&A...564A..93M}{M14}	 	 
\newcommand{\Menu}{\citetalias{2014A&A...564A..93M}}

\shorttitle{The bulk density distribution of the \TW \ disk from SPHERE observations}
\shortauthors{SPHERE GTO team}

\newcommand{\mum}{\,$\mu$m}
\newcommand{\degree}{$^{\circ}$}
\newcommand{\Mstar}{$M_*$}
\newcommand{\Lstar}{$L_*$}

\newcommand{\Msun}{M$_{\odot}$}
\newcommand{\Lsun}{L$_{\odot}$}
\newcommand{\Rsun}{R$_{\odot}$}
\newcommand{\AU}{au}

\newcommand{\msunyr}{M$_{\odot}$yr$^{-1}$}
\newcommand{\cms}{cm\,s$^{-1}$}

\newcommand{\Mearth}{M$_{\oplus}$} 

\newcommand{\Mjup}{M$_{\rm{Jup}}$} 
\newcommand{\Rjup}{R$_{\rm{Jup}}$}

\newcommand{\simil}{$\approx$}
\newcommand{\Mdotacc}{$\dot M_{\rm{acc}}$}

\newcommand{\HCOp}{HCO$^+$}

\newcommand{\rsq}{$R^2$}
\newcommand{\fwhm}{$fwhm$}

\newcommand{\fig}{Fig.}
\newcommand{\fsep}{\,}
\newcommand{\figs}{Figs.}
\newcommand{\sek}{Section}
\newcommand{\seks}{sections}
\newcommand{\seksep}{\,}
\newcommand{\tab}{Table}
\newcommand{\tabsep}{\,}
\newcommand{\eq}{equation}
\newcommand{\eqsep}{\,}

\newcommand{\lamR}{0.63} 
\newcommand{\lamI}{0.79} 
\newcommand{\lamJ}{1.24} 
\newcommand{\lamH}{1.62} 

\newcommand{\Lp}{L$^{\prime}$}

\newcommand{\Rone}{85}
\newcommand{\Rtwo}{21}
\newcommand{\Rthree}{6}

\newcommand{\Rp}{R$^{\prime}$}
\newcommand{\Ip}{I$^{\prime}$}

\newcommand{\dpc}{54}
\newcommand{\amin}{$a_{\rm{min}}$}

\newcommand{\Qphi}{$Q_{\phi}$}
\newcommand{\Uphi}{$U_{\phi}$}

\newcommand{\ds}{\,}
\newcommand{\du}{\,}

\begin{document}

\title{Three radial gaps in the disk of TW~Hydrae imaged with SPHERE}
\author{
R.~van~Boekel\altaffilmark{1},       
Th.~Henning\altaffilmark{1},         
J.~Menu\altaffilmark{1,2},           
J.~de~Boer\altaffilmark{3,4},        
M.~Langlois\altaffilmark{5,6},       
A.~M\"uller\altaffilmark{4,1},       
H.~Avenhaus\altaffilmark{7},         
A.~Boccaletti\altaffilmark{8},       
H.\,M.~Schmid\altaffilmark{9},       
Ch.~Thalmann\altaffilmark{9},        
M.~Benisty\altaffilmark{10,11},      
C.~Dominik\altaffilmark{12},         
Ch.~Ginski\altaffilmark{3},          
J.\,H.~Girard\altaffilmark{4,10,11}, 
D.~Gisler\altaffilmark{9,13},        
A.~Lobo~Gomes\altaffilmark{14},      
F.~Menard\altaffilmark{15,7},        
M.~Min\altaffilmark{16,12},          
A.~Pavlov\altaffilmark{1}            
A.~Pohl\altaffilmark{1},             
S.\,P.~Quanz\altaffilmark{9},        
P.~Rabou\altaffilmark{10,11},        
R.~Roelfsema\altaffilmark{17}        
J.-F.~Sauvage\altaffilmark{18}       
R.~Teague\altaffilmark{1},           
F.~Wildi\altaffilmark{19}, and       
A.~Zurlo\altaffilmark{20,6,7}}       

\altaffiltext{1}{Max Planck-Institut f\"ur Astronomie, K\"onigstuhl 17, D-69117 Heidelberg, Germany}
\altaffiltext{2}{Instituut voor Sterrenkunde, KU Leuven, Celestijnenlaan 200D, 3001 Leuven, Belgium}
\altaffiltext{3}{Leiden Observatory, Leiden University, PO Box 9513, 2300 RA Leiden, The Netherlands}
\altaffiltext{4}{European Southern Observatory, Alonso de C\'ordova 3107, Casilla 19001 Vitacura, Santiago 19, Chile}
\altaffiltext{5}{Centre de Recherche Astrophysique de Lyon, CNRS, Universite\'e Lyon 1, 9 avenue Charles Andr\'e, 69561 Saint-Genis-Laval Cedex, France}
\altaffiltext{6}{Aix Marseille Universit\'e, CNRS, LAM (Laboratoire d'Astrophysique de Marseille) UMR 7326, 13388 Marseille, France}
\altaffiltext{7}{Departamento de Astronom\'ia, Universidad de Chile, Casilla 36-D, Santiago, Chile}
\altaffiltext{8}{LESIA, Observatoire de Paris, CNRS, Universit\'e Paris Diderot, Universit\'e Pierre et Marie Curie, 5 place Jules Janssen, 92190 Meudon, France}
\altaffiltext{9}{Institute for Astronomy, ETH Z\"urich, Wolfgang-Pauli-Strasse 27, 8093 Z\"urich, Switzerland}
\altaffiltext{10}{Universit\'e Grenoble Alpes, IPAG, F-38000 Grenoble, France}
\altaffiltext{11}{CNRS, IPAG, F-38000 Grenoble, France}
\altaffiltext{12}{Anton Pannekoek Institute for Astronomy, University of Amsterdam, Science Park 904, 1098 XH Amsterdam, The Netherlands}
\altaffiltext{13}{Kiepenheuer-Institut f\"ur Sonnenphysik, Sch\"oneckstrasse 6, 79104 Freiburg, Germany}
\altaffiltext{14}{Instituto de Astrof\'isica, Pontificia Universidad Catolica de Chile, Av. Vicun ̃a Mackenna 4860, Macul, Santiago de Chile}
\altaffiltext{15}{UMI-FCA, CNRS/INSU, France (UMI 3386)}
\altaffiltext{16}{SRON Netherlands Institute for Space Research, Sorbonnelaan 2, 3584 CA Utrecht, The Netherlands}
\altaffiltext{17}{NOVA Optical-Infrared Instrumentation Group at ASTRON, Oude Hoogeveensedijk 4, 7991 PD Dwingeloo, The Netherlands}
\altaffiltext{18}{ONERA, BP72, 29 avenue de la Division Leclerc, 92322 Ch\^atillon Cedex, France}
\altaffiltext{19}{Observatoire astronomique de l'Universit\'e de Gen\`eve, 51 ch. des Maillettes, 1290 Versoix, Switzerland}
\altaffiltext{20}{N\'ucleo de Astronom\'ia, Facultad de Ingenier\'ia, Universidad Diego Portales, Av. Ejercito 441, Santiago, Chile}

\email{boekel@mpia.de}

\setcounter{footnote}{0}

\begin{abstract}
We present scattered light images of the TW Hya disk performed with SPHERE in PDI mode at \lamR, \lamI, \lamJ \ and \lamH\du\mum. We also present H2/H3-band ADI observations. Three distinct radial depressions in the polarized intensity distribution are seen, around \simil\ds\Rone, \simil\ds\Rtwo, and $\lesssim$\ds\Rthree\,\AU\footnote{\label{gaia_footnote}Throughout this work we have assumed a distance of 54\du pc to \TW. This is $\approx$\ds10\% less than the new GAIA distance of $59.5^{+0.96}_{-0.93}$\du pc \citep{2016arXiv160904172G}. We discuss the implications of the new, somewhat larger distance in \sek\seksep\ref{sec:gaia_distance}.}. The overall intensity distribution has a high degree of azimuthal symmetry; the disk is somewhat brighter than average towards the South and darker towards the North-West. The ADI observations yielded no signifiant detection of point sources in the disk.

Our observations have a linear spatial resolution of 1 to 2\du\AU, similar to that of recent ALMA dust continuum observations. The sub-micron sized dust grains that dominate the light scattering in the disk surface are strongly coupled to the gas. We created a radiative transfer disk model with self-consistent temperature and vertical structure iteration and including grain size-dependent dust settling. This method may provide independent constraints on the gas distribution at higher spatial resolution than is feasible with ALMA gas line observations.

We find that the gas surface density in the ``gaps'' is reduced by $\approx$\ds50\% to $\approx$\ds80\% relative to an unperturbed model. Should embedded planets be responsible for carving the gaps then their masses are at most a few 10\du\Mearth. The observed gaps are wider, with shallower flanks, than expected for planet-disk interaction with such low-mass planets. If forming planetary bodies have undergone collapse and are in the ``detached phase'', then they may be directly observable with future facilities such as METIS at the E-ELT.
\end{abstract}


\section{Introduction}
\label{sec:introduction}

\setcounter{footnote}{1}

The distribution of gaseous and solid material in the circumstellar disks around young stars, the physical and chemical properties of this material, and the temporal evolution of these quantities provide important boundary conditions for modeling of the formation of planets and other bodies in the solar system and other planetary systems. However, the small angular scales involved pose great observational challenges. Recent ALMA dust continuum observations of \TW, the nearest young system with a gas-rich disk, provide a detailed map of the spatial distribution of large, $\approx$\ds mm-sized dust, at a linear spatial resolution of $\approx$\ds1\du\AU \ \citep{2016ApJ...820L..40A}. Studying the distribution of gas at the same spatial scales is not feasible with ALMA but on larger scales the gas distribution can be mapped \citep[e.g.][]{2016ApJ...823...91S} using trace species like CO isotopologues as a proxy for the bulk mass of H$_2$ and He gas, which is not directly observable. 

In this work we apply a qualitatively different method to constrain the bulk gas density distribution. We use the observed optical and near-infrared scattered light surface brightness distribution observed with SPHERE instrument to infer the radial bulk gas surface density profile. This requires detailed physical modeling of the disk, at the heart of which lies the strong dynamical coupling between the gas and the $\approx$\ds0.1\du\mum \ particles that dominate the scattered light, and the assumption that the disk is a passive irradiated structure in vertical hydrostatic equilibrium. Our SPHERE observations have a linear spatial resolution of 1 to 2\du\AU, which is similar to that of the ALMA dust continuum observations.

Due to its close proximity\footnotemark[\ref{gaia_footnote}] \citep[\dpc\du pc][]{1998A&A...336..242H,2007A&A...474..653V} the \TW \ system has been the focus of many observational studies employing in particular high spatial resolution facilities. Its disk is seen nearly face-on \citep[$i \approx 7$\,\degree][]{2004ApJ...616L..11Q} and it is the only such object for which a direct measure of the bulk gas mass could be obtained so far. Despite its relatively high age \citep[3$-$10\du Myr][]{1998A&A...336..242H,2006A&A...459..511B,2011ApJ...732....8V} the disk is still relatively massive with $M_{\rm{gas}}$\,$\gtrsim$\,0.05\du\Msun \ \citep{2013Natur.493..644B}.

Based on its small near-infrared excess \TW \ was identified as a ``transition disk'' with an inner hole of $\approx$\ds4\du\AU \ in the optically thick dust distribution, with additional optically thin dust at smaller radii \citep{2002ApJ...568.1008C}. The latter was indeed detected with interferometric observations around 2.2\du\mum \ \citep{2006ApJ...637L.133E} and 1.6\du\mum \ \citep{2015A&A...574A..41A}. Interferometry in the N-band (8$-$13\,\mum) with MIDI at the VLTI suggested a much smaller inner hole of $\approx$\ds0.6\du\AU \ \citep{2007A&A...471..173R}. A combined analysis of the SED and interferometric data at infrared (MIDI), 1.3\du mm (SMA) and 9\du mm (eVLA) wavelengths resulted in a refined model where the disk has a peak surface density at $\approx$\ds2\du\AU, with a rounded-off inner inner rim between 0.35 and 2\du\AU \ \citep{2014A&A...564A..93M}.

Scattered light observations at optical and near-infrared wavelengths probe the distribution of small ($\lesssim$\ds1\du\mum) dust suspended in the disk atmosphere high above the midplane. These grains are extremely well coupled to the gas and can therefore be used to probe the gas distribution. The scattered light observations are thus complementary to millimeter interferometry which is mostly sensitive to much larger grains ($\gtrsim$100\,\mum \ to \simil cm sizes) which settle to the midplane and are prone to radial drift in the direction of increasing gas pressure.

Scattered light observations with the HST \citep{2000ApJ...538..793K,2002ApJ...566..409W,2005ApJ...622.1171R,2013ApJ...771...45D,2016ApJ...819L...1D} and from the ground with VLT/NACO \citep{2004A&A...415..671A}, Subaru/HiCiao \citep{2015ApJ...802L..17A}, and Gemini/GPI \citep{2015ApJ...815L..26R} have yielded an increasingly clear view of the disk surface. Radial depressions in the surface brightness around 80\du\AU \ and around 23\du\AU \ from the star were found. These have been interpreted as radial depressions in the gas surface density \citep[e.g.][]{2013ApJ...771...45D,2015ApJ...802L..17A,2015ApJ...815L..26R}, possibly caused by forming planets embedded in the disk, though caution is needed as there is no direct evidence yet for any planet in these gaps. Also non-ideal MRI-based simulations yield radial depressions in the gas surface density \citep{2015A&A...574A..68F,2016A&A...590A..17R}. Dust evolution \citep{2015ApJ...813L..14B}, increased coagulation efficiency after ice condensation fronts \citep{2015ApJ...806L...7Z}, and a sintering-induced change of agglomeration state \citep{2016ApJ...821...82O} may all lead to ringed structures in disks.

At millimeter wavelengths the larger grain population near the midplane is seen in the continuum, while the gas and its kinematics can be traced using molecular rotational lines. In the pre-ALMA era such observations have been highly challenging in terms of both spatial resolution and sensitivity. Nonetheless, the distribution of large dust could be resolved \citep{2000ApJ...534L.101W,2007ApJ...664..536H} and it was shown to be centrally concentrated with a sharp outer edge around 60\du\AU \ using the SMA \citep{2012ApJ...744..162A}. The gas is much more extended and is observed out to at least 230\du\AU \ in CO line emission \citep[e.g.][]{2012ApJ...744..162A}. Using the SMA \cite{2015ApJ...799..204C} detected a source of \HCOp \ emission at $\approx$\ds0\farcs43 South-West of the central star, i.e., located in the ``gap'' seen in scattered light around 23\du\AU \ \citep{2015ApJ...802L..17A,2015ApJ...815L..26R}. Early observations of N$_2$H$^+$ with ALMA showed that the CO iceline in the midplane is located at $\approx$\ds30\du\AU \ \citep{2013Sci...341..630Q}. Early ALMA observations revealed a radial depression in the distribution of large dust around 25\du\AU, accompanied by a ring at $\approx$\ds41\du\AU \ where the large dust has accumulated \citep{2016ApJ...819L...7N}. Using new ALMA continuum observations around  870\du\mum \ with dramatically improved spatial resolution \cite{2016ApJ...820L..40A} could reveal a highly structured radial intensity distribution with a multitude of bright and dark rings, in an azimuthically highly symmetric disk. \cite{2016arXiv160500289T} complemented this study using ALMA observations at multiple wavelengths, albeit with somewhat lower spatial resolution, allowing measurement of the mm spectral index and revealing a deficit of large grains in the 22\du\AU \ gap region.

\vspace{1.25mm}

In this work we present new optical and near-infrared scattered light observations obtained with SPHERE at the VLT as part of the Guaranteed Time Observations (GTO), which surpass all previous studies in terms of contrast and inner working angle. Our scientific focus is the radial distribution of the bulk gas, using the small dust particles that we observe as a tracer.

\section{Observations and data reduction}
\subsection{Observations}
\label{sec:methods:observations}

\subsubsection{Polarimetric disk imaging}
\label{sec:methods:PDI}

Polarimetric observations of \TW \ were conducted with the Spectro-Polarimetric High-contrast Exoplanet REsearch instrument \citep[SPHERE,][]{2008SPIE.7014E..18B}, within the SPHERE GTO program. Two of the sub-instruments of SPHERE were used: the Z\"urich Imaging POLarimeter \citep[ZIMPOL,][]{2012SPIE.8446E..8YS} in field stabilized (P2) mode, and the Infra-Red Dual-beam Imaging and Spectroscopy instrument \citep[IRDIS,][]{2008SPIE.7014E..3LD,2014SPIE.9147E..9PL} in Dual-band Polarimetric Imaging (DPI) mode.

The ZIMPOL/P2 observations were performed during the night of March 31, 2015, simultaneously in the \Rp \ ($\lambda_c$\eqsep$=$\eqsep626.3\du nm; $\Delta\lambda$\eqsep$=$\eqsep148.6\du nm; where $\lambda_c$ denotes the central wavelength and $\Delta\lambda$ denotes the full width at half maximum of the filter transmission curve) and \Ip \ ($\lambda_c$\eqsep$=$\eqsep789.7\du nm; $\Delta\lambda$\eqsep$=$\eqsep152.7\eqsep nm) photometric bands. No coronagraph was used. These observations, with integration times of 10\du s per frame and 4 frames per file, are divided in polarimetric cycles. Each cycle contains observations at 4 Half Wave Plate (HWP) angles, 0\degree, 45\degree, 22.5\degree, and 67.5\degree. At each HWP position the two orthogonal polarization states are measured simultaneously\footnote{Thus, eight images per polarimetric cycle are obtained, corresponding to the Stokes components ($I \pm Q$)$/2$, ($I \mp Q$)$/2$, ($I \pm U$)$/2$, and ($I \mp U$)$/2$ respectively.}. Subtraction of one orthogonal state from the other at each of the HWP angles yields the $+Q$, $-Q$, $+U$ and $-U$ Stokes components, respectively. A total of 33 polarimetric cycles were recorded, yielding a total exposure time of 88\du min.

During the same night we also observed \TW \ with IRDIS/DPI in the H-band ($\lambda_c$\eqsep$=$\eqsep1625.5\du nm; $\Delta\lambda$\eqsep$=$\eqsep291\du nm) using an apodized Lyot coronagraph with a radius of 93\du mas. The integration time was 16\du s per frame, and 4 frames were combined before saving. We observed 25 polarimetric cycles with the same 4 HWP angles as for ZIMPOL, resulting in a total exposure time of 107\du min.

In addition, non-coronagraphic IRDIS/DPI observations in the J-band ($\lambda_c$\eqsep$=$\eqsep1257.5\du nm; $\Delta\lambda$\eqsep$=$197\du nm) were performed during the night of May 7, 2015. To minimize saturation, the shortest possible integration time of 0.837\du s was used and 24 frames were combined per saved image. We recorded 16 polarimetric  cycles for a total exposure time of 21.4\du min.

\begin{figure*}[t!]
\hspace{-0.52cm}
\begin{tabular}{cc}
\includegraphics[width=1.06\columnwidth,trim=0 0 0 0, clip]{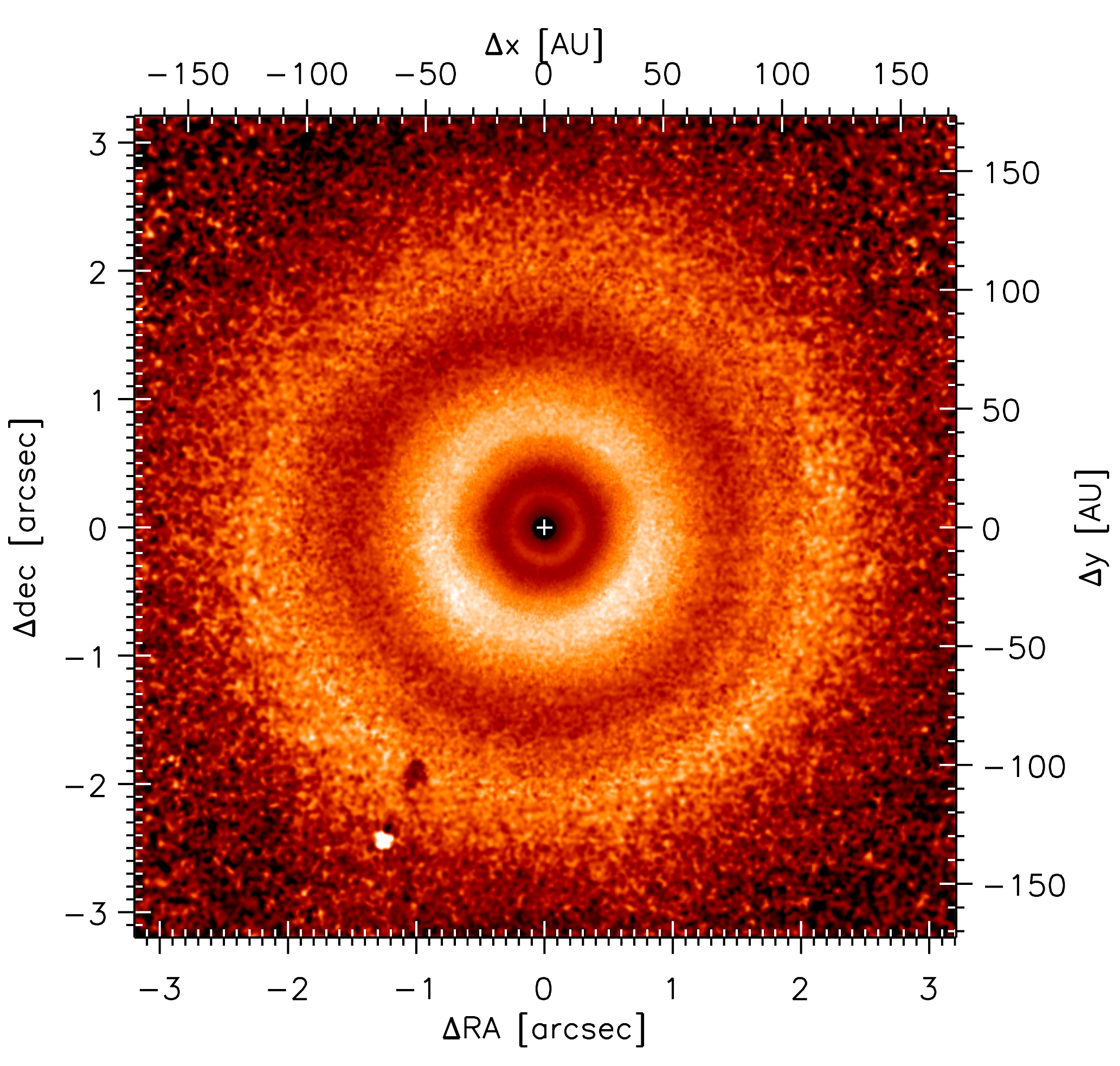} &
\includegraphics[width=1.06\columnwidth,trim=0 0 0 0, clip]{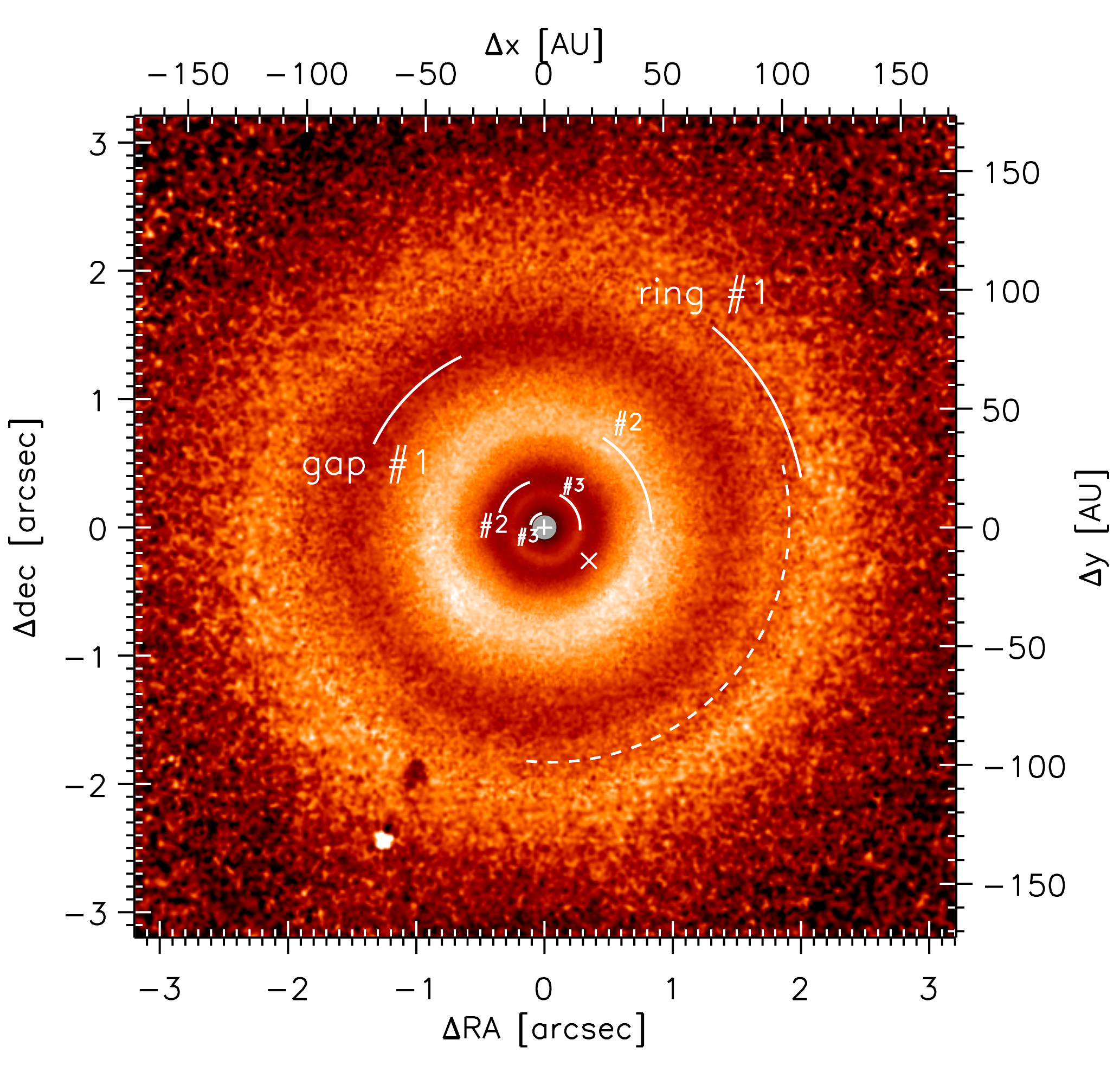} \\
\end{tabular}
\caption{\label{fig:legend}\emph{Left panel}: The \TW \ disk in polarized intensity at \lamH\du\mum, scaled by \rsq. The position of the star is denoted by the $+$ sign. A distance of \dpc\du pc has been adopted. The dark and bright patch near [-1,-2] arcseconds are artefacts. \emph{Right panel}: the same image with annotations. The adopted nomenclature of bright ``rings'' and radial ``gaps'' has been indicated. The region under the coronagraphic mask of 93\du mas radius has been greyed out. The ``dark spiral'' is indicated with a dashed line. The $\times$ symbol denotes the position of the compact \HCOp \ source found by \cite{2015ApJ...799..204C}.}
\end{figure*}

\subsubsection{Angular differential imaging}
\label{sec:adi_observations}

During the course of the night of February 3, 2015, TW Hya was observed during its meridian passage with IRDIS in dual-band imaging \citep[DBI][]{2010MNRAS.407...71V} mode in the atmospheric H-band using an apodized pupil Lyot coronagraph. The pupil stabilized observations allow for angular differential imaging (ADI) observations and were performed using the $H$23 filter pair ($\lambda_c$(H2)\eqsep$=$\eqsep$1588.8$\du nm; $\Delta\lambda$(H2)\eqsep$=$\eqsep$53.1$\du nm; $\lambda_c$(H3)\eqsep$=$\eqsep$1667.1$\du nm; $\Delta\lambda$(H2)\eqsep$=$\eqsep$55.6$\du nm).
The integration time was 64\du s per frame for 64 frames in total. A field rotation of almost 77\degree / was achieved. In addition to the coronagraphic images 15 frames of the unsaturated star (i.e. PSF images) for photometric calibrations, and three sky frames were recorded. A detailed description of the IRDIS/DBI observing sequence can be found in \cite{2016A&A...587A..56M}.

\subsection{Data Reduction}
\label{sec:methods:data_reduction}

\subsubsection{ZIMPOL/PDI \Rp \ and \Ip band}
\label{sec:method:data_reduction:ZIMPOL}
	
The ZIMPOL data were reduced using the Polarimetric Differential Imaging (PDI) pipeline described by de Boer, Girard et al.~(in prep.). The charge shuffling of ZIMPOL allows the quasi-simultaneous detection of two orthogonal polarization states by the same pixels on the detectors \citep{2012SPIE.8446E..8YS}. These authors describe how for subsequent ZIMPOL frames within each file (called the $0$ \& $\pi$ frames) the order is reversed in which the orthogonal polarization components are stored. After dark and flat correction, the first differential images are created for each $0$ \& $\pi$ frame by subtracting the orthogonal polarization components. The two resulting single difference images are both aligned using through fitting a Moffat function, before the $\pi$ single difference image is subtracted from the $0$ single difference. From the mean of the differential images of the first file of each polarimetric cycle the mean differential image of the second file is subtracted to create the Stokes $Q$ image, while the equivalent subtraction of the differential images of the fourth from the third file yields the Stokes $U$ image.

We discard one polarimetric cycle during which the AO loop opened during the exposure and take the median over the $Q$ and $U$ images of the remaining 32 cycles. The stacked $Q$ and $U$ images are corrected for instrumental and (inter-) stellar polarization and, separately, for sky background polarisation with the correction methods described by \cite{2011A&A...531A.102C}. Finally, the azimuthal Stokes components $Q_\phi$ \& $U_\phi$ are computed according to \cite{2014ApJ...781...87A}:
\begin{eqnarray}
Q_\phi  &=&  +Q \cos(2\phi) + U \sin(2\phi), \label{eq:Qphi} \\
U_\phi  &=&  -Q \sin(2\phi) + U \cos(2\phi),  \label{eq:Uphi}
\end{eqnarray}
with
\begin{eqnarray}
\label{eq:theta}
\phi   &=&  \arctan \frac{x-x_{star}}{y-y_{star}} +\theta.
\end{eqnarray}
The angle $\theta$ is the offset of $\phi$\eqsep$=$\eqsep$0$ with the $y$-axis of the image, which is 0 for the ZIMPOL observations. The \Qphi \ image  represents the radial (0\degree, negative signal)  and tangential (90\degree, positive signal) components of the linearly polarized intensity. The \Uphi \ image represents the amplitude of the polarized intensity in the direction 45\degree \ offset from the radial vector. In the idealized case of a face-on disk and single scattering events, the \Qphi \ image will contain only positive signal and be equal to the total linearly polarized intensity image whereas the \Uphi \ image will contain zero signal. For inclined disks, or if a substantial fraction of the photons is scattered more than once, the \Uphi \ image contains substantial signal \citep[e.g.,][]{2015A&A...582L...7C}.

\subsubsection{IRDIS/DPI $J$ \& $H$ band}
\label{sec:method:data_reduction:IRDIS}
After dark subtraction and flat-fielding, the two orthogonally polarized beams are extracted from each individual frame. The two beams are centered and subtracted per frame, after which the median is taken over all frames of one file. Similar to the ZIMPOL data reduction, the differential images of the first and second file of each polarimetric cycle yield the $Q$ image, while the difference of the third and fourth yield the $U$ image.	Both $Q$ and $U$ are corrected for instrumental polarization and sky background polarization.

Field stabilization in the IRDIS/DPI observations is achieved by the adjustment of a derotator in the common path of SPHERE. While reducing this dataset we found that the polarimetric efficiency was strongly affected by crosstalk induced by the derotator. This causes the characteristic ``butterfly pattern'' of the Q and U images to rotate on the detector during an observation. Because the observations are performed in field-tracking mode this rotation shows that the effect is instrumental in nature, as any astrophysical signal would remain stationary. The characterization of this crosstalk is described, together with a detailed description of the reduction of this dataset, by de Boer, van Holstein et al.~(in prep.).  A summary of the approach is given in this section.

The instrumental crosstalk can be corrected by applying the proper value of $\theta$ in \eq~\ref{eq:theta}. If the astrophysical signal has a purely tangential polarization then the \Uphi \ signal reduces to zero at the correct choice of $\theta$. Hence, determining the proper value of $\theta$ by minimizing the resulting \Uphi \ signal in an annulus around the star is a commonly adopted technique. If the astrophysical polarized signal has a non-tangential component then this minimization will generally not yield an ``empty'' \Uphi \ image, because the spatial signature of the astrophysical signal \citep[e.g.,][]{2015A&A...582L...7C} differs from that of the instrumental crosstalk. Conversely, if a \Uphi \ image with approximately zero signal can be obtained by an appropriate choice of $\theta$, then this means the non-tangential component of the astrophysical polarized signal must be very close to zero.

We selected the 18 cycles with the best signal to noise ratio in the H-band observations to be combined into a final image, and used the entire data set of 16 cycles for the $J$-band image. For each polarimetric cycle $i$ we determined the appropriate value for $\theta$ by minimization of the \Uphi \ signal. After each $\theta_i$ was determined, $Q_{\phi,i}$ \& $U_{\phi,i}$ were computed for each polarimetric cycle. The final \Qphi \ and \Uphi \ images were obtained by median combination of all cycles.

\subsection{Data quality and PSF}
\label{sec:results:psf}

The Strehl ratio of the polarimetric observations, defined as the fraction of the total flux that is concentrated in the central Airy disk, relative to the corresponding fraction in a perfect diffraction-limited PSF ($\approx$\ds80\% for a VLT pupil), is approximately 13\% in the \Rp-band and approximately 69\% in the H-band in our data. The PSF core in the \Rp-band is substantially less ``sharp'' than that of a perfect Airy pattern due to the combination of moderate conditions and the relatively low flux of \TW \ in the R-band, where the wavefront sensor operates. The central PSF core is slightly elongated with a \fwhm \ of 56$\times$48\,mas with the major axis oriented $\approx$\ds127\degree \ E of N, and contains $\approx$\ds30\% of the total flux. In comparison, a perfect Airy disk would have a \fwhm \ of 16\,mas. The H-Band data have a nearly diffraction-limited PSF shape with \fwhm\,$\approx$\ds48.5 mas, compared to 41.4 mas for a perfect Airy disk.

\subsection{H-band angular differential imaging}
\label{sec:methods:IRDIS_DPI}
The cosmetic reduction of the coronagraphic images included subtraction of the sky background, flat field and bad pixel correction. One image out of 64 had to be discarded due to an open AO loop. Image registration was performed based on ``star center'' frames which were recorded before and after the coronagraphic observations. These frames display four crosswise replicas of the star, which is hidden behind the coronagraphic mask, with which the exact position of the star can be determined. The modeling and subtraction of the PSF is performed using a principal component analysis (PCA) after \cite{2013A&A...559L..12A}, which is in turn based on \cite{2012ApJ...755L..28S}. We apply the following basis steps: (1) Gaussian smoothing with half of the estimated FWHM; (2) intensity scaling of the images based on the measured peak flux of the PSF images; (3) PCA and subtraction of the modeled noise; (4) derotation and averaging of the images.

\section{Results}
\label{sec:results}

\subsection{A first look at the images}
\label{sec:first_look}

\begin{figure*}[t]
\hspace{-0.5cm}
\includegraphics[width=1.05\textwidth,trim=0 0 0 0, clip]{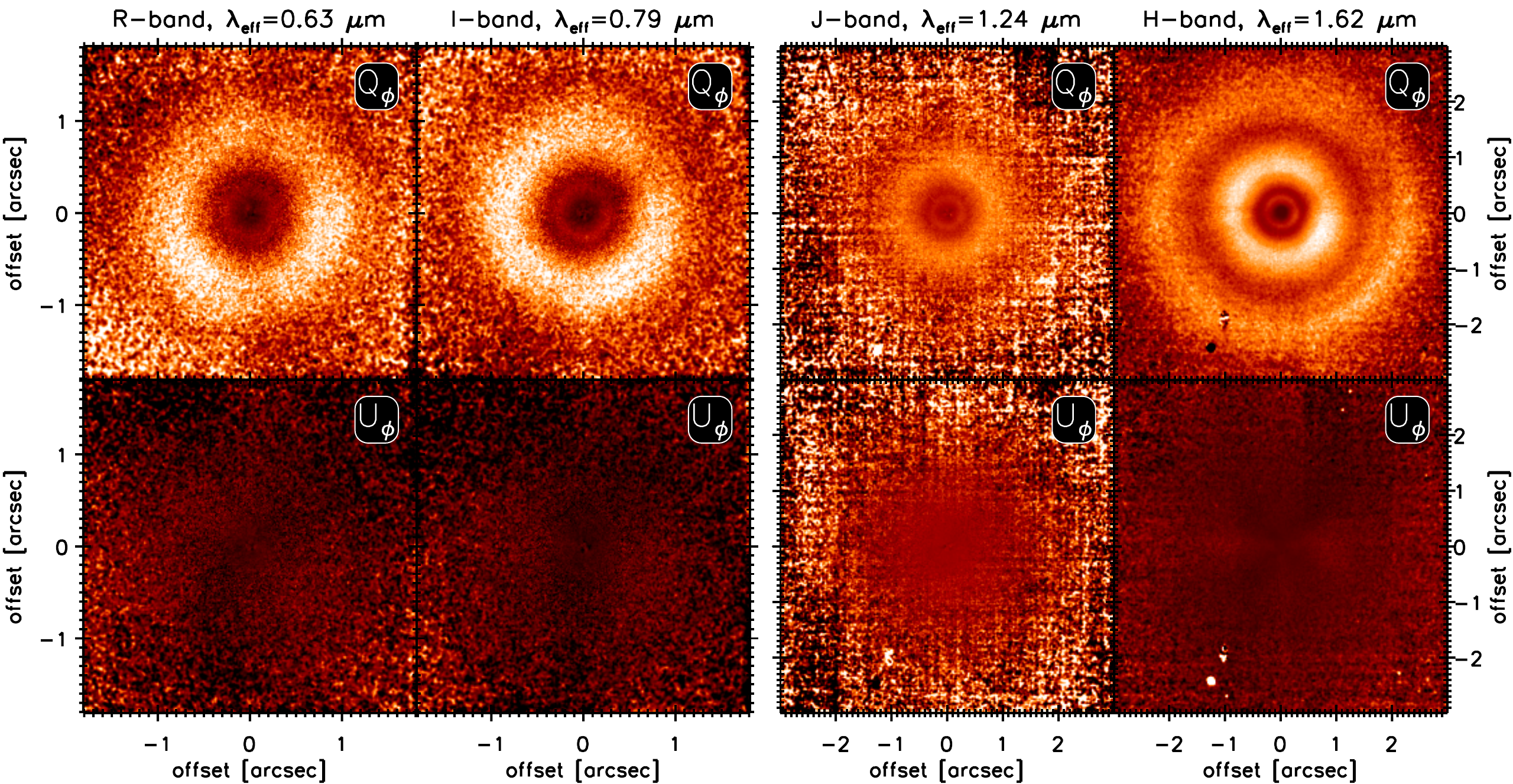}
\caption{\label{fig:overview} \Qphi \ and \Uphi \ images of the \TW \ disk. North is up and East is to the left. The images have been scaled by $R^2$ to correct for the dilution factor of the stellar irradiation and each pair of \Qphi \ and \Uphi \ images in a given band is displayed on the same linear stretch. }
\end{figure*}

In the left panel of \fig\fsep\ref{fig:legend} we show the H-band \Qphi \ image of the \TW \ disk, with an effective wavelength of \lamH\du\mum. In the right panel we show an annotated version of the same image to graphically illustrate the adopted nomenclature and identification of various structural features in the \TW \ disk. The \Qphi \ image contains only positive signal and the corresponding \Uphi \ image (shown in \fig\fsep\ref{fig:overview}) contains approximately zero signal. Thus our data show the expected signature of an approximately face-on disk dominated by single scattering, and the \Qphi \ image of the \TW \ disk closely approximates the total linearly polarized intensity image (see also \sek\seksep\ref{sec:method:data_reduction:IRDIS}). The intensity in each image has been multiplied by $R^2$, where $R$ denotes the projected distance to the central star, in order to correct for the radial dependence of the stellar radiation field. Thus, the displayed image shows how effectively stellar light is scattered into our direction, i.e. approximately directly ``upward'', at each location in the disk.

The intensity distribution is azimuthally very symmetric. The most striking structural features are radial intensity variations in the form of three bright and three dark rings, which we will refer to as ``rings'' \#1 to \#3 and ``gaps'' \#1 to \#3 in this paper. In the $R^2$-scaled \Qphi \ image the intensity of ring \#1 peaks around 115\du\AU, that of ring~\#2 shows a bright plateau between 42 and 49\du\AU, and that of ring~\#3 peaks at 14\du\AU. The gap~\#1 region is faintest around 81\du\AU \ and gap~\#2 around 20\du\AU. Of gap~\#3 we see only the outer flank in the H-band image; the intensity steadily decreases between the peak of ring~\#3 at 14\du\AU \ and the edge of our coronagraphic mask at $\approx$\ds5\du\AU, suggesting that its intensity minimum lies at $\lesssim$\ds6\du\AU. The intensity distribution is substantially affected by PSF convolution, whose effects are largest at small angular separations from the star. We model this in detail (see \sek\seksep\ref{sec:methods:convolution}) and show that gap~\#3 is \emph{not} an artifact of PSF convolution. Convolution does decrease the apparent brightness contrast between the gaps and rings, particularly the inner ones, and conversely the true intensity contrast is higher than that in our images. We choose to not de-convolve our observed images in our analysis, but instead convolve our model images before comparing them to the observations. 

Ring~\#2 is the substantially brighter than the other two rings. In the gap regions the surface brightness is lower than in the rings, but it does not approach zero, indicating that the gap regions are not empty. The southern half of the disk is somewhat brighter than the northern half. A dark, spiral-like feature is seen in ring~\#1, starting roughly 100\du\AU \ south of the star and winding outward counter-clockwise, with a very shallow pitch angle of $\approx$\ds1.5~degrees.

\vspace{0.15cm}

The full SPHERE data set is displayed in \fig\fsep\ref{fig:overview}, where the \Qphi \ and \Uphi \ images taken through the R, I, J, and H filters are shown. Note that the ZIMPOL images (R- and I-band) have a smaller field of view than the IRDIS images (J- and H-band). The images have not been flux calibrated and our analysis focuses on the shape of the intensity distribution. The apparent brightness of the disk increases with wavelength. This is largely due to the red spectral energy distribution (SED) of the central star which leads to approximately 4 times as many photons reaching the disk surface per unit time in the H-band as in the R-band (integrated over the respective bands).
The SNR in the SPHERE data increases with increasing wavelength accordingly, with the exception of the J-band observation which was taken without a coronagraph in order to explore the innermost disk regions, leading to the outer regions being read-out noise limited.
 The scattering behavior of the actual dust particles in the disk is approximately neutral, to slightly blue in the outer disk $R\gtrsim100$\du\AU \ \citep[see also, e.g.,][]{2013ApJ...771...45D}.

\subsection{Radial intensity profiles}
\label{sec:radial_intensity_profiles}

\begin{figure}[t!]
\includegraphics[width=\columnwidth,trim=0 0 0 0, clip]{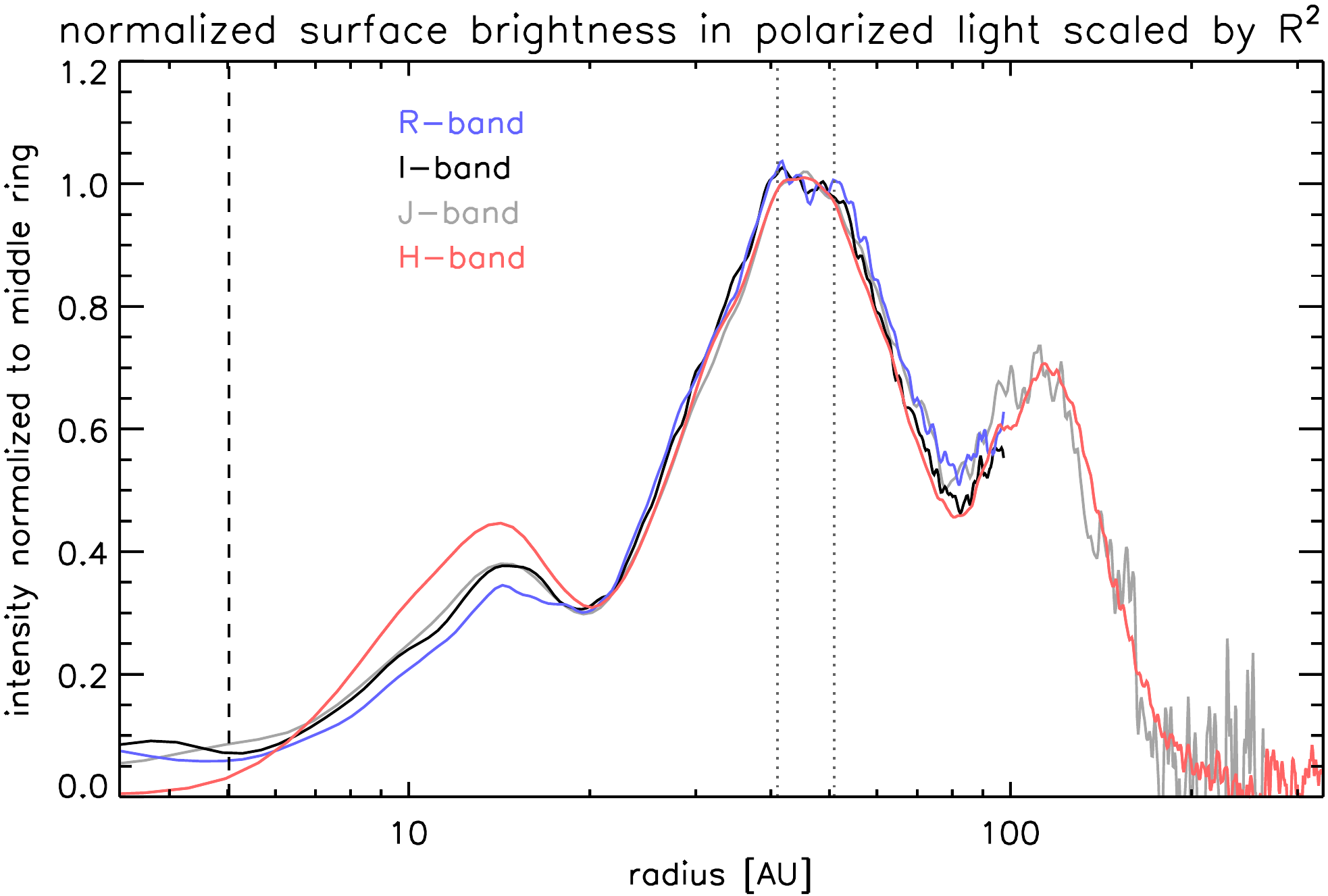}
\caption{\label{fig:radial_profiles} Azimuthally averaged polarized intensity profiles at \lamR, \lamI, \lamJ, and \lamH\du\mum, scaled by \rsq \ and normalized to the average value between 41 and 51\,\AU \ (dotted lines). A distance of \dpc\du pc has been assumed.}
\end{figure}

In \fig\fsep\ref{fig:radial_profiles} we show the radial profiles of the \TW \ disk in polarized intensity, obtained by taking the azimuthal average of our images after scaling by \rsq \ to approximately account for the spatial dilution of the incident stellar radiation reaching the disk surface. The curves have been normalized to the average value between 41 and 51\du\AU, where ring~\#2 has its ``plateau''. The IRDIS H-band data have the highest SNR, followed by the ZIMPOL I-band data. The IRDIS J-band data have a comparatively low SNR at larger radii because they were optimized for the inner disk region, with a setup that is readout noise-limited at larger radii. The ZIMPOL R-band data have lower SNR because \TW \ has a very red spectrum and there are simply fewer photons compared to the longer wavelengths. 

A first inspection of \fig\fsep\ref{fig:radial_profiles} shows that the surface brightness of the three detected bright rings, after scaling by \rsq, is profoundly different: the middle ring (\#2) is brightest, followed by the outer ring (\#1), and the innermost detected ring (\#3) is faintest. When accounting for PSF convolution (see \sek\seksep\ref{sec:methods:convolution}), ring~\#1 and \#3 are approximately equally bright in the underlying true intensity distribution, each at $\approx$\ds60\% of the surface brightness of ring~\#2 after $R^2$ scaling. Furthermore, the polarized intensity profiles show little wavelength dependence. The outer ring (\#1) appears to be slightly brighter towards shorter wavelength, i.e. it is slightly ``blue'' in scattering behavior. The apparent ``red'' color of ring~\#3 is an artefact of PSF convolution and due to the higher Strehl ratio at longer wavelengths.

\begin{figure}[t!]
\hspace{-0.03\columnwidth}
\includegraphics[width=1.03\columnwidth,trim=0 0 0 0, clip]{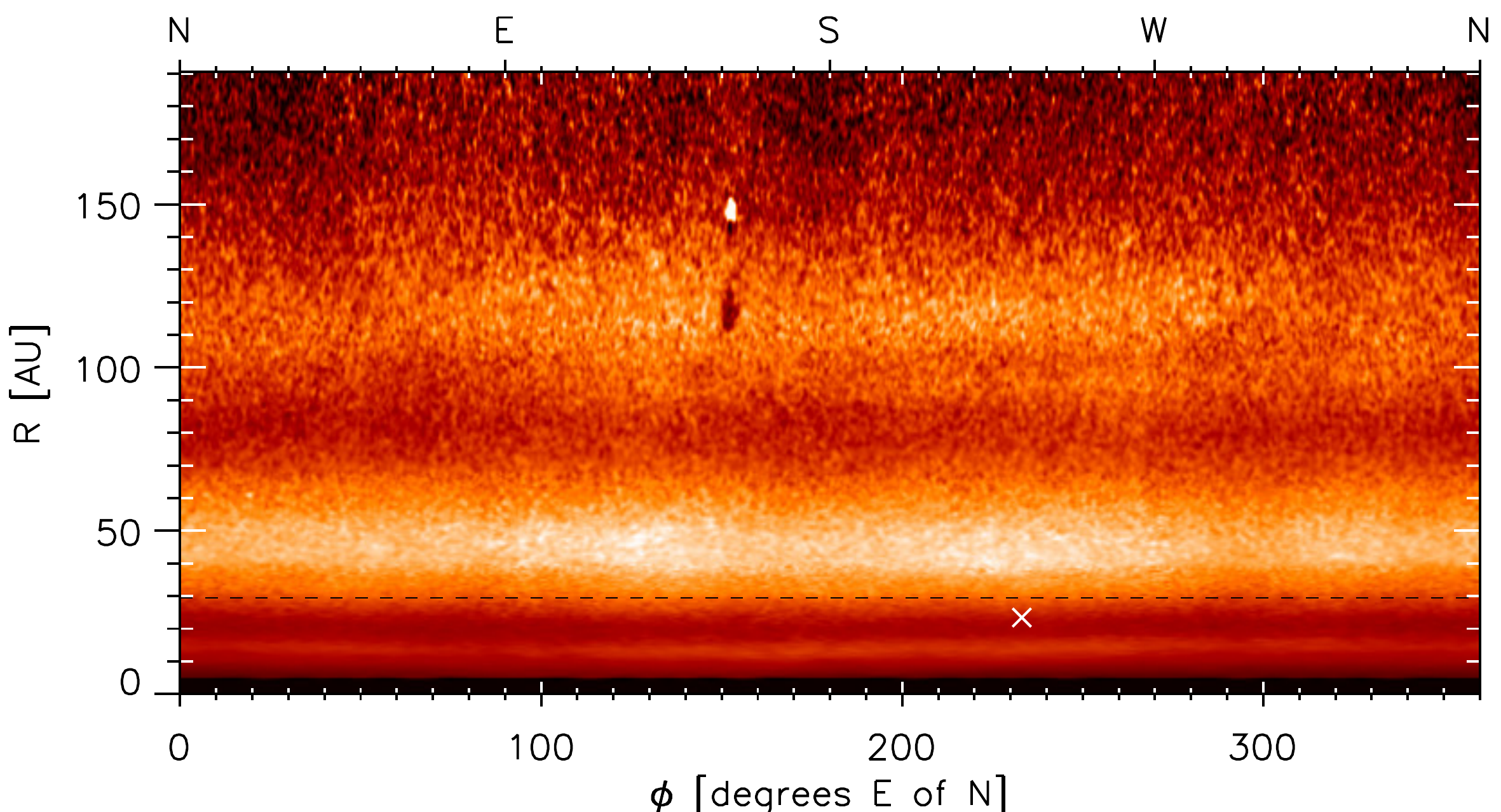}
\caption{\label{fig:irdis_polar} H-band \Qphi \ image of the \TW \ disk in polar projection. The black dashed line indicates the position of the CO iceline in the midplane \citep{2013Sci...341..630Q}, the $\times$ symbol denotes the position of the compact \HCOp \ source detected by \cite{2015ApJ...799..204C}. A distance of \dpc\du pc has been assumed.}
\end{figure}

\begin{figure*}[t!]
\hspace{-0.1cm}
\begin{tabular}{cc}
\includegraphics[width=0.40\textwidth,trim=0 0 0 0, clip]{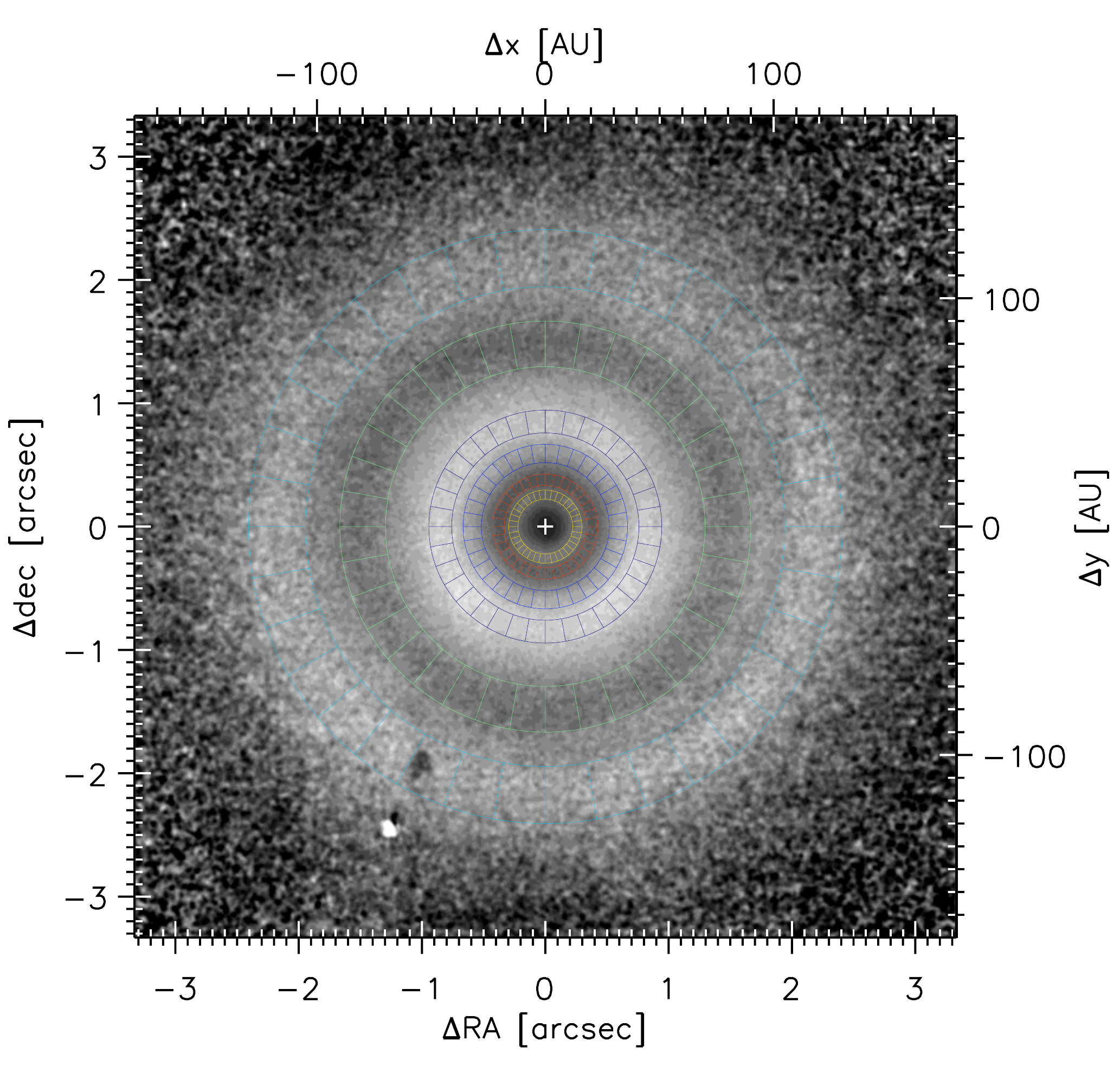} &
\includegraphics[width=0.57\textwidth,trim=0 0 0 0, clip]{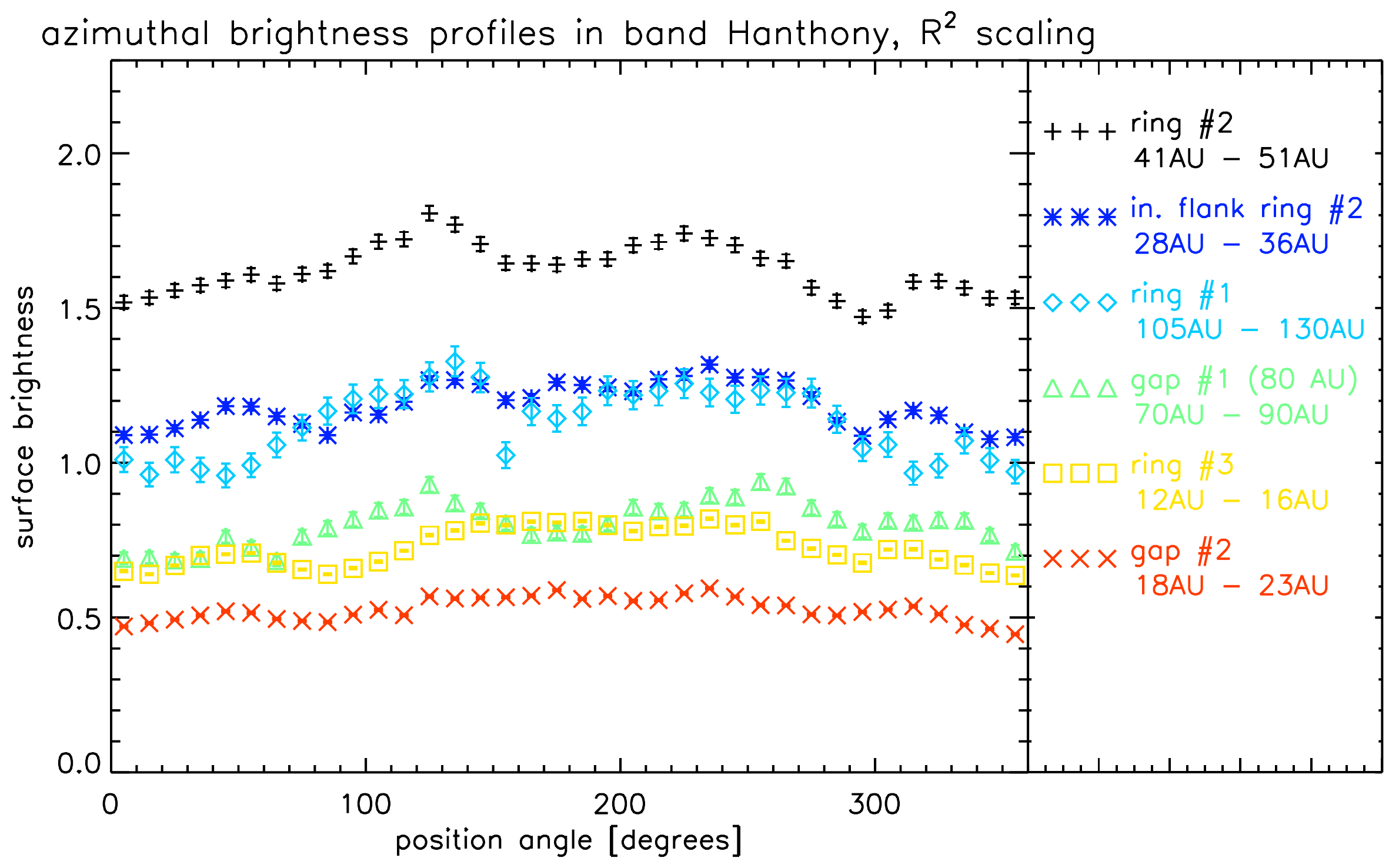} \\
\end{tabular}
\caption{\label{fig:azimuthal_profiles} Azimuthal intensity profiles in the H-band, radially integrated over the gap and ring regions as indicated in the image. A distance of \dpc\du pc has been assumed.}
\end{figure*}

\subsection{Azimuthal intensity profiles}
\label{sec:azimuthal_structure}

The intensity distribution of the \TW \ disk in scattered light shows a high degree of azimuthal symmetry. The strong sub-structure seen in some other disks, such as spiral arms, is not present in \TW. Nonetheless, there are significant azimuthal brightness variations. These can be seen in both the ZIMPOL and IRDIS data in \fig\fsep\ref{fig:legend} and \fig\fsep\ref{fig:overview}, and are also illustrated in \fig\fsep\ref{fig:irdis_polar} where we show the H-band image in polar projection, as well as in \fig\fsep\ref{fig:azimuthal_profiles} where we show azimuthal brightness profiles covering six annuli that correspond to the 3 ``rings'' and 3 ``gaps'' we identified.

The Southern half of the disk is brighter than Northern half, and the disk is faintest towards the North-West. This is seen at multiple wavelengths, most clearly in the \Ip- and H-band data (the \Rp- and J-band data have lower SNR). The disk shows particularly bright regions towards the South-West and towards the South-East. The darker and brighter regions are most clearly seen in ring\,\#2 because the disk is brightest there, but they can be seen over much of the radial extent of the disk. The azimuthal brightness profiles extracted at the various radii are similar.

\subsection{Material in the inner few \AU?}
\label{sec:results:innerdisk}

\begin{figure*}[t!]
\begin{tabular}{cc}
\hspace{-0.08\columnwidth}
\includegraphics[width=0.6\textwidth,trim=0 0 0 0, clip]{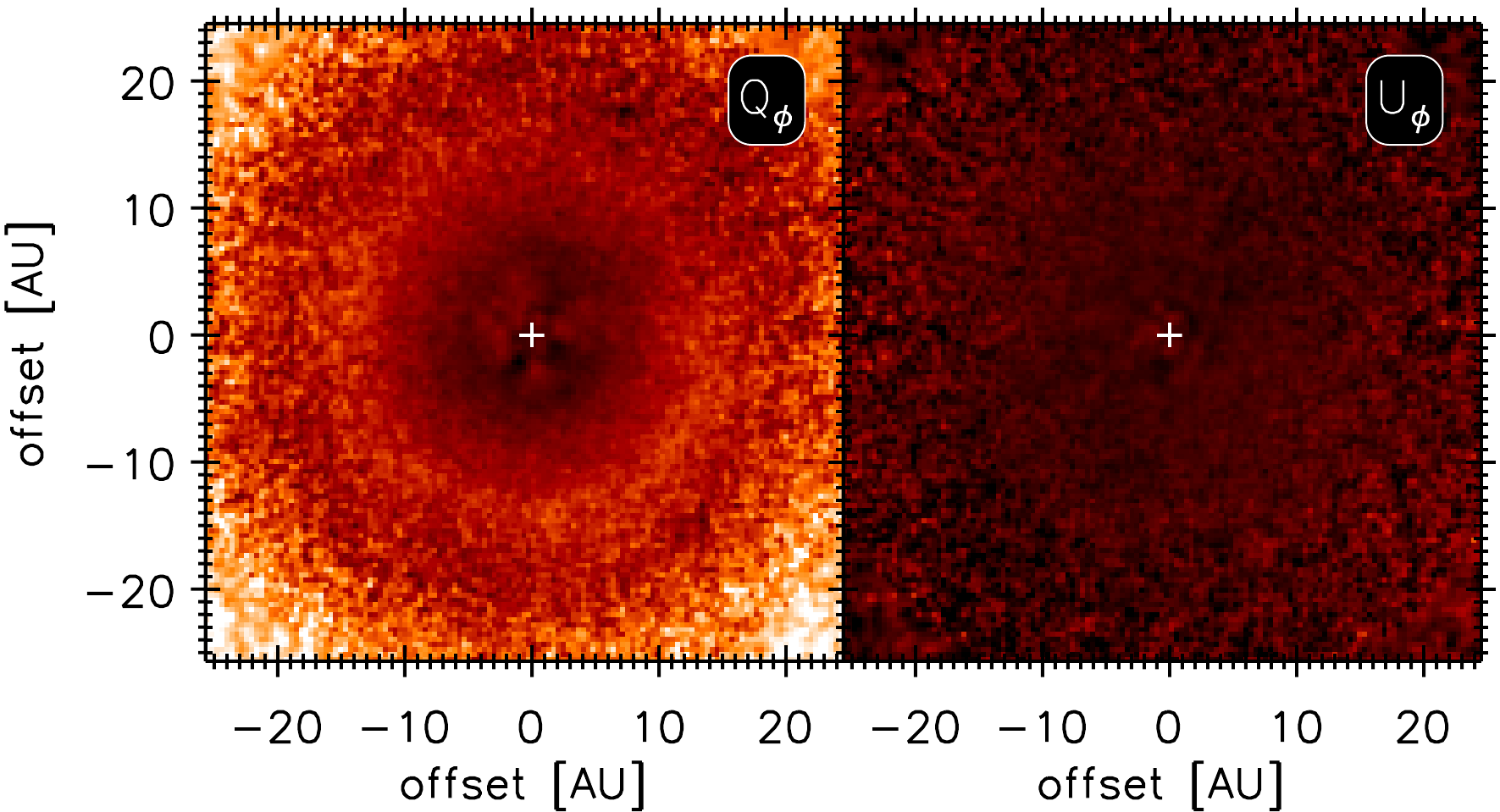} &
\includegraphics[width=0.4\textwidth,trim=0 0 0 0, clip]{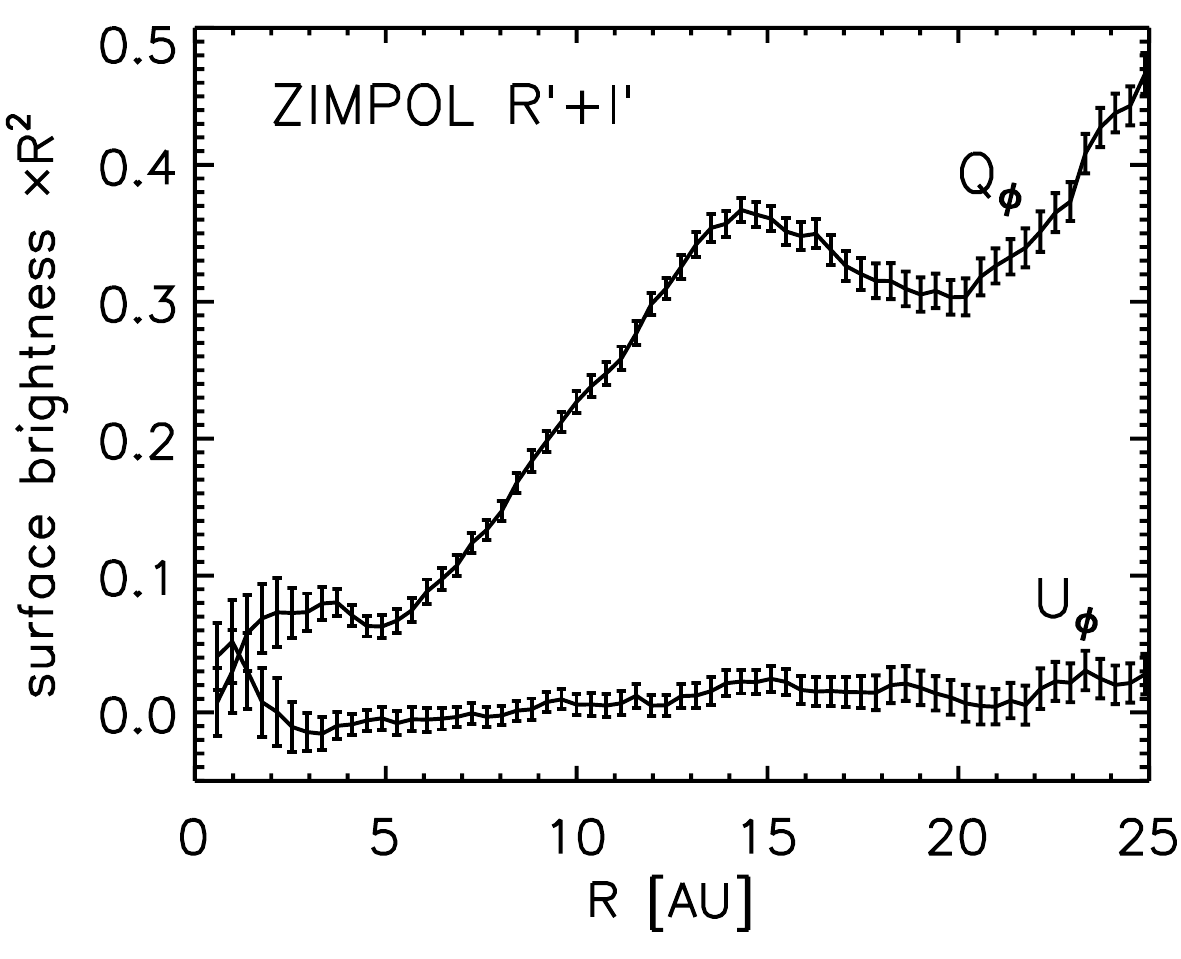}
\end{tabular}
\caption{\label{fig:inner_disk} Zoom-in on the central disk region as observed with ZIMPOL. We show the sum of the \Rp \ and \Ip \ data, scaled by \rsq. A distance of \dpc\du pc has been assumed. \emph{Left panel}: the \Qphi \ and \Uphi \ images on the same linear stretch. The position of the star is marked with the $+$ sign. \emph{Right panel}: the corresponding radial intensity profiles. A distance of \dpc\du pc has been assumed.}
\end{figure*}

In \fig\fsep\ref{fig:inner_disk} we show the ZIMPOL data of the innermost disk regions. We have combined the \Rp- and \Ip-band data to achieve optimum SNR, and show the resulting \Qphi \ and \Uphi \ images on the same linear scale so they can be directly compared. Part of ring \#2 is seen near the edge of the FOV, gap \#2 and ring \#3 are completely within the field. In the innermost regions that are covered by the coronagraph in the H-band data ($R<5$\du\AU) there is some signal in the \Qphi \ image, even though this signal does not look like the nice smooth rings seen at larger radii. In the \Uphi \ image there is no signal, except in the central 1 or 2~\AU. In the right panel of \fig\fsep\ref{fig:inner_disk} we show azimuthally averaged brightness profiles of the \Qphi \ and \Uphi \ images. Here the emission is seen more clearly and, taking the strength of the signal in the \Uphi \ image as a measure for the uncertainty, the signal in the \Uphi \ image is significant down to about 2\du\AU. This supports that we have indeed detected the inner disk, but further observations are needed to constrain the intensity profile more accurately.

\subsection{Comparison to earlier scattered light observations}
\label{sec:results:comparison_previousWork}

\TW \ has been the subject of quite a number of scattered light imaging experiments, both in total intensity with the HST as well as in polarized light using ground-based facilities, and both at optical and near-infrared wavelengths. 

A comprehensive overview of the total intensity observations was presented by \cite{2013ApJ...771...45D}, later complemented by a refined analysis of a subset of these data \citep{2016ApJ...819L...1D}. Combined, these papers present images and radial intensity profiles at 7 wavelengths between 0.48\du\mum \ and 2.22\du\mum \ obtained with the NICMOS and STIS instruments onboard the HST, probing radii from 22 to 160\du\AU. A comparison of these total intensity profiles to our polarized light profiles is insightful. 

The filter sets of our study and that of \cite{2013ApJ...771...45D} are different. Our H-band filter and the NICMOS F160W filter provide a very close match in central wavelength, with the F160W filter having a slightly larger bandwidth. Our J-band filter is closest to the NICMOS F110W filter and is of similar bandwidth but has a central wavelength of \lamJ\du\mum, compared to 1.10\du\mum \ for the F110W filter. The STIS data have a very broad band that encompasses both our R- and I-band filters. The STIS central wavelength is quoted as 0.58\du\mum \ but due to the very red spectrum of \TW \ the effective wavelength of the STIS image is longer, and the image can be compared to our \Rp \ and \Ip \ data.

The shape of the SPHERE \Rp \ and \Ip \ radial intensity profiles of the polarized light matches that of the STIS profile in total intensity closely. There is therefore no substantial radial dependence of the fractional polarization of the scattered light at these optical wavelengths. The ``plateau'' in ring~\#2 seen in all four SPHERE bands is not seen in the STIS data, which instead show a much more round peak of ring~\#2, and thus possibly indicate a locally lower degree of polarization.

The comparison of the SPHERE H-band and NICMOS F160W data yields somewhat less clear results. The overall shapes are similar, but the contrast between the ring and gap regions is substantially lower in the NICMOS data. In particular ring~\#2 appears substantially less bright in the F160W data and its peak appears truncated. The NICMOS F222 data presented by \cite{2013ApJ...771...45D} and \cite{2016ApJ...819L...1D} yield a much better match to the SPHERE and STIS data, as well as to the K-band data of \cite{2015ApJ...815L..26R}. Given the good agreement between the data obtained with STIS, ground-based facilities, and with NICMOS at longer wavelength, it appears likely that the F160W data are affected by systematic effects from imperfect subtraction of the stellar light. The latter is much more challenging in total intensity observations than in polarimetric imaging.

\vspace{0.15cm}
In addition to the HST observations \TW \ has been a popular target for ground-based, polarimetric imaging. \cite{2015ApJ...802L..17A} used the HiCIAO system on the Subaru telescope in the H-band, detecting a radial depression in the surface brightness of light scattered off the disk around 20\du\AU \ (``gap~\#2'' in our nomenclature). The H-band intensity profiles derived from the SPHERE and HiCIAO data are similar, though the contrast between gaps and rings is substantially higher in the SPHERE data. The profile of the HiCIAO data at radii beyond $\approx$\ds50\du\AU \ is noisy and the match between the HiCIAO and SPHERE data is less good. When the flux levels of the SPHERE H-band data are matched to those in the HiCIAO data and compared to the NICMOS F160W total intensity data, a fractional polarization of $\approx$\ds35\% is obtained.

\cite{2015ApJ...815L..26R} presented polarimetric images in the J- and K-bands obtained with the GPI instrument \citep{2014SPIE.9148E..0JM}. These data have a strongly improved contrast performance compared to the study by \cite{2015ApJ...802L..17A}, and confirm the 20\du\AU \ gap. \cite{2015ApJ...815L..26R} compare their measured profiles to predictions of hydrodynamical calculations of planet-disk interaction \citep{2015ApJ...809...93D} and conclude the observed profile is consistent with that expected from a $\approx$\ds0.2\du M$_{\rm{Jup}}$ planet embedded in the disk. They see signs of gap~\#3 in their data but could not establish its reality, arguing instead that they are likely dealing with an instrumental artifact. 

Our new SPHERE data present the next step forward, securely detecting the disk to both smaller and larger radii than the \cite{2015ApJ...815L..26R} data, at better contrast, and at down to a factor of 2 shorter wavelength. The radial intensity profile we measure in the H-band is very similar to that of the J- and K-band curves by \cite{2015ApJ...815L..26R}, except that the gaps in our data are notably ``deeper'' due to the AO performance of SPHERE and the associated contrast performance. 

The reality of gap~\#3 inward of 10\du\AU \ is unquestionable in our SPHERE data: we independently detect it in all four photometric bands, both with (H-band) and without (\Rp-,\Ip-, and J-band) a coronagraph, and with consistent profiles after the wavelength dependence of the PSF is taken into account.

\subsection{Comparison to ALMA observations}
\label{sec:alma_coparison}

\begin{figure}[t!]
\includegraphics[width=\columnwidth,trim=0 0 0 0, clip]{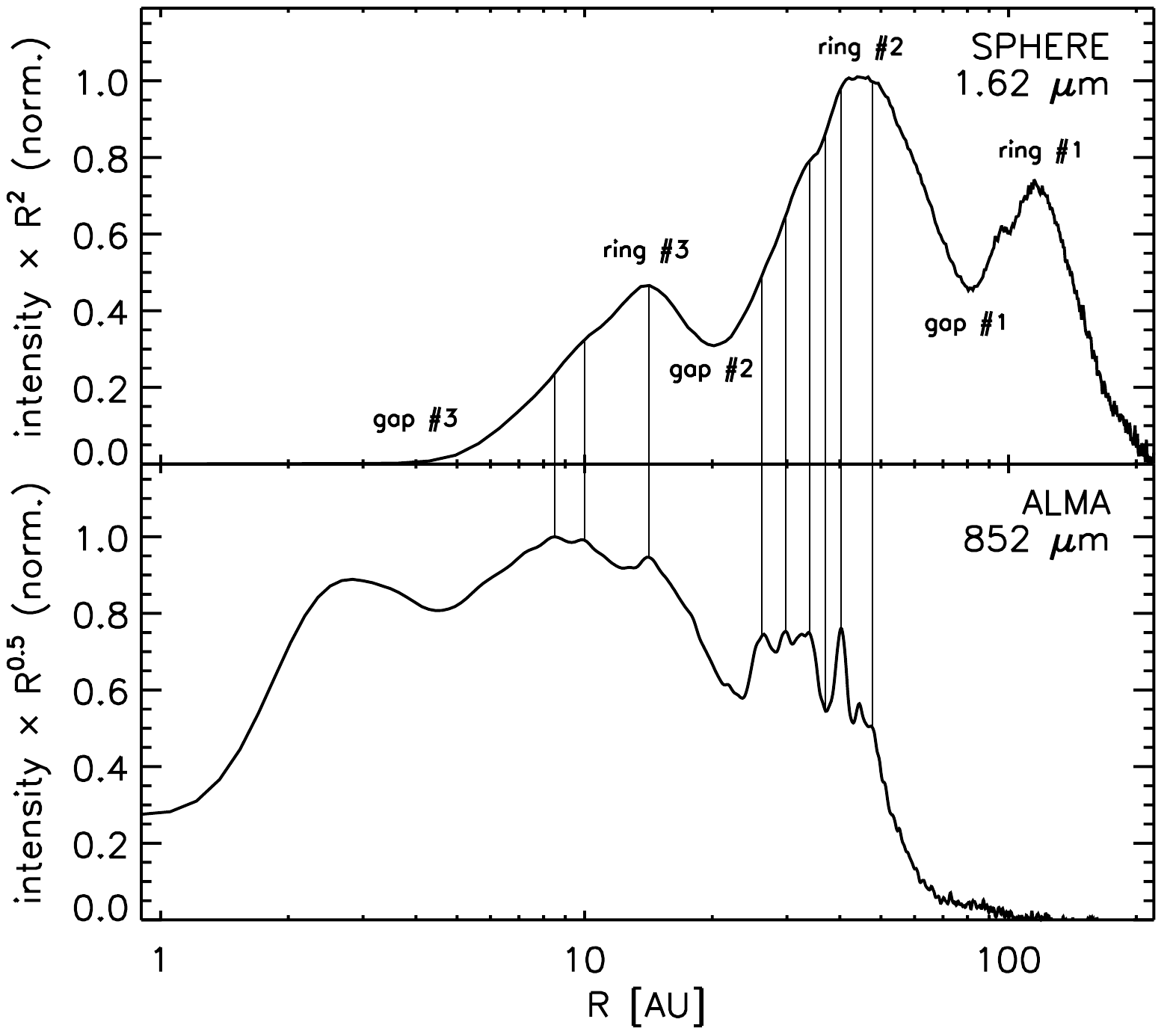}
\caption{\label{fig:alma_comparison} SPHERE H-band scattered light profile scaled with \rsq \ (top panel) and ALMA 852\du\mum \ continuum intensity profile scaled with $\sqrt R$. See \sek\seksep\ref{sec:alma_coparison} for a discussion. A distance of \dpc\du pc has been assumed.}
\end{figure}

\cite{2016ApJ...820L..40A} recently observed the \TW \ disk with ALMA in Band~7 with a mean frequency and bandwidth of 352 \ and 6.1\du GHz (852 and 14.8\du\mum), respectively. These observations trace the dust continuum emission which at these wavelengths is expected to be dominated by a population of roughly millimeter-sized grains that are concentrated near the midplane of the disk. Our SPHERE observations, on the other hand, probe the sub-micron sized dust population in the disk surface that, as we will argue in \sek\seksep\ref{sec:methods:gas_dust_coupling}, closely follow the bulk gas density. Here we do a first comparison of the distributions of both components.

In \fig\fsep\ref{fig:alma_comparison} we show the radial intensity distribution of the SPHERE H-band image (top panel) and the 852\du\mum \ dust continuum emission seen with ALMA (bottom panel). The SPHERE intensity curve has the usual \rsq \ scaling to correct for the dilution factor of the disk irradiation. The ALMA curve has been scaled with $\sqrt R$; in this way the observed intensity distribution more closely approximates the large dust surface density distribution\footnote{If the temperature distribution near the midplane would follow an $R^{-1/2}$ profile, the dust would be vertically optically thin, and the dust opacity would be constant with radius, then the $\sqrt R$-scaled curve in \fig\fsep\ref{fig:alma_comparison} would directly correspond to the surface density distribution. None of these assumptions holds exactly, but this simple scaling is nevertheless helpful in the current qualitative comparison. We properly model the temperature and optical depth effects later on in this work with radiative transfer calculations, but also there we do not consider a radially variable dust opacity.}. The region at $R<5$\du\AU \ is hidden behind the coronagraph in the SPHERE data. The linear resolution of the ALMA data is $\approx$\ds1.1\du\AU \ and that of the SPHERE H-band data is $\approx$\ds2.6\du\AU. Whereas the small grains are seen out to 200\du\AU, in agreement with the disk radius as seen in CO lines, the emission from large grains is mostly limited to the central 60\du\AU \ of the disk. 

The ALMA data show a number of local variations in the radial intensity profile which are azimuthally highly symmetric and form nearly perfect rings \citep[cf. Figs.\fsep 1 and 2 of][]{2016ApJ...820L..40A}. We draw vertical lines between a number of these features in the ALMA intensity profile in \fig\fsep\ref{fig:alma_comparison} and the SPHERE intensity curve at the corresponding locations.

The ALMA data show a distinct shoulder at 48\du\AU, beyond which the intensity steeply drops off before falling below the noise level at around 100\du\AU. A distinct ring is seen at 40\du\AU, and this ring and shoulder approximately coincide with the inner and outer radius of the plateau of ring~\#2 in the SPHERE data. The intensity minimum around 37\du\AU \ in the ALMA profile roughly coincides with a small ``dip'' in the SPHERE profile in the inner flank of ring~\#2. Note that the SPHERE data show a qualitatively similar but more profound dip in the inner flank of ring~\#1 at 100\du\AU. The ALMA data show a plateau between 26 and 34\du\AU, with some sub-structure. The outer shoulder of this plateau coincides with the inner edge of the aforemetioned dip in the SPHERE profile; the rest of the plateau aligns with the outer flank of gap~\#2 in the SPHERE profile where no sub-structure is seen. The minimum in the ALMA curve around 24\du\AU \ lies substantially further out than the minimum of gap~\#2 in the SPHERE profile at 20\du\AU. The distinct shoulder in the ALMA profile at 14\du\AU \ coincides nearly perfectly with the peak of ring~\#3 in the SPHERE profile. A less prominent shoulder in the ALMA profile at 10\du\AU \ coincides with a subtle shoulder in the SPEHRE profile at the same location. The minimum in the ALMA profile at 4-5\du\AU \ plausibly coincides with a minimum in the SPHERE ZIMPOL observations (see \fig\fsep\ref{fig:inner_disk}), though the ZIMPOL detection in the central few \AU \ of the disk remains tentative at this point (see \sek\seksep\ref{sec:results:innerdisk}).

Overall, there is quite some correspondence between the structural features seen in the ALMA and SPHERE data, even though they probe vertically very different disk regions. There are large differences, too. In particular the outer disk appears devoid of large dust as shown by the complete absence of SPHERE ring~\#1 in the ALMA profile. As a general observation, the brightness contrast between the large scale gaps and rings is higher in the SPHERE data, whereas the smaller scale features are more profound in the ALMA data.

\subsection{Upper limits on the brightness of point sources}
\label{sec:methods:ADI_non_detection}

\begin{figure*}[t!]
\begin{center}
\begin{tabular}{ccc}
\hspace{-0.4cm}
\includegraphics[height=5.6cm,trim=0 0 0 0, clip]{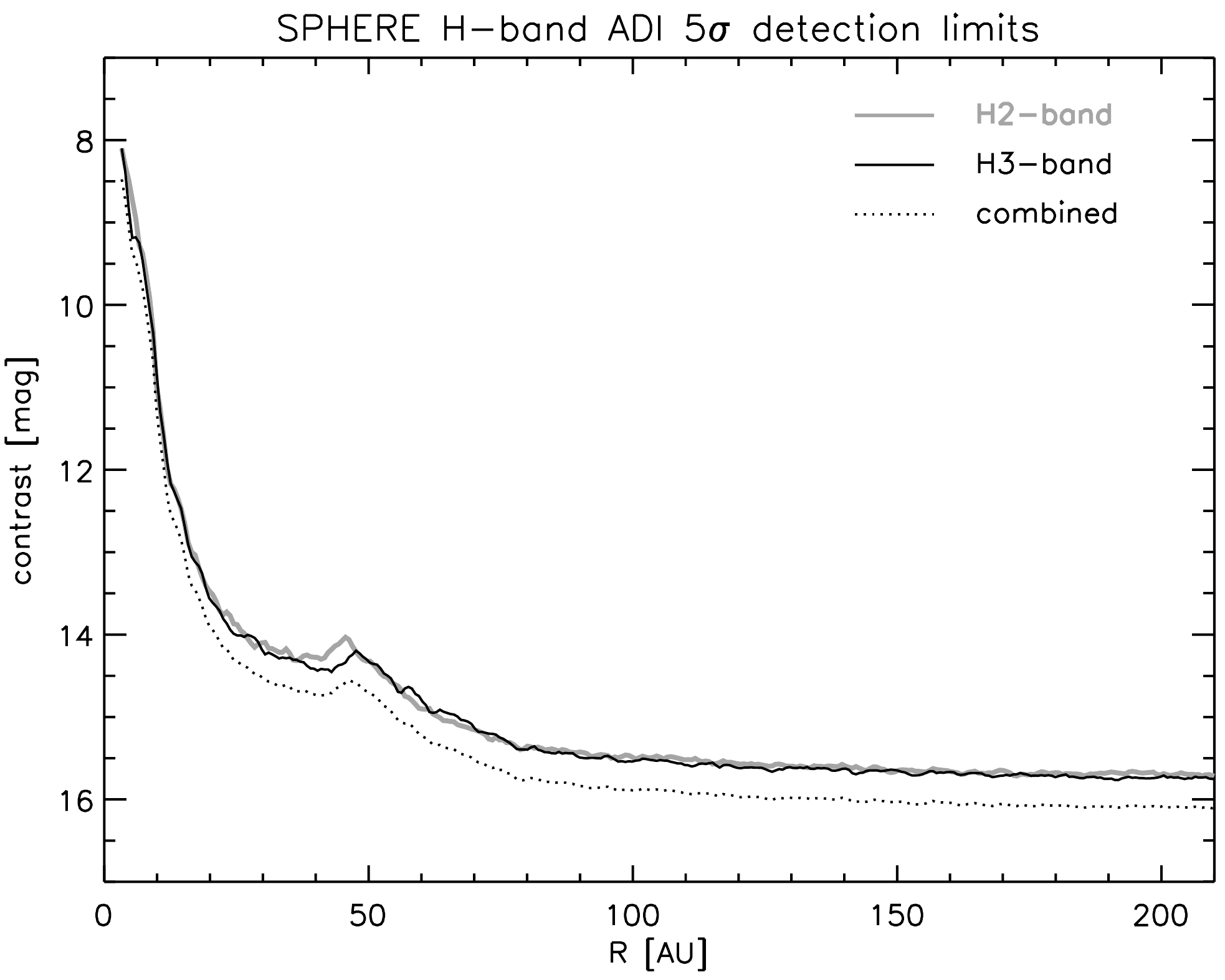} &
\includegraphics[height=5.4cm,trim=0 0 143 0, clip]{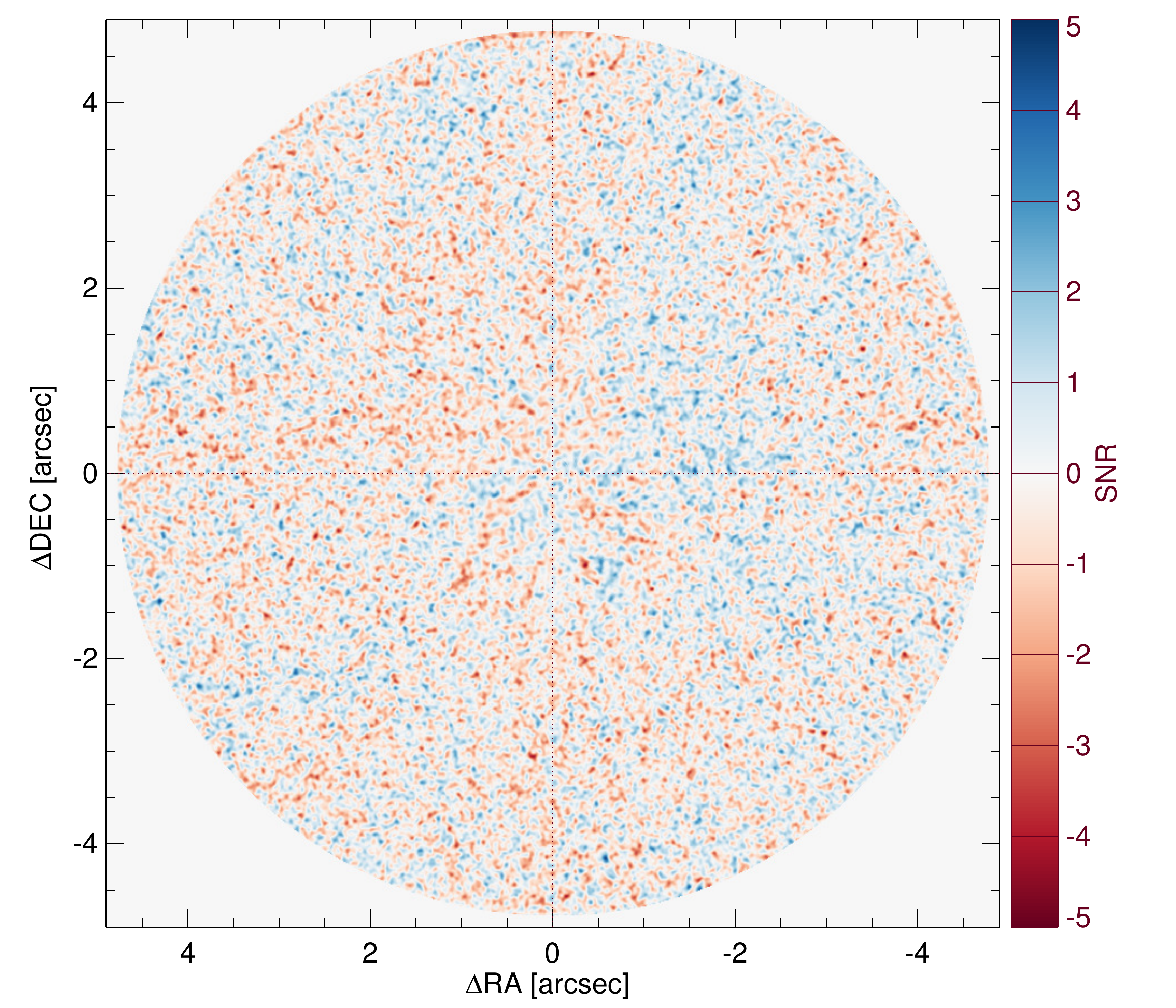} &
\hspace{-0.4cm}
\includegraphics[height=5.4cm,trim=85 0 0 0, clip]{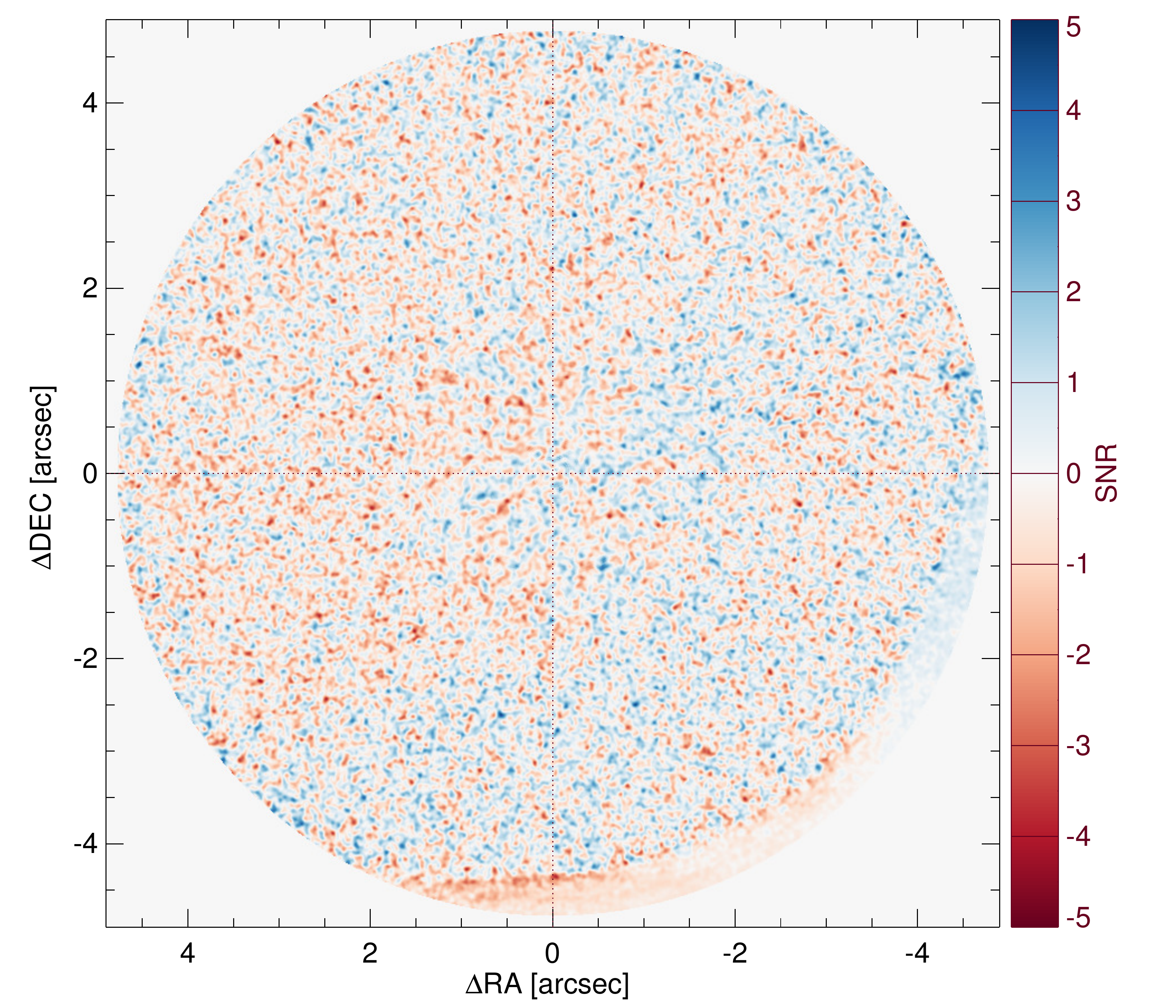} \\
\end{tabular}
\end{center}
\caption{\label{fig:adi_contrast} Summary of the IRDIS H2/H3-band angular differential imaging experiment. \emph{Left}: achieved 5$\sigma$ contrast curves. \emph{Right} SNR maps obtained in the H2 and H3 filters. A distance of \dpc\du pc has been assumed.}
\end{figure*}

No point sources were detected in the disk of \TW \ in the angular differential imaging experiment we performed. In \fig\fsep\ref{fig:adi_contrast} we show the achieved 5$\sigma$ contrast curves obtained in the H2 and H3 photometric bands of IRDIS. These were obtained by measuring the radial noise profiles. At each radial distance we placed apertures of the size of one resolution element and measured the noise inside these apertures. We applied a correction to the values following \cite{2014ApJ...792...97M} to account for small sample statistics, relevant at small radial distances. In addition, we injected artifical planetary signals at three different position angles and at all radii in the images to quantify the effect of self-subtraction in the ADI processing, and corrected the contrast curves accordingly.

Also shown in \fig\fsep\ref{fig:adi_contrast} are the ``detection maps'' in the H2 and H3 bands. These show the signal after ADI processing, relative to the local noise level, at each location around the star. 

\section{Physical disk modeling}
\label{sec:methods:RT}

Motivated by the strong radial variations in the surface brightness and the high degree of azimuthal symmetry in our observations, we focus our analysis on explaining the radial variations. In the following we will argue that the sub-micron sized dust grains that dominate the scattered light are well coupled to that and can thus be used as a tracer to address the central question ``what is the bulk surface density profile of the \TW \ disk?''.

To this purpose we developed a radiative transfer (RT) model of the \TW \ disk with the following prime characteristics: (1) self-consistent, iterative temperature and vertical structure calculation; (2) grain size-dependent dust settling; (3) full non-isotropic scattering; (4) independent distributions of coupled gas and small dust as traced by the SPHERE observations on the one hand, and large dust as traced by the ALMA observations on the other.

\subsection{Gas-dust coupling of small grains}
\label{sec:methods:gas_dust_coupling}

In this section we review the dynamical coupling of dust particles to the gas in a disk. Our goal here is to ensure the validity of our fundamental assumption that the small dust particles dominating the scattered light signal are well coupled to the gas. The Stokes number $\mathrm{St}$ of a dust particle can be expressed as:
\begin{eqnarray}
\label{eq:Stokes_number}
\mathrm{St}  & =  & t_{\rm{stop}} \Omega  \\
t_{\rm{stop}} & = & \frac{\rho_{\rm{b,d}} a}{\rho_{\rm{g}} v_{\rm{therm}}} \\
v_{\rm{therm}} & = & c_{\rm{s}} \sqrt{8/\pi}
\end{eqnarray}
where $t_{\rm{stop}}$ is the particle's stopping time, $\Omega$ is the orbital frequency, $\rho_{\rm{b,d}}$ is the bulk density of the particles ($\rho_{\rm{b,d}} = \rho_{\rm{m,d}}(1-\phi_{\rm{p}})$, where $\rho_{\rm{m,d}}$ is the material density which typically is $\approx$\ds3~g\,cm$^{-1}$ and $\phi_{\rm{p}}$ is the porosity, i.e. the volume fraction of vacuum within the particle), $a$ is the radius of the assumed spherical particle, $\rho_{\rm{g}}$ is the local gas density, $v_{\rm{therm}}$ is the thermal velocity of gas particles, and $c_{\rm{s}}$ is the sound speed.

The radial drift velocity $v_{\rm{R}}$ of a spherical particle of radius $a$ can be approximated by:
\begin{eqnarray}
\label{eq:radial_drift}
v_\mathrm{R} & = & -2 \eta v_\mathrm{k}/(\mathrm{St} + \mathrm{St}^{-1}\,(1+\epsilon)^2) \\
\eta         & = & - \frac{c_{\rm{s}}^2}{2 v_{\rm{k}}^2} \frac{d \mathrm{ln} P}{d \mathrm{ln} R} \\
\label{eq:epsilon}
\epsilon     & = & \rho_{\rm{d}} / \rho_{\rm{g}}
\end{eqnarray}
Here, $v_{\rm{k}}$ is the kepler speed, $P$ is the gas pressure, $\rho_{\rm{d}}$ and $\rho_{\rm{g}}$ are the local dust and gas densities. In this section we evaluate these expressions for small dust particles in the surface layer at all radii. We explore both the case of compact grains with $a$\ds$=$\ds0.1\du\mum \ as well as fractal aggregates with a volume equivalent radius\footnote{The volume equivalent radius $a_{\rm{V}}$ is the radius that the particle would have if it were compact, i.e. the actual particle radius $a$\ds$=$\ds$a_{\rm{V}} / \sqrt[\leftroot{-2}\uproot{3}3]{(1-\phi_{\rm{p}})}$.} of $a_{\rm{V}}$\ds$=$\ds1.0\du\mum. In the discussion section we will apply the same formalism to investigate dynamical effects on larger particles near the disk midplane.

\begin{figure}[t!]
\includegraphics[width=\columnwidth,trim=0 0 0 0, clip]{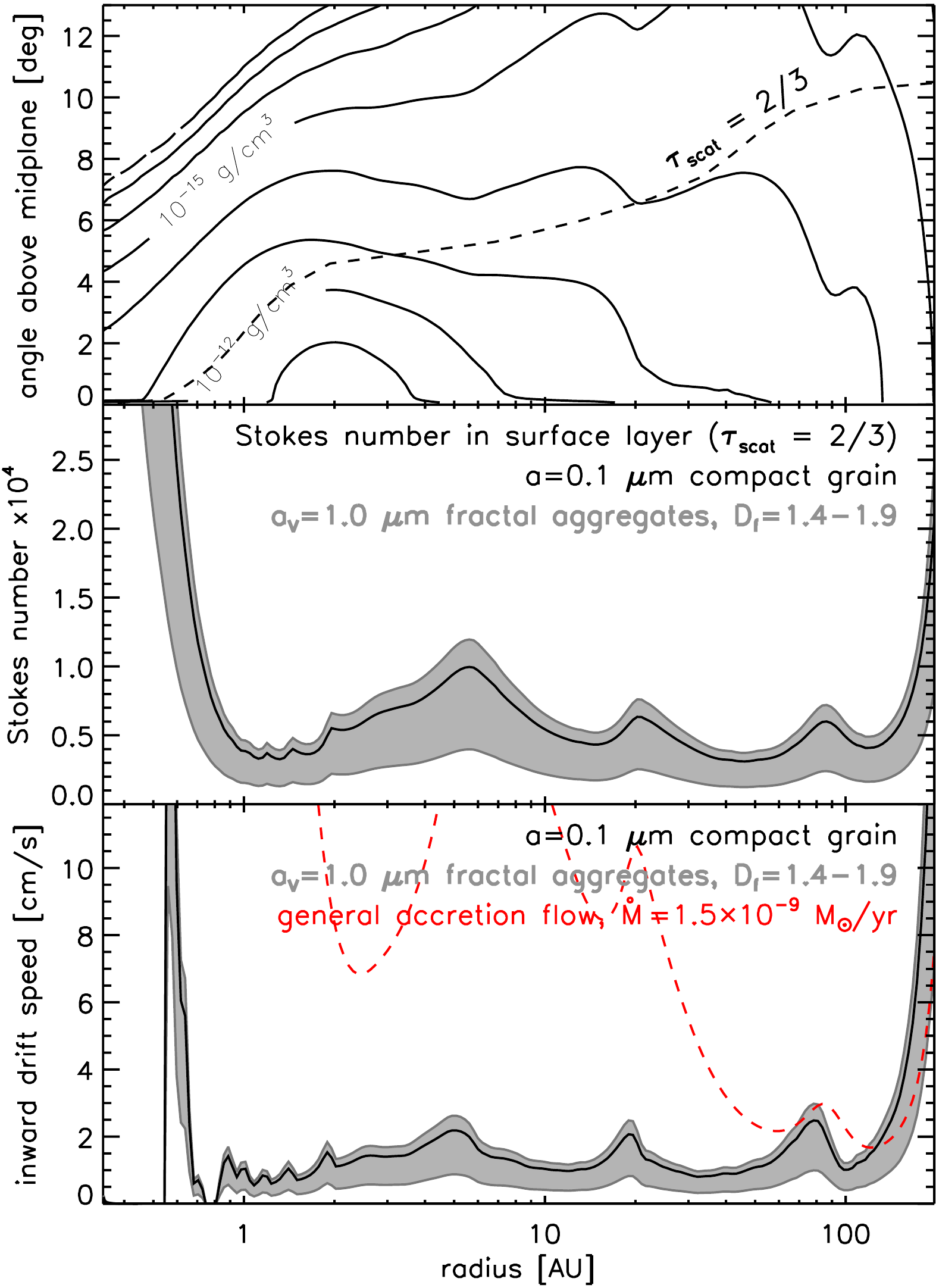}
\caption{\label{fig:gas_dust_coupling} Gas-dust coupling in the disk surface as probed by SPHERE. \emph{Top panel}: density contours and H-band scattering optical depth $\tau_{\rm{scat}}=2/3$ surface; \emph{middle panel}: Stokes number for $a=0.1$\du\mum \ compact grains and $a_{\rm{V}}=1.0$\du\mum \ fractal aggregates, \emph{lower panel}: inward radial drift speed in a laminar disk and speed of general gas accretion flow (red-dashed). A distance of \dpc\du pc has been assumed}.
\end{figure}

In \fig\fsep\ref{fig:gas_dust_coupling} we summarize our findings. The Stokes numbers are at the $\lesssim$\ds$10^{-4}$ level everywhere, except near the inner and outer edges of the disk due to the tapering of the gas density distribution. For the aggregates we investigated particles with a range of fractal dimensions from $D_{\rm{f}}$\eqsep$=$\eqsep$1.4$ \citep[coagulation under Brownian motion,][]{2000PhRvL..85.2426B,2004PhRvL..93b1103K,2006Icar..182..274P} to $D_{\rm{f}}$\eqsep$=$\eqsep$1.9$ \citep[coagulation in a turbulent gas,][]{1998Icar..132..125W}. The corresponding values for the porosity are $\phi_{\rm{p}}$\eqsep$\approx$\ds$0.96$ and $\phi_{\rm{p}}$\eqsep$\approx$\ds$0.88$, respectively \citep[e.g.,][]{2006A&A...445.1005M}.

In the lower panel of \fig\fsep\ref{fig:gas_dust_coupling} we show the resulting radial drift speed of particles in the disk surface. We also show the speed at which the gas would move inward to sustain the current accretion rate, for which we adopt \Mdotacc\ds$=$\ds\ds1.5\ds$\times$\ds$10^{-9}$\du\msunyr \ \citep{2008ApJ...681..594H} but note that this rate is temporally variable. Thus, even in a laminar disk, the motion of the small dust grains would be at most similar to that of the general accretion flow at most radii. For reference, at a drift speed of 2\du\cms \ it takes about 240.000 years to drift inward by 1\du\AU. Moreover, in a real disk there is turbulent vertical mixing supplying small dust from the disk interior to the surface, compensating for the drift effect. We conclude that relative radial motion between small dust particles and the gas can have at most a minor effect on the presence and abundance of small dust particles in the disk surface, i.e. our assumption that the small dust grains closely follow the gas distribution is justified.

\subsection{Radiative transfer code and modeling approach}
\label{sec:methods:approach}

We employed the 2D version of the radiative transfer code \verb MCMax \ by \cite{2009A&A...497..155M}, which was also used by \citet[hereafter \Menu]{2014A&A...564A..93M}. The \Menu \ model has a continuous radial distribution of dust, with the small dust grains (particle radius $a \leq$\,100\du\mum) following a $\Sigma_{{a \leq 100~\mu m}} \propto R^{-0.5}$ surface density distribution, and the larger grains ($a >$\,100\du\mum) being more centrally concentrated with a $\Sigma_{{a > 100~\mu m}} \propto R^{-1.4}$ distribution. It has a tapering at the inner edge (inside $\approx$\ds2\du\AU, i.e. on scales smaller than probed by the SPHERE data, see \fig\fsep\ref{fig:radial_profiles}  and \Menu \ for details) leading to a surface density maximum at 2.5\du\AU, agreeing remarkably well with the central bright ring seen in the ALMA data of \cite{2016ApJ...820L..40A}.

Our new model is based on the \Menu \ model but guided by the new observational data we made a number of adaptations:

\begin{enumerate}
\item we extended the outer radius from 60 to 200\du\AU \ (\Menu \ modeled the available infrared and mm interferometry data, but did not model the optical/near-infrared scattered light distribution),

\item we explicitly include the gas as a separate component with a surface density distribution following  $\Sigma_{\rm{gas}} \propto R^{-3/4}$ and a fixed total mass of 0.05\du\Msun \ \citep{2013Natur.493..644B}. The small dust follows this distribution. The large dust has a fully independent radial distribution that is adapted to match the ALMA observations by \cite{2016ApJ...820L..40A},

\item the small dust follows an MRN size distribution with grain radius $a_{\rm{min,s}} \leq a < a_{\rm{max,s}}$, the large dust also has an MRN size distribution with $a_{\rm{min,l}} \leq a < a_{\rm{max,l}}$,

\item we introduced parameterized ``radial depressions'' in the gas surface density distribution and adapted the small dust distribution accordingly such that the gas to small dust mass ratio is the same everywhere in the disk.

\end{enumerate}

Furthermore, in our modeling we make the following basic assumptions:

\begin{enumerate}

\item \label{assumption:Tstructure} the disk is in vertical hydrostatic equilibrium and the temperature structure is governed by irradiation from the central star,

\item \label{assumption:grain_mixture} the dust grains are a mixture of amorphous silicates and amorphous carbon, and they are compact, 

\item \label{assumption:segregation} there is vertical segregation of the various grain sizes due to dust settling, but no radial segregation, and

\item \label{assumption:constant_alpha} the disk viscosity can be approximated using the  ``alpha prescription'' \citep{1973A&A....24..337S} and the value of $\alpha$ is the same at every location in the disk.

\end{enumerate}

In this framework we then self-consistently calculated the temperature and vertical density structure of the disk, using the built-in iterative solver of \verb MCMax . We compared the resulting scattered light intensities to the observations, after which the surface density perturbations were refined until a good match between the model and the observations was obtained. Our goal was to find a surface density distribution that results in the observed surface brightness distribution, within the framework of our model, but not to comprehensively search for all models that are compliant with the observations. 

The scattered light intensity distribution of a given model results from a combination of the disk vertical structure and the local dust properties. The disk will be bright where the incidence angle of stellar radiation impinging upon the disk is steep, where the scattering cross sections of the grain mixture in the disk surface are favorable compared to the absorption cross sections, and where the scattering phase function directs a large fraction of the scattered light towards the observer, which is close to 90\degree \ for the nearly pole-on disk of \TW. Because the measured radial intensity profiles are so similar in shape between the various wavelengths of our observations (see \fig\fsep\ref{fig:radial_profiles}), a radial variation of the intrinsic dust scattering properties appears unlikely as the underlying mechanism for the radial brightness variations. Radial bulk density variations, on the other hand, are expected for a transition disk and naturally lead to variations in the scattered light brightness.

The disk vertical structure, and hence the incidence angle of the stellar radiation on the disk, is not a free parameter in our model; it is calculated self-consistently. It does depend on the choice of the viscosity parameter $\alpha$, where very low values ($\alpha <$ few $10^{-5}$) yield poor gas-dust coupling, resulting in stronger dust settling and a more weakly flared outer disk geometry.

\vspace{0.15cm}

In the following we comment on some of the assumptions adopted in our modeling. With respect to assumption \ref{assumption:Tstructure} we note that, while one may consider ``local'' heating mechanisms in the interior of the disk, such as heating by shock waves induced by embedded protoplanets, such mechanisms typically lead to deviations from azimuthal symmetry (spiral waves) which are not observed in \TW. Therefore such mechanisms, if at work in the \TW \ disk, play only a minor role regarding the temperature and hence vertical structure of the disk. Also, at the current accretion rate of \Mdotacc\ds$=$\ds\ds1.5\ds$\times$\ds$10^{-9}$\du\msunyr \ \citep{2008ApJ...681..594H} the irradiation term for the heating of the disk interior dominates over the accretion term by a factor of $\approx$\ds$5$\ds$\times$\ds$10^2$ ($R$\ds$=$\ds1\du\AU) to $\approx$\ds$2$\ds$\times$\ds$10^4$ ($R$\ds$=$\ds50\du\AU). Therefore also accretion has a negligible effect on the temperature and hydrostatic structure of the disk at the radii probed by the SPHERE observations.

Concerning assumption \ref{assumption:segregation}: size-dependent dust settling is incorporated using the formalism of \cite{2004A&A...421.1075D} which was implemented in \verb MCMax \ by \cite{2012A&A...539A...9M}. The dust is divided in to a range of size bins and for each size bin the absorption cross sections and full scattering matrices are calculated according to the adopted size distribution $n(a) \propto a^{-3.5}$. The individual bins are treated as separate dust components, each having their own vertical distribution according to their coupling to the gas. The disk material is assumed to be in thermal contact and hence dust grains of all sizes as well as the gas have the same temperature. The larger grains are more susceptible to settling and are hence more concentrated towards the disk midplane than the smaller ones. This effect depends strongly on the adopted turbulence strength, parameterized with $\alpha$ \citep{1973A&A....24..337S}: for weak turbulence (low $\alpha$) there is strong segregation and only the smallest grains are present in the disk surface, with strong turbulence the various sizes are better mixed with also larger grains are present in the disk surface where they can contribute to the scattering. 

Because the disk viscosity has a strong influence on the scattered light profile assumption \ref{assumption:constant_alpha}, that $\alpha$ is the same throughout, the disk is a central one. Decreasing the viscosity at some location would locally lead to stronger settling of the larger grains and a disk surface layer that is dominated by the smallest grains in the distribution. Depending on the smallest grain size \amin \ in the adopted dust distribution this may lead to a strong decrease in the scattered light surface brightness. Therefore, it should in principle be possible to construct a ``radial viscosity profile'' $\alpha(R)$, together with a suitable choice of dust properties, then do a self-consistent calculation of the temperature and vertical structure in a similar way to our approach, and arrive at a solution that matches the observations. We choose to take the surface density profile $\Sigma(R)$ as our ``free fit variable'' rather than the viscosity profile $\alpha(R)$. This is motivated by the observation that radial surface density variations are common in disks (with transition disks being the extreme example) and there are less direct constraints on location-dependent viscosity in disks, i.e. multiple MRI ``dead zones'', although modeling suggests multiple dead zones are possible \citep[][]{2013ApJ...765..114D}. Moreover, any dead zones in a disk are expected to be in the disk interior, with turbulent layers on top. It is unclear whether in such a situation, all but the smallest ($<$~few~$10^{-2}$\du\mum) grains could be removed from the upper layers. That would, however, be required in order to explain large differences in scattered light surface brightness through this mechanism.

\subsubsection{Model parameters}
\label{sec:methods:parameters}

The free parameters in our radiative transfer disk model are:

\begin{enumerate}
\item the radial surface density profile $\Sigma(R)$
\item the global disk viscosity parameter $\alpha$
\item the smallest and largest sizes in the population of small grains $a_{\rm{min,s}}$ and $a_{\rm{max,s}}$
\end{enumerate}

The following parameters are kept fixed (see also \tab\tabsep\ref{tab:model_parameters}): (1) the stellar parameters; (2) the dust composition which is adopted from \Menu: 80\% amorphous magnesium-iron olivine-type silicates (MgFeSiO$_4$) and 20\% amorphous carbon, by mass (the optical constants of taken form \cite{1995A&A...300..503D} for the silicates and from \cite{1993A&A...279..577P} for the Carbon); (3) the functional form of the grain size distribution which is $n(a) \propto a^{-3.5}$; (4) the gas/dust ratio for the small dust of 100 by mass\footnote{For the density of small dust, an MRN distribution of grains with radii from $10^{-2}$\du\mum \ to 10\du mm is set up, with a total mass equal to the gas mass divided by the gas/dust ratio. Only the grains with sizes between $a_{\rm{min,s}}$ and $a_{\rm{max,s}}$ are kept, which contain only about 10\% of the mass of the original MRN distribution. The remainder is assumed to have coagulated.}; (5) the disk inclination $i$\,$=$\,0\degree \ adopted for fitting the radial intensity profiles, motivated by the high degree of azimuthal symmetry; (6) the disk inner radius of 0.34\du\AU \ and outer radius of 200\du\AU; (7) the smallest and largest grains sizes in the large grain distribution $a_{\rm{min,l}} = 1$\du mm and $a_{\rm{max,l}} = 10$\du mm; (8) the ``unperturbed'' surface density profile $\Sigma_0(R)$ which has the form 
\begin{equation}
\label{eq:sigma0}
\Sigma_0(R) = \Sigma_{\rm{exp}} \left(\frac{R}{R_{\rm{exp}}}\right)^{-p} \Gamma_{\rm{in}}(R)
\end{equation}
where $\Gamma(R)$ is an exponential tapering in the inner disk of the form 
\begin{equation}
\label{eq:inner_taper}
\Gamma_{\rm{in}}(R)={\rm exp}\left(-\left(\frac{1-R/R_{\rm{exp}}}{w}\right)^3\right)
\end{equation}
at $R<R_{\rm{exp}}$, and with unity value at $R \geq R_{\rm{exp}}$ (see \Menu). An overview of these parameters is given in \tab\tabsep\ref{tab:model_parameters}.

\vspace{0.2cm}

Parameterized radial depressions are introduced by multiplying the initial density distribution by a radial depletion function $f(R)$, with values between 0 and 1, so that the surface density distribution becomes
\begin{equation}
\label{eq:sigma}
\Sigma(R)=f(R)\Sigma_0(R)
\end{equation}
Guided by the 3 ``gaps'' we see in the data, we introduce 3 depressions $f_{\rm i}$. We find that asymmetric gaussians of the form 
\begin{equation}
\label{eq:gap_shape}
f_{\rm i}(R)=\left(1-d_{\rm i} \ {\rm exp}\left(-\frac{(R-R_{\rm c,i})^2}{2\sigma_{\rm{in/out,i}}^2}\right)\right)
\end{equation}
allow to construct a density profile that provides a good match to the observed scattered light profiles in the self-consistent calculations. Here $d_{\rm i}$ is the ``depth'' of depression i, $R_{\rm c,i}$ is its central radius, and the width is given by $\sigma_{\rm{in,i}}$ in the inner flank ($R<R_{\rm c,i}$) and by $\sigma_{\rm{out,i}}$ in the outer flank ($R>R_{\rm c,i}$). In addition to the 3 ``gaps'', a gaussian tapering $\Gamma_{\rm{out}}(R)$ at the outer edge of the disk is needed which has the value:
\begin{equation}
\label{eq:outer_taper}
\Gamma_{\rm{out}}(R)={\rm exp}\left(-\frac{(R-R_{\rm{o}})^2}{2\sigma_{\rm{o}}^2}\right)
\end{equation}
at $R \geq R_{\rm{o}}$ and unity at smaller radii. The $f(R)$ curve by which we multiply the surface density $\Sigma_0$ of \Menu \ model then is:
\begin{equation}
\label{eq:fR}
f(R)=\left(\prod_{i} f_{\rm i}(R)\right)\Gamma_{\rm{out}}(R)
\end{equation}
We thus use a total of 14 parameters, 4 per gap and 2 for the tapering at the outer disk edge, to describe $f(R)$; see \tab\tabsep\ref{tab:fr_parameters} for a summary. This prescription is not meant to be unique and we have not tried to find other ways of describing $f(R)$ that require fewer parameters. Our goal is to find an approximate density profile that is consistent with our observations, not to exhaustively explore all possible profiles that match the data.

\subsection{Stellar spectrum}
\label{sec:stellar_spectrum}

We created a model photospheric spectrum of \TW \ based on that by \cite{2013ApJ...771...45D}, who find that a combination of an M2- and K7-type model with total contributions of 55\% and 45\% to the total flux provides the best match to their HST spectrum. We adopt this spectral shape, using PHOENIX model atmospheres at 3600\du K and 3990\du K, respectively, for both components, and require a stellar radius of 1.18\du\Rsun \ to match the 2MASS H- and K-band fluxes for an assumed distance of \dpc\du pc, and after accounting for a small near-infrared excess of $\Delta$H\,$=$\,0.035\du mag due to scattered light. This yields a stellar luminosity of 0.256\du\Lsun.

\subsection{PSF convolution}
\label{sec:methods:convolution}

We performed detailed modeling of the SPHERE data in the \Rp- and H-bands, i.e. at the extreme wavelengths of our data set. We used the actual SPHERE data of the science observation to produce point spread functions (PSFs). This is possible because the light scattered off the disk contributes only a few percent to the total radiation received. Therefore, the observations after basic reduction but before subtracting the image pairs recorded through orthogonal linear polarizers such that the (mostly unpolarized) direct photospheric flux is not canceled, are a good measure of the PSF. We could thus obtain good approximations to the actual PSF through a 0\degree \ and 90\degree \ linear polarizer (the ``+Q'' and ``-Q'' images), as well as through a 45\degree \ and 135\degree \ polarizer (the ``+U'' and ``-U'' images). These were nearly identical and we used their average as our final PSF, in each spectral filter.

Because a coronagraph was employed in the H-band observations, the central part of the PSF is missing in the science data. We used short integrations taken without a coronagraph, directly prior and after the coronagraphic observations to obtain the central part of the PSF. At radii far from the star these shallow observations are dominated by readout noise and have a low signal to noise ratio (SNR) compared to the coronagraphic observations. Therefore we matched the flux levels in the shallow observations to those of the coronagraphic observations in the wings of the PSF, and combined the central 15 pixels (183 mas, radius) of the shallow observations with the complementary part of the coronagraphic observation into the final PSF. The \Rp \ observations were performed without a coronagraph and could be used directly.

In order to compare the RT model images with our data we simulated the process of observing the source: we produced +Q, -Q, +U, and -U images and convolved them with the PSF. We then produced the Q and U images by subtracting the respective pairs. Then, the resulting radial and tangential components of the polarized flux were calculated according to equations \ref{eq:Qphi} and \ref{eq:Uphi}. Because of \TW's nearly face on inclination, the observed high degree of azimuthal symmetry, and our prime goal of constraining the \emph{radial} bulk density distribution, we ignored inclination effects in the RT modeling.

The PSF convolution naturally reduces the contrast in the images because part of the light arising from any location gets spread over a much larger region. This applies to all regions of all images. However, the effect is particularly dramatic for polarized light coming from regions that are not spatially resolved by the observations, \emph{and} have approximately zero \emph{net} polarization. This situation occurs at the inner edge of the dusty disk, which lies at $\approx$\ds0.34\du\AU \ in our model. This region is extremely bright in scattered light compared to the rest of the disk\footnote{The surface brightness of the inner edge of our model is approximately 3$\times$10$^4$ times higher than that of ring~\#2 at 45\du\AU.}, but because this region is small compared to the PSF core the signal recorded in the orthogonal polarization directions cancels almost perfectly, leaving no net signal in the central area. This creates a ``central gap'' in the images that is an artifact due to the limited spatial resolution. 

\subsection{Radial density profile}
\label{sec:results:radial_density_profile}

We have constructed a radiative transfer model of the \TW \ disk with a self-consistent temperature and vertical structure and including grain size-dependent dust settling, as described in \sek\seksep\ref{sec:methods:RT}. Our objective was to find a radial density distribution $\Sigma(R)$ which, when implemented in such a model, yields a match to the observed radial intensity curves in (polarized) scattered light.

\subsubsection{Scattered light contrast vs. gap depth}
\label{sec:results:contrast_vs_gapDepth}

\begin{figure}[t!]
\includegraphics[width=\columnwidth,trim=0 0 0 0, clip]{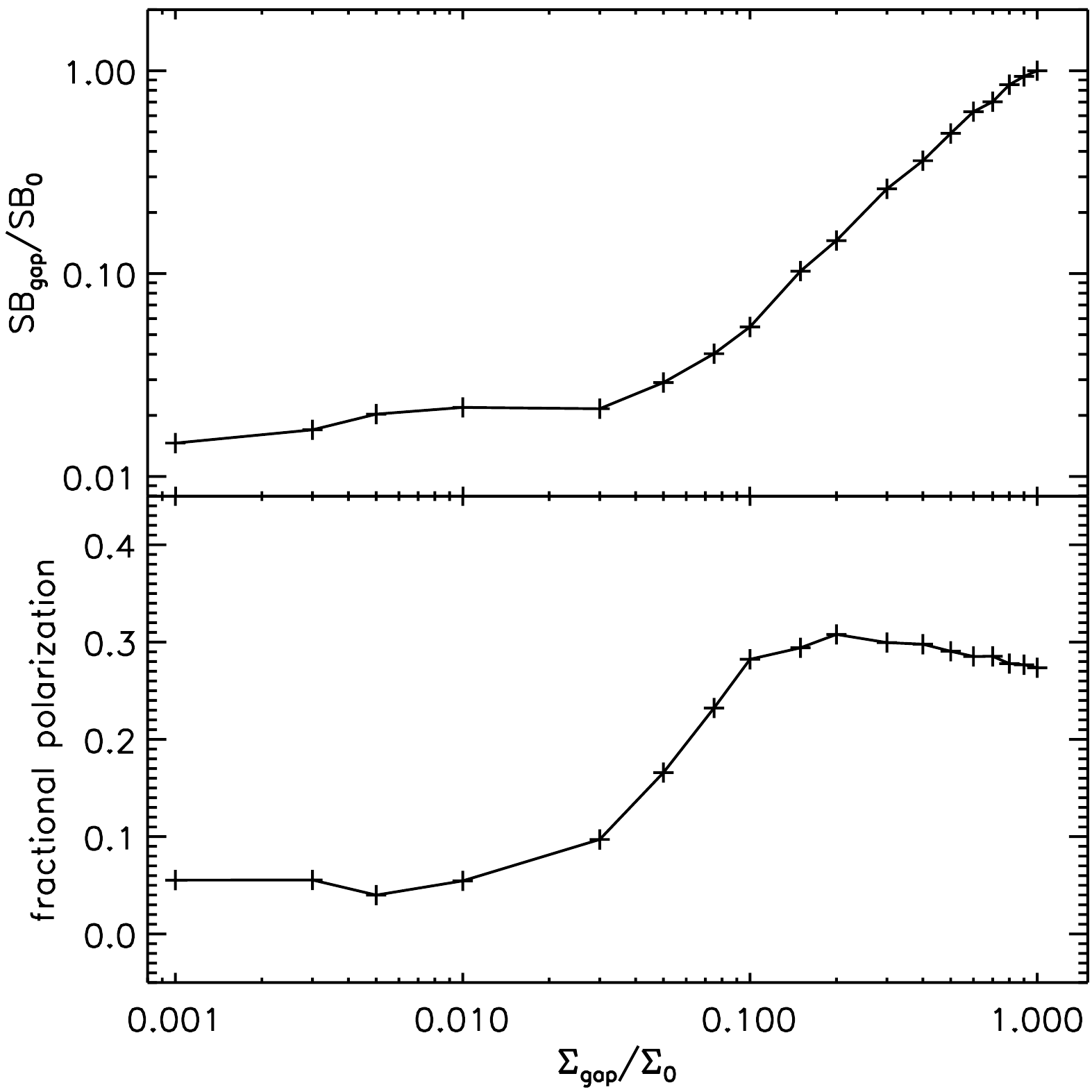}
\caption{\label{fig:contrast_vs_gapDepth} The contrast in scattered light total intensity (``Stokes I'', top panel) and fractional polarization (bottom panel) vs. the ``depth'' of the radial surface density depression, for a generic gap at the radial location of gap~\#2.}
\end{figure}

We first investigated how, for a generic location and shape of a radial depression in the surface density, the observed scattered light surface brightness contrast depends on the density contrast (i.e. ``gap depth''). To this purpose, we took the unperturbed model of \Menu \ and introduced a radial density depression with a range of gap depths at the location of gap~\#2. We then compared the resulting scattered light brightness in the middle of the gap region relative to that in the unperturbed model, as a function of gap depth. The result is shown in the top panel of \fig\fsep\ref{fig:contrast_vs_gapDepth}. The bottom panel shows the resulting polarization fraction.

As the surface density in the gap region decreases the surface brightness of the scattered light decreases, too. Initially the surface brightness in both total and polarized intensity decrease approximately linearly with the decrease in surface density. In this regime, the signal in the gap region is dominated by photons that are scattered once in the disk surface and then reach the observer. When the surface density is decreased by a factor $\gtrsim$\,10, the scattered light signal in the gap region becomes dominated by photons that are scattered twice, first down into the gap region and then back up towards the observer. This has two effects: (1) the fractional polarization becomes much smaller, and (2) the total intensity gradually becomes independent of the surface density (as long as the scattering optical depth of the material in the gap remains $\gtrsim$\,1). 

\begin{table}[t!]
\begin{center}
\caption{\label{tab:model_parameters}Parameters of radiative transfer disk model fitting the radial polarized intensity profiles.\vspace{0.15cm}}
\begin{tabular}{lcc}
\tableline
\multicolumn{3}{c}{fixed parameters:} \\
\tableline
disk inner radius                    & $R_{\rm{in}}$       & 0.34~\AU \\
disk outer radius                    & $R_{\rm{out}}$      & 200~\AU \\
$\Sigma_0$ normalization$^{\rm{a}}$   & $\Sigma_{\rm{exp}}$ & 94.7 g\,cm$^{-2}$ \\
$\Sigma_0$ power law exponent$^{\rm{a}}$  & $p$            & 0.75 \\
$\Sigma_0$ tapering radius$^{\rm{a}}$ & $R_{\rm{exp}}$      & 2.7~\AU \\
$\Sigma_0$ tapering width$^{\rm{a}}$  & $w$                & 0.45~\AU \\
disk inclination                     & $i$                & 0\degree \\
stellar mass                         & \Mstar             & 0.87\,\Msun    \\
stellar luminosity                   & \Lstar             & 0.256\,\Lsun   \\
spectral type                        & \multicolumn{2}{c}{see \sek\seksep\ref{sec:stellar_spectrum}} \\
distance                             & $d$                & 54\,pc \\
\tableline
\tableline
\multicolumn{3}{c}{free parameters:} \\
\tableline
viscosity parameter                  & $\alpha$            & $2\times10^{-4}$ \\
minimum small grain size                  & $a_{\rm{min,s}}$   & 0.1\,\mum \\
maximum small grain size                  & $a_{\rm{max,s}}$   & 3.0\,\mum \\
minimum large grain size                  & $a_{\rm{min,s}}$   & 1\,mm \\
maximum large grain size                  & $a_{\rm{max,s}}$   & 10\,mm \\
depletion profile$^{\rm{b}}$     & $f(R)$ &  see \tab\tabsep\ref{tab:fr_parameters} \\ 
\end{tabular}
\end{center}
\vspace{-0.2cm}
$^{\rm{a}}$\footnotesize{$\Sigma_0$ denotes the ``unperturbed'' radial surface density profile adopted from \Menu \ before the introduction of parameterized radial density depressions $f(R)$.}
$^{\rm{b}}$\footnotesize{The bulk surface density distribution is given by the unperturbed profile multiplied by the mass depletion profile, i.e. $\Sigma(R)=\Sigma(R) f(R)$.}
$^{\rm{c}}$\footnotesize{The total disk mass is not an independent parameter, it follows from $f(R)$ and the mass of the unperturbed model $M_{\rm{disk,0}} = 7.15\times10^{-2}$\,\Msun. See \sek\seksep\ref{sec:gaia_distance} for a discussion of how the various parameters are affected by the new GAIA distance that is $\approx$\ds10\% larger than the value assumed in this work.}
\end{table}

\begin{table}[htb]
\begin{center}
\caption{\label{tab:fr_parameters}Parameters describing the shape of the gas surface density depletion profile $f(R)$ (see \eq s \ref{eq:sigma}$-$\ref{eq:fR}).\vspace{0.15cm}}
\begin{tabular}{lcccc}
         & $d_{\rm i}$ & $R_{\rm{c,i}}$ & $\sigma_{\rm{in,i}}$ & $\sigma_{\rm{out,i}}$ \\
         &            &  [\AU]        &   [\AU]            &  [\AU] \\
\tableline
gap \#1  & 0.44 & 85  & 15  & 16 \\
gap \#2  & 0.61 & 21  & 3.75 & 18 \\
gap \#3  & 0.81 & 6 & 3 & 13 \\
\end{tabular}
\vspace{0.2cm}
\end{center}
\vspace{-0.35cm}
$^{\rm{a}}$\footnotesize{In addition, a tapering at the outer disk edge is applied according to \eq\eqsep\ref{eq:outer_taper}, with parameters $R_{\rm{o}}$\eqsep$=$\eqsep105\du\AU \ and $\sigma_{\rm{o}}$\eqsep$=$\eqsep55\du\AU}
\end{table}

\begin{figure}[t!]
\includegraphics[width=\columnwidth,trim=0 0 0 0, clip]{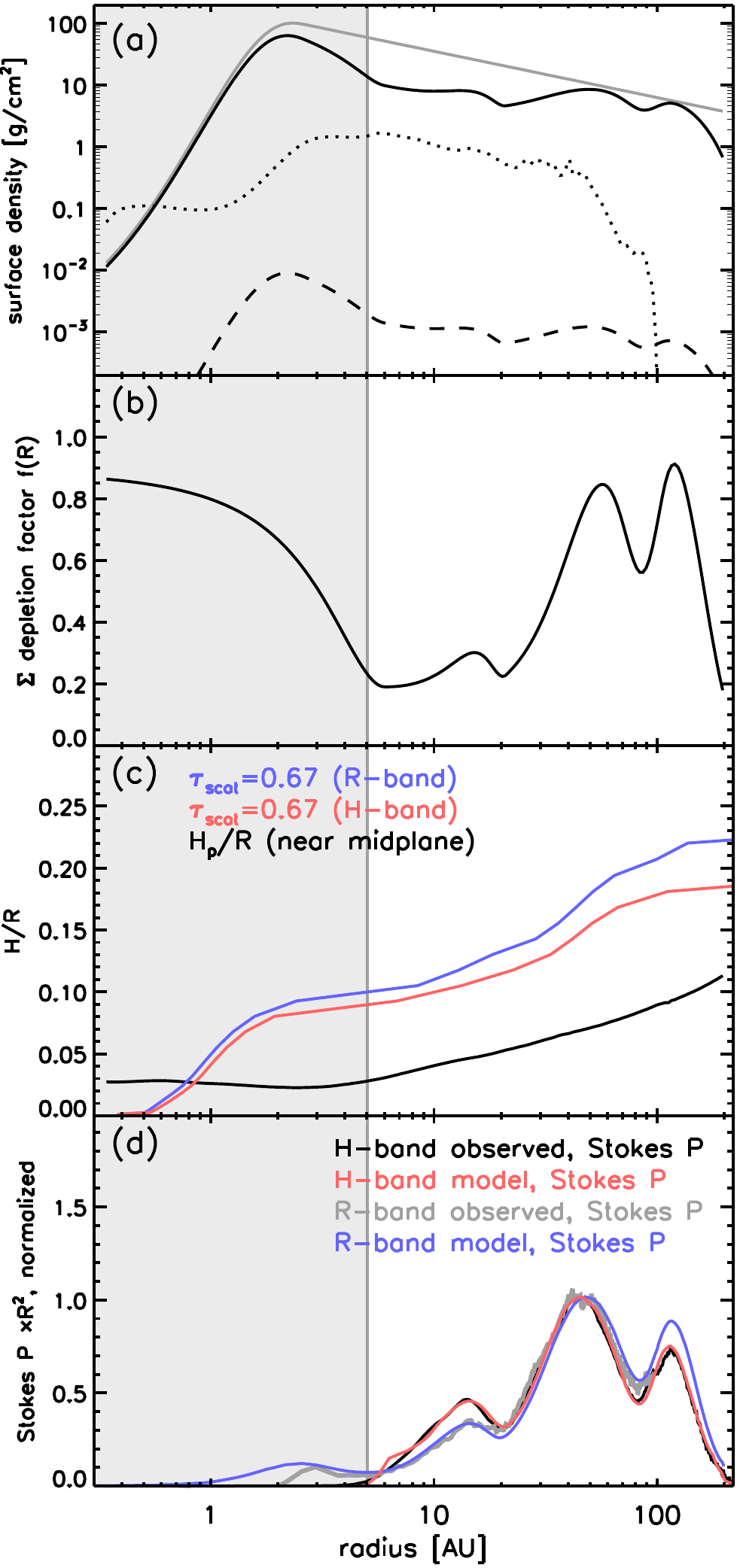}
\caption{\label{fig:results_overview} Overview of our radiative transfer model. Bottom panel: Black: H-band data, grey: \Rp-band data, red: convolved H-band model, blue: convolved \Rp-band model. The grey region at $R\lesssim$\ds$5$\du \AU \ lies behind the coronagraph in the H-band observations. See \sek\seksep\ref{sec:results:radial_density_profile} for details. A distance of \dpc\du pc has been assumed.}

\end{figure}

\subsubsection{Derived surface density distribution}

In \fig\fsep\ref{fig:results_overview} we summarize the main characteristics of the derived solution for the density profile. Panel (a) shows the ``unperturbed'' surface density profile $\Sigma_0(R)$ from \Menu \ in grey and the final profile $\Sigma(R)$ in black. Panel (b) shows the depletion profile $f(R) = \Sigma(R)/\Sigma_0(R)$. Panel (c) shows $H_{\rm p}/R$, where $H_{\rm p}$ is the pressure scale height in the disk interior (between $z=0$ and $z=H_{\rm p}$), as well as the ``surfaces'' where the radial scattering optical depth $\tau_{\rm{scat}}$ as seen from the central star equals 2/3 in the \Rp-band (blue) and H-band (red). Panel (d) shows the resulting surface brightness plots in polarized intensity (scaled by $R^2$) in the \Rp \ and H bands using the same color scheme, as well as the observed profiles (\Rp \ in grey, H-band in black).

The physical parameters of this model are summarized in \tab\tabsep\ref{tab:model_parameters}, the parameters describing the depletion profile $f(R)$ are listed separately in \tab\tabsep\ref{tab:fr_parameters}.

\begin{figure}[t]
\includegraphics[width=\columnwidth,trim=0 0 0 0, clip]{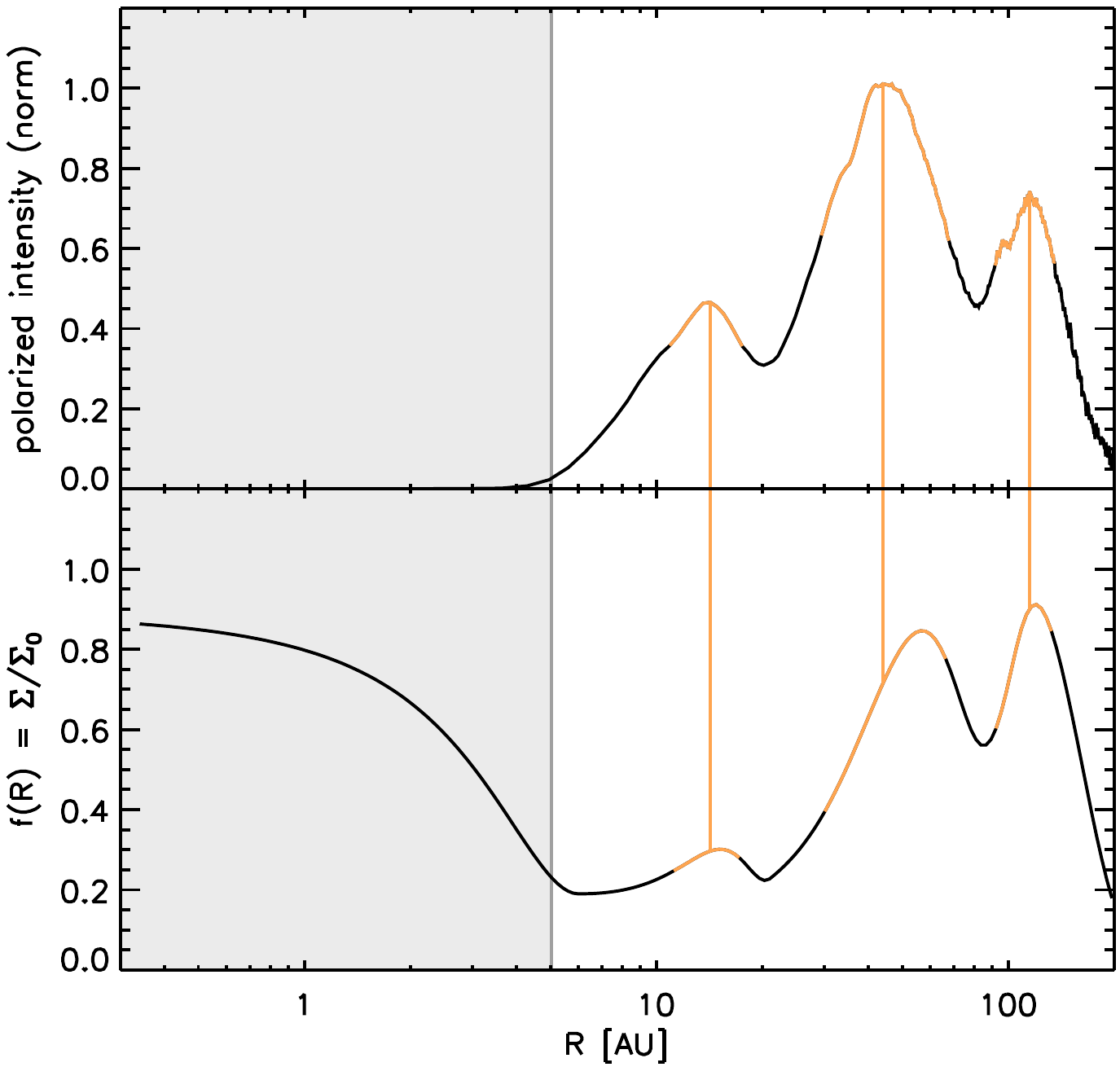}
\caption{\label{fig:sigma_stokesP} \emph{Top panel:} observed polarized intensity, scaled with $R^2$ and normalized to ``ring \#2''  \emph{Bottom panel:} surface density multiplication factor $f(R)$. A distance of \dpc\du pc has been assumed.}
\end{figure}

In \fig\fsep\ref{fig:sigma_stokesP} we show the observed H-band polarized light intensity profile and the inferred radial surface density multiplication factor $f(R)$, to illustrate that the bright rings seen in scattered light correspond to the outer flanks of the surface density depressions. The disk is brightest where $f(R)$ is increasing radially outward, hence the peaks in the density distribution lie at larger radii than the scattered light brightness peaks.

\subsubsection{Radial depressions in surface density}
\label{sec:results:radial_depressions}
The radial depressions that we need to apply in order to match the observed surface brightness profiles (see \fig\fsep\ref{fig:results_overview}b and \tab\tabsep\ref{tab:fr_parameters}) have two prime characteristics:
\begin{enumerate}
\item their ``depths'' are modest, $\approx$\ds45\% to $\approx$\ds80\%,
\item they are very wide, with shallow flanks.
\end{enumerate}
An immediate consequence of the modest depth is that in the ``gap'' regions a large fraction of the material from the original, unperturbed disk model remains. This follows from the scattered light surface brightness which remains much larger than zero in the gap regions, contrary to other well known transition disks with much ``deeper'' gaps.
The corresponding high surface density in the gap regions causes the disk to remain optically thick at infrared wavelengths in the vertical direction. This has consequences for the detectability of potential embedded (proto-) planets, as will be further discussed in \sek\seksep\ref{sec:discussion:planet_brightness}. 

The width and shape of the implied radial depressions is interesting: the profile implied from our data differs substantially from typical profiles obtained from planet-disk interaction calculations, as further discussed in \sek\seksep\ref{sec:discussion:gap_profiles}. Gap\,\#2 and gap\,\#3 appear more like a single, very broad gap ranging from $\lesssim$\ds6\du\AU \ to $\approx$\ds60\du\AU, with a broad and shallow outer flank between $\approx$\ds20 and $\approx$\ds60\du\AU. There is only a modest ``bump'' in the surface density multiplication factor $f(R)$ peaking around 15\du\AU, which is responsible for ring\,\#3 in our data. At the location of this bump the value of $f(R)$ is $\approx$\ds0.30, compared to $\approx$\ds0.19 in the deepest parts. In the actual surface density profile (\fig\fsep\ref{fig:results_overview}~a) this bump is not a peak but rather a ``shoulder'' at the outer edge of a plateau of roughly constant surface density between $\approx$\ds6 and $\approx$\ds15~\AU. Nevertheless, this surface density distribution naturally leads to a very distinct observational feature.

\begin{figure}[t!]
\includegraphics[width=\columnwidth,trim=0 0 0 0, clip]{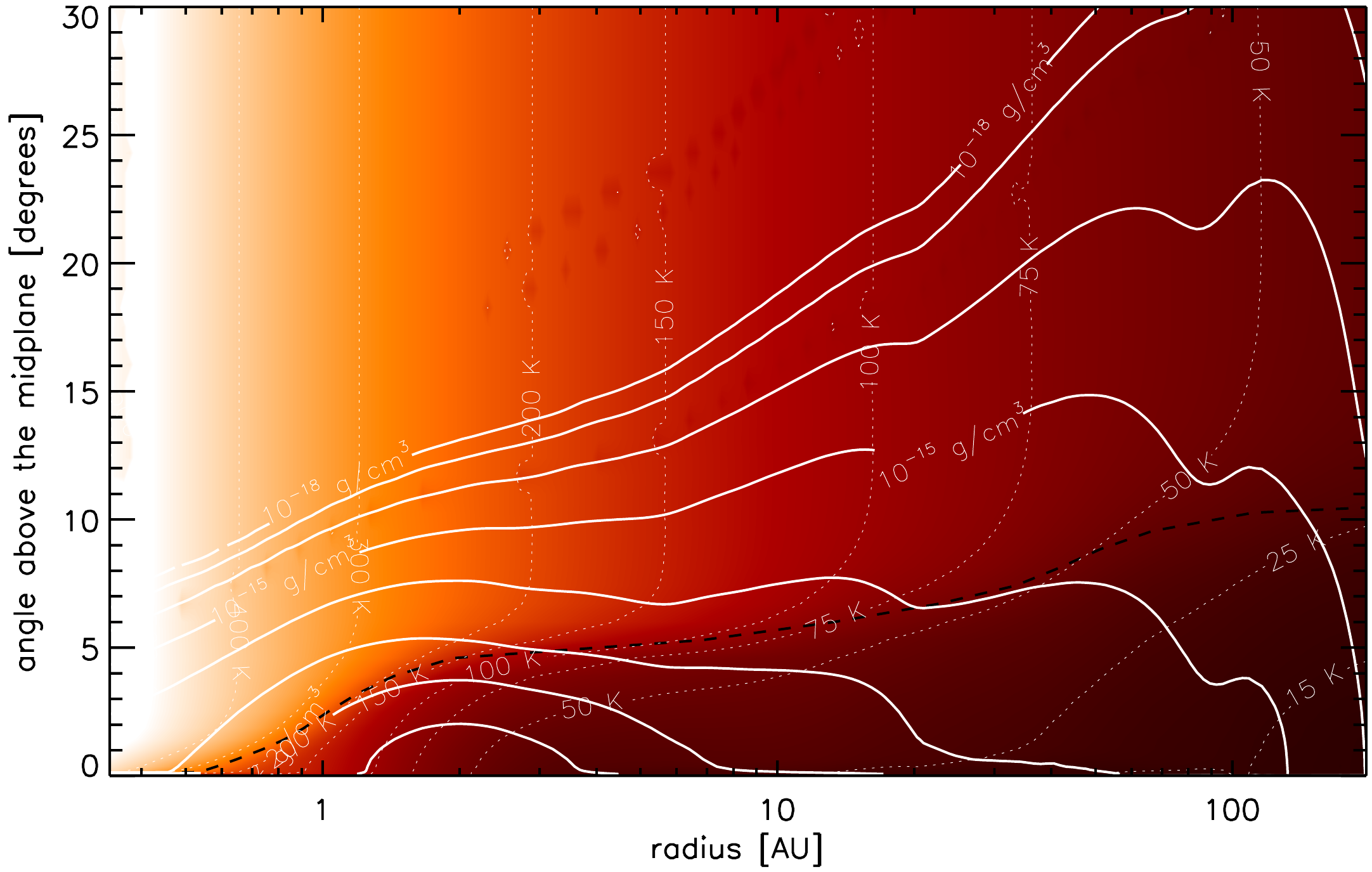}
\caption{\label{fig:model_denstemp}  Density (full contours) and temperature (colors and dotted contours) of radiative transfer model. The black dashed line shows the $\tau_{\rm{scat}} = 2/3$ surface in the H-band. A distance of \dpc\du pc has been assumed.}
\end{figure}

\subsection{Disk temperature structure}
\label{sec:results:temperature_structure}
In \fig\fsep\ref{fig:model_denstemp} we show the temperature and density structure that results from our self-consistent calculation. The radial depressions we introduced in the surface density are clearly seen in 2D density contours. They also affect the resulting temperature distribution, though their effect is less evident in \fig\fsep\ref{fig:model_denstemp}. High above the disk the temperature corresponds to the optically thin temperature for the smallest grains in the dust distribution and the radial surface density depressions have no effect; this region is easily identified by the vertical temperature contours. In the disk surface and deeper in the disk the depressions make the disk cooler in the gap regions because less stellar light is absorbed and the heating of the disk interior is locally reduced. In the outer flanks of the gaps the temperatures are higher than in the unperturbed model, because the disk intercepts more radiation per unit area.

\subsection{Spectral energy distribution}
\label{sec:SED}

\begin{figure}[t]
\includegraphics[width=\columnwidth,trim=0 0 0 0, clip]{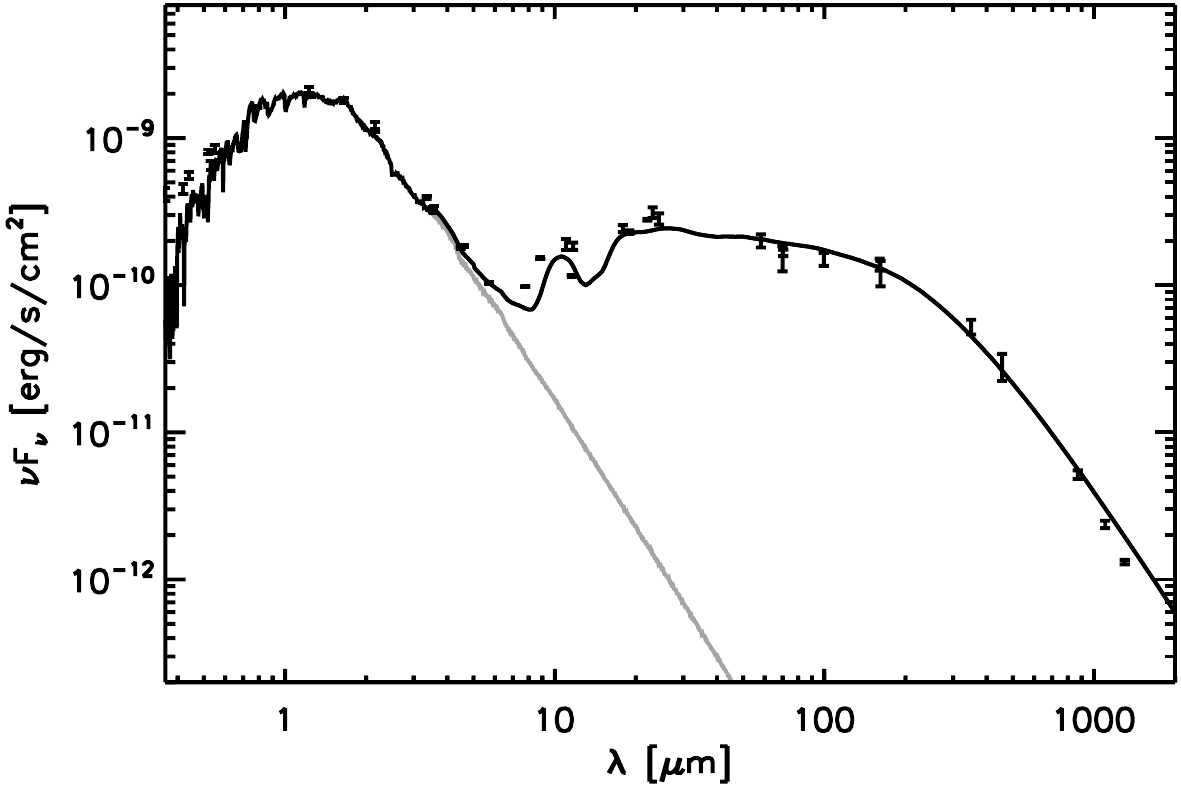}
\caption{\label{fig:SED} Spectral energy distribution of the radiative transfer model of the \TW \ disk compared with observations.}
\end{figure}

In \fig\fsep\ref{fig:SED} we show the observed and modeled SED. The model matches the data reasonably well, except for an $\approx$\ds25\% flux deficit in the 10 micron spectral region. While based on the \Menu \ model, whose SED more closely matches the data at these wavelengths, our model has a different dust distribution: the large grain component is constrained directly by the ALMA data, and the small grain component contains only dust up to 3\du\mum \ in size, contrary to the \Menu \ model, in which a continuous grain distribution up to 100\du\mum \ was adopted. If we include grains up to 100\du\mum \ in our distribution then our model yields too much far-infrared flux (a factor of $\approx$\ds2 around 100\du\mum). \Menu \ adopted a very low turbulence level ($\alpha = 10^{-5}$), leading to an only weakly-flared disk geometry, to circumvent this problem. However, models with such low turbulence fail to reproduce the scattered light data: they are too faint. Our bi-model dust distribution does simultaneously match the scattered light data and the SED. It is slightly deficient in warm dust in the central $\approx$\ds1\du\AU \ whence the flux in the 10 micron spectral region arises. We did not include additional warm dust; this would increase the number of free parameters further whereas this region is too close to the star to be probed by our SPHERE data.

\section{Discussion}
\label{sec:discussion}

We will now discuss the derived bulk gas radial surface density profiles in the framework of planet-disk interaction. Alternative explanations may exist, such as density variations caused by the magneto-rotational instability \citep{2015A&A...574A..68F}, but exploring these is beyond the scope of the present work. Gravitational instability (GI) in a massive disk may also lead to structure, but in the case of \TW \ the disk is likely not sufficiently massive for this to occur; moreover GI normally leads to strong spiral arms which are in stark contrast to the high degree of azimuthal symmetry we observe \citep[e.g.][]{2015MNRAS.453.1768P}. Radial variations in dust properties \citep{2015ApJ...813L..14B} may in principle explain the brightness variations but would require a scenario accounting for multiple bright and dark rings. The same is true for radial variations in the turbulent strength. We do not go into these scenarios here.

\subsection{Embedded planets?}
Here we explore the scenario where embedded planets are responsible for the creation of such partial gaps, and we estimate the corresponding limits of the planet masses following \cite{2013ApJ...771...45D}.

\subsubsection{Planetary mass estimates}
\label{sec:discussion:planet_masses}

\begin{figure}[t]
\includegraphics[width=\columnwidth,trim=0 0 0 0, clip]{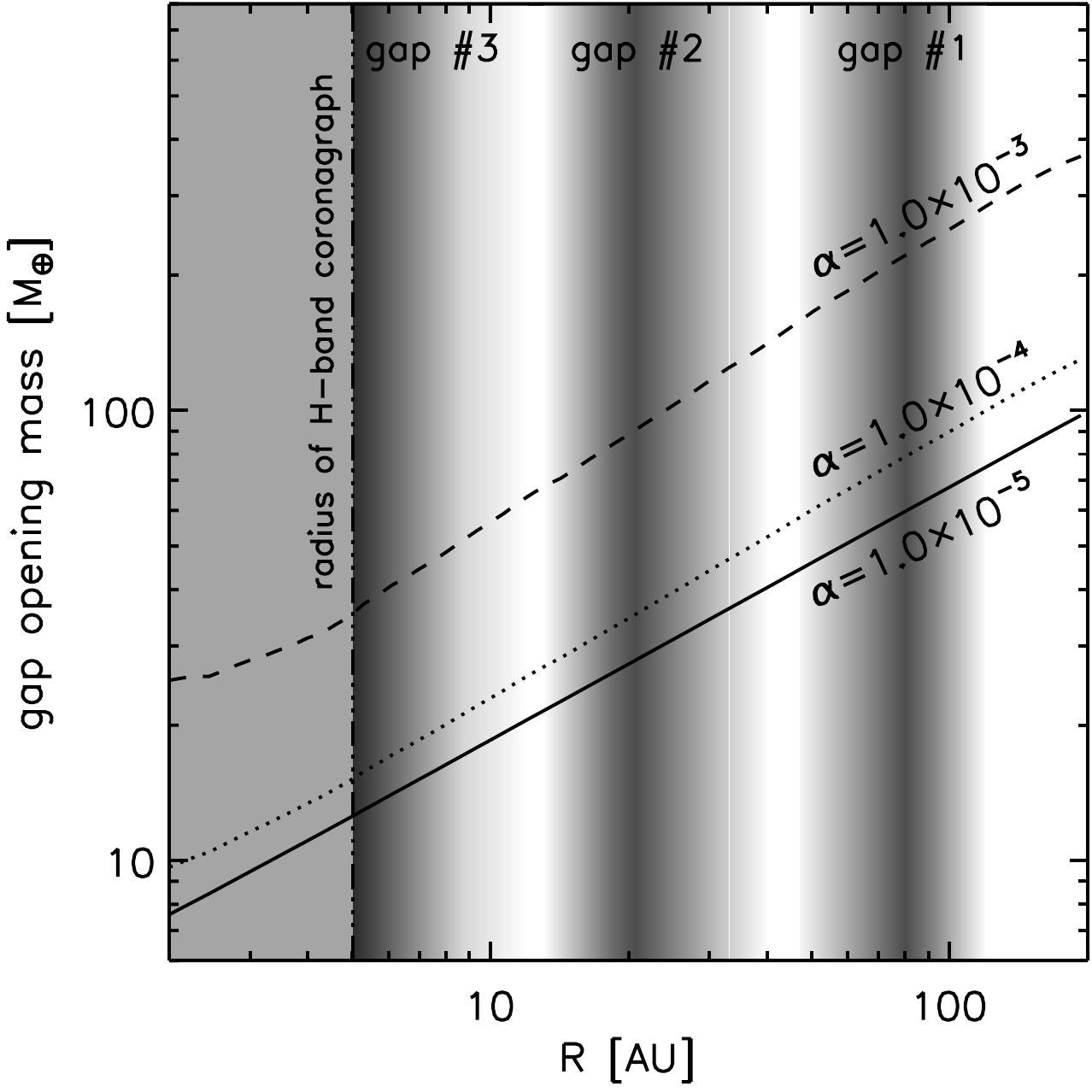}
\caption{\label{fig:jonathan_GOM} Gap opening masses for \TW \ disk parameters. A distance of \dpc\du pc has been assumed.}
\end{figure}

The ``gap opening mass'' of a forming planet embedded in a viscous disk can be expressed as \citep{2006Icar..181..587C}:
\begin{equation}
\label{eq:viscgapcrit}
\frac{3H_{\rm{p}}}{4r_{\mbox{\scriptsize Hill}}} + \frac{50}{q {\cal R}} \lesssim 1
\end{equation}
\noindent
Here, $H_{\rm{p}}$ is the pressure scale height of the disk, $r_{\mbox{\scriptsize Hill}}$ is the radius of the planet's Hill sphere, $q=m_p/M_*$ is the planet/star mass ratio, and $\cal R$ is the Reynolds number.  The pressure scale height is given by $H_{\rm{p}} = \frac{c_s}{v_{\phi}}a$, where $c_s=\sqrt{kT/\mu}$ is the sound speed in the gas with mean molecular mass $\mu$ at temperature $T$ around the disk midplane, $v_{\phi}=\sqrt{GM_*/a}$ is the planet's orbital speed at the orbital radius $a$. The radius of the planet's Hill sphere is given by \(r_{\mbox{\scriptsize Hill}} = a \left(\frac{m_p}{3M_*}\right)^{1/3}\). The Reynolds number is given by ${\cal R}\equiv r^2\Omega_P/\nu_v$, where $\Omega_P$ is the angular velocity of the planet and $\nu_v$ is the viscosity, which in the $\alpha$ prescription \citep{1973A&A....24..337S} is given by $\nu_v=\alpha c_s H_{\rm{p}}$.

If a planet is massive enough to fulfill the above criterion it will create an approximately azimuthally symmetric annular depression in the gas surface density around the planet's orbital radius. The surface density in this region is decreased by several orders of magnitude, leading to a practically ``empty'' gap. For the \TW \ disk the gap opening mass as a function of radius is illustrated in \fig\fsep\ref{fig:jonathan_GOM}, for several choices of the turbulence parameter $\alpha$.

In the \TW \ disk the gaps are only partially cleared which implies that, if planet-disk interaction is the underlying physical mechanism, the embedded planets have masses that are substantially below the gap opening mass. \cite{2015ApJ...807L..11D} provides an analytic recipe for the depth and shape of radial surface density depressions created by relatively low mass planets, that approximates results of numerical simulations of planet-disk interaction. In this regime, the planet mass can be found from \citep[see equations 9 and 10 of][]{2015ApJ...807L..11D}:
\begin{equation}
\label{eq:planet_masses}
q^2=\frac{3 \pi \alpha d}{(1-d) f_0 \mathcal{M}^5}
\end{equation}
\noindent
where $q$\eqsep$=$\eqsep$m_p/M_*$ denotes the planet/star mass ratio, $\alpha$ is the viscosity parameter, $d$\eqsep$=$\eqsep$1-\Sigma_p/\Sigma_0$ is the gap depth, $\mathcal{M}$\eqsep$=$\eqsep$R/H_{\rm{p}}$ is the Mach number whose value is highest in the inner disk ($\mathcal{M}$\eqsep$\approx$\ds37 at $R$\eqsep$<$\eqsep6\du\AU) and gradually decreases outward to $\mathcal{M}$\eqsep$\approx$\ds10 at 200\du\AU \ for our \TW \ disk model (see also \fig\fsep\ref{fig:results_overview}~c). $f_0$\eqsep$=$0.45 is a dimensionless constant whose value \cite{2015ApJ...807L..11D} derived for his analytical prescription to best match the outcome of numerical simulations.

If we take the approximate amplitudes $d$ of our parameterized radial depressions in the surface density (\tab\tabsep\ref{tab:fr_parameters}) at face value and apply equation\,\ref{eq:planet_masses} then we find masses of approximately 34\du\Mearth, 15\du\Mearth, and 6.4\du\Mearth \ for the assumed planets at \Rone, \Rtwo, and \Rthree\du\AU, respectively.

These mass estimates depend on other parameters of the disk model, in particular on the choice of $\alpha$. If $\alpha$ increases then the mass required to reach a given gap depth $d$ is higher (\eq\eqsep\ref{eq:planet_masses}). If the viscosity parameter is doubled to $\alpha = 4\times10^{-4}$ then the corresponding mass estimates are 49\du\Mearth, 22\du\Mearth, and 9\du\Mearth, respectively. If, on the other hand, the viscosity parameter is decreased to $\alpha = 1\times10^{-4}$ then the mass estimates become 24\du\Mearth, 11\du\Mearth, and 4.5\du\Mearth, respectively. 

\subsubsection{Gap profiles}
\label{sec:discussion:gap_profiles}

\begin{figure}[t]
\includegraphics[width=\columnwidth,trim=0 0 0 0, clip]{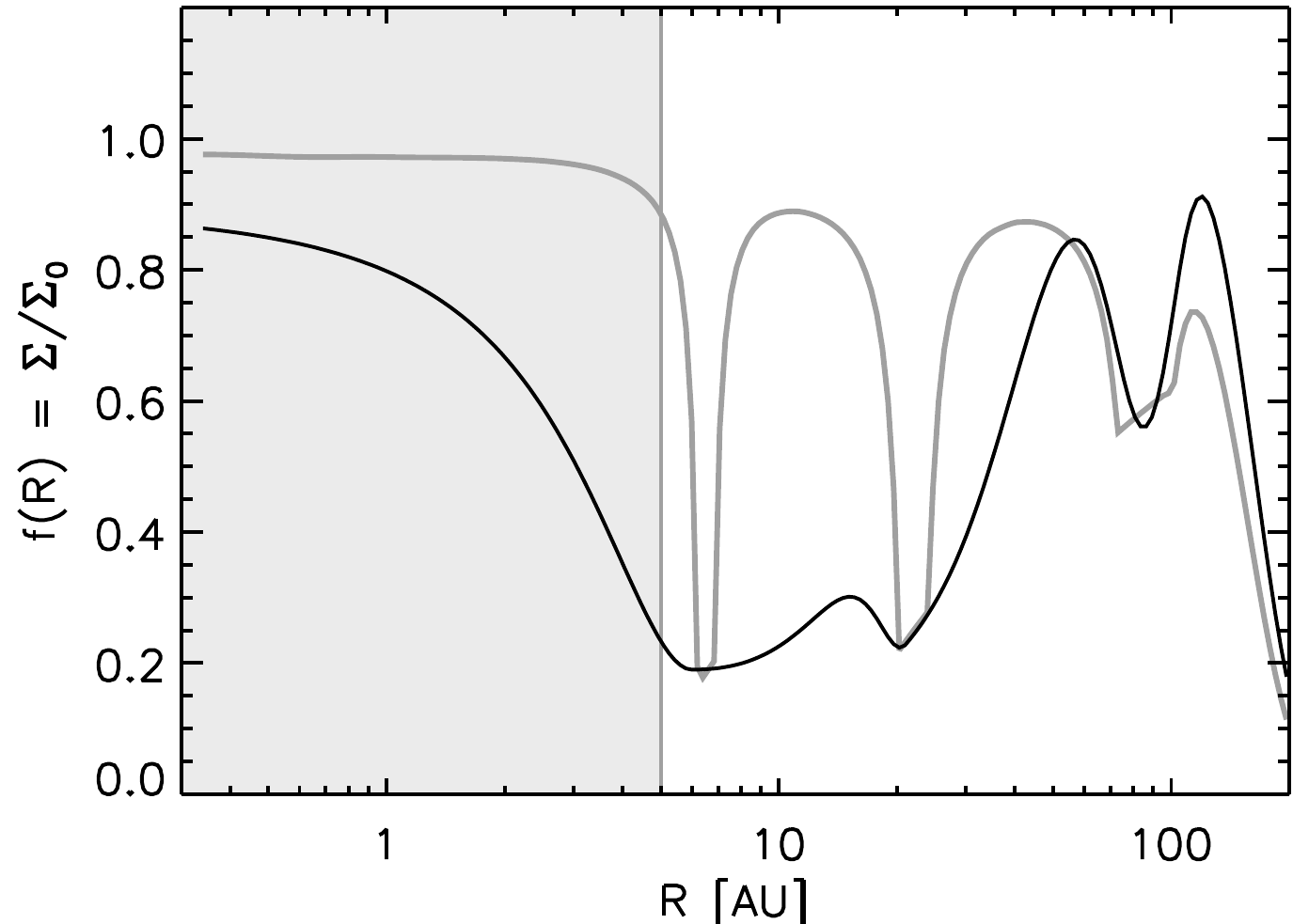}
\caption{\label{fig:gap_profiles} Comparison between our derived radial surface density depletion factor ($f(R)$, black curve) and an implementation of the model of \cite{2015ApJ...807L..11D} with three planets, approximately matching the depth of the gaps (grey curve). The innermost disk regions that are not well probed with our observations are masked. A distance of \dpc\du pc has been assumed.}
\end{figure}

\begin{figure}[t!]
\includegraphics[width=\columnwidth,trim=0 0 0 0, clip]{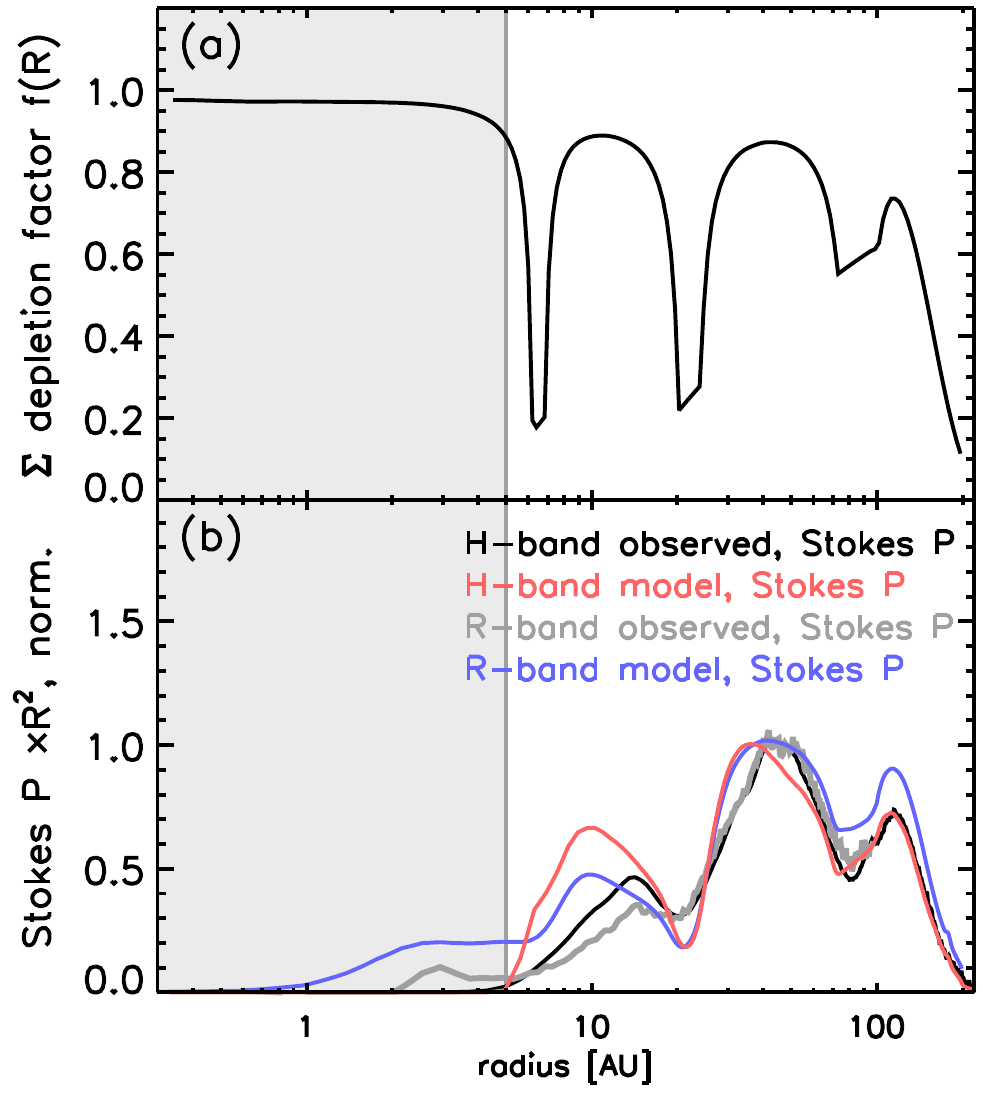}
\caption{\label{fig:overview_duffell} Equivalent to \fig\fsep\ref{fig:results_overview}, but for the profile of the Duffell model in \fig\fsep\ref{fig:gap_profiles}. A distance of \dpc\du pc has been assumed.}
\end{figure}

In \fig\fsep\ref{fig:gap_profiles} we compare the radial surface density depletion factor $f(R)$ as derived from our modeling to an implementation of the analytical gap model of \cite{2015ApJ...807L..11D} with three planets, approximately matching the depth of the gaps. As an additional illustration we show what the disk would look like in scattered light if the surface density profile would follow this Duffell model in \fig\fsep\ref{fig:overview_duffell}.

The match between the analytical model and the $f(R)$ profile derived from observations is poor. In particular the inner gaps \#2 and \#3 are much narrower in the analytical model. This is primarily due to the radial dependence of the pressure scale height in our flared disk, with $H_{\rm{p}}/R$ being much smaller in the inner disk than at larger radii (see also \sek\seksep\ref{sec:discussion:planet_masses}). Only for the outer gap around 85\du\AU \ the width and depth of the gap in the analytical model approximately match the gap in the $f(R)$ curve derived from our observations, though in detail the curves differ.

The mismatch between the $f(R)$ curve derived from our scattered light observations through radiative transfer modeling (see \seks\seksep\ref{sec:methods:RT} and \ref{sec:results:radial_density_profile}) and the results of planet-disk interaction models raises questions:

\begin{enumerate}
\item is the radial surface density distribution we derived realistic?
\item if so, are embedded (proto-) planets responsible for radial surface density variations in the \TW \ disk?
\end{enumerate}
 
\noindent
The first question lies at the core of our approach: we have \emph{assumed} that variations in the radial surface density of the gas (and with it, of population of small dust grains) are the cause of the observed brightness variations, and that a physical disk model as ours correctly yields the corresponding disk structure and scattered light distribution. There may be other ways to achieve a match to the observations, such as spatially variable intrinsic dust scattering properties and/or spatially variable gas-dust coupling through a variable viscosity parameter $\alpha(R)$. We consider variations in the surface density $\Sigma(R)$ to be the most obvious parameter to explore and leave the modeling of other plausible mechanisms underlying the brightness variations to future studies.

One may consider whether a closer match to the derived density profile can be obtained with a model with a larger number of planets. This should be possible with an ad-hoc solution with more planets in the 5$-$30\du\AU \ region, but without more direct observational evidence, and proof that a situation with many planets close together in the inner disk is a dynamically stable configuration, this is mere speculation.

The prescription of \cite{2015ApJ...807L..11D} provides a good approximation of the results of numerical planet-disk interaction calculations in the regime where relatively low-mass planets create partial gaps. For more massive planets that create deep, nearly empty gaps the numerical models create broader gaps whose flanks are shallower \citep[e.g.][]{2014ApJ...782...88F}. However, also these profiles do not provide a good match to the $f(R)$ profile we derived because their flanks are not sufficiently shallow and their $f(R)$ goes very close to zero around the planet, in contrast to our results.
 
\emph{If} planet-disk interaction is responsible for the radial surface density profile and the number of planets in the disk is two or three\footnote{One planet causing the gap around 85\du\AU \ and one or two planets causing the very broad gap that reaches from $\lesssim$\ds6\du\AU \ to $\approx$\ds50\du\AU.} then this would imply that the explored planet-disk interaction models do not include all physical effects that govern the shape of the gas distribution.

\subsection{Radial concentrations of large grains}
\label{sec:discussion:rings_large_grains}

Radial variations in the gas density may lead to radial drift of dust particles that are large enough to no longer be perfectly coupled to the gas, yet small enough to still be influenced by it (i.e. particles with Stokes numbers not largely different from unity). For our large grain component we used particles in the size range of 1 to 10~mm. These experience substantial headwind from the gas in consequence migrate inward by 1\du\AU \ every $10^3$ to $10^4$ years, depending on grain size and radial location in the disk. Note that this migration rate is about 2 orders of magnitude larger than the migration rate for the small particles in the disk surface.

Our physical disk model yields the gas and dust densities and the temperature. From this we can calculate the radial pressure profile in the midplane, where the strongly vertically settled large dust population resides. Using eqns.~\ref{eq:Stokes_number} to \ref{eq:epsilon} we can then calculate the radial drift velocity $v_{\rm{R}}$ in the same we as we did for the small dust grains in the disk surface in \sek\seksep\ref{sec:methods:gas_dust_coupling}. We expect the local surface density of large grains to scale with $v_{\rm{R}}^{-1}$; a simple ``toy'' model for the resulting density profile may have the form $\Sigma_{\mathrm{large}}(R)=\Sigma_{\mathrm{0,large}} v_{\rm{R}}^{-1}$. 

\begin{figure}[t]
\includegraphics[width=\columnwidth,trim=0 0 0 0, clip]{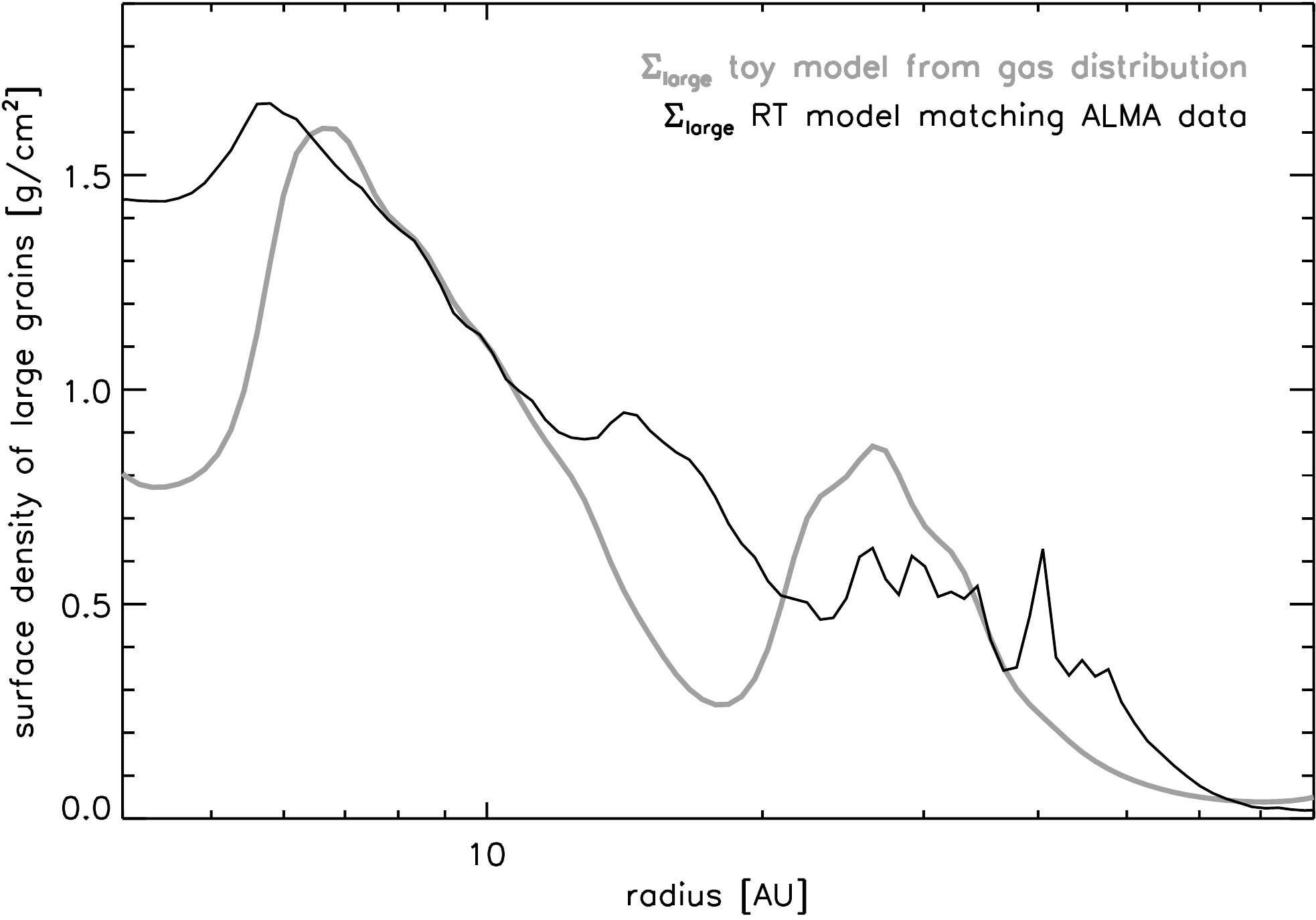}
\caption{\label{fig:radial_drift} Toy model for the radial distribution of large dust derived from the gas distribution in our radiative transfer model (grey) compared to the radial distribution of large grains that reproduces the ALMA data (black).  A distance of \dpc\du pc has been assumed.}
\end{figure}

In \fig\fsep\ref{fig:radial_drift} we compare this toy model for an initial distribution $\Sigma_{\mathrm{0,large}} \propto R^{-1}$ to the density profile derived from the ALMA data (hereafter called the ``data''). Note that the absolute level of the surface density of large grains in the data is not well determined; with our assumed large grain sizes of 1-10\du mm we obtain an opacity of $\approx$\ds1~cm$^2$/g for our large grains, but if we take a distribution of 0.1$-$1\du mm grains we obtain a surface density distribution similar in shape but with a factor 4-5 lower mass. Grains will concentrate in locations where the radial pressure gradient is small and will move away from regions with a high radial pressure gradient. The match between the simple model and the data is modest, we qualitatively reproduce the higher density plateau at 25-35\du\AU \ in the data and the gradient towards smaller radii, but the model is not able to reproduce the overall profile well\footnote{Note that the small scale structure in the data cannot be reproduced per design, because the gas distribution of the RT model contains no structure on these small scales.}. Considering the low complexity of the model, this may be expected. The model ignores any kind of dynamical instabilities. In particular, the mm-sized grains that we use for the large dust component are highly susceptible to the streaming instability. The large grains are concentrated near the midplane, their vertical scale height at 2\du\AU \ is $\approx$\ds8\% of that of the gas, at 40\du\AU \ it is $\approx$\ds2.5\%. The large dust dominates the total mass near the midplane, with the local dust/gas ratio reaching values of $\approx$\ds2. Therefore, the assumption that the gas velocity (relative to the kepler speed) is governed by the radial gas pressure gradient in the midplane is not accurate; in the midplane the dust will be more effective in forcing the gas towards the kepler speed than the gas in forcing the dust away from the kepler speed. More realistic modeling of the gas-dust dynamical interaction is clearly needed in order to connect the large dust distribution derived from the ALMA continuum data to the gas distribution derived from the SPHERE data.

\subsection{The compact \HCOp \ emission}
\label{sec:discussion:HCOp_emission}

\cite{2015ApJ...799..204C} observed the \TW \ disk in the \HCOp(3$-$2) line at 268\du GHz. They detect extended emission from the disk out to at least $\approx$\ds100\du\AU, this component is approximately centered on the central star. This emission is expected to arise mostly between about 2 and 5 scale heights above the midplane \citep[e.g.][]{2015A&A...574A.137T}. They also detected a compact source of \HCOp emission approximately 0\farcs43 South-West of the central star (see their \fig\fsep1b; the approximate position of the detected source is also indicated with a $\times$ sign in our \figs\fsep\ref{fig:legend} and \ref{fig:irdis_polar}), i.e. at approximately 23\du\AU \ from the star. This coincides with radial location of the minimum surface density in gap~\#2, which lies at 22\du\AU. This close match may be by chance, but it justifies a closer look at a possible relation between both observations.

\cite{2015ApJ...799..204C} briefly discuss possible origins of the local excess \HCOp \ emission, distinguishing between origins near the disk surface (local scale height enhancement, energetic particles from a stellar coronal mass ejection impinging on the disk) and near the disk midplane (accreting protoplanet embedded in the disk). Based on our SPHERE data we can exclude the local scale height enhancement scenario: in order to get the large local increase in absorbed stellar ionizing radiation needed for the strong enhancement in \HCOp \ production, the local scale height increase would need to be large and this would result in a correspondingly large increase in scattered light surface brightness\footnote{At the spatial resolution of the SMA data of \cite{2015ApJ...799..204C}, the intensity of the \HCOp \ line emission is approximately twice as high at the position of the point source compared to its immediate surroundings. Since the source is not spatially resolved in the SMA data, this is a lower limit on the actual contrast. Assuming that the \HCOp \ emission scales linearly with the amount of radiation locally intercepted by the disk, this would require a scale height enhancement that roughly doubles the angle under which the stellar light impinges on the disk surface. This would lead to approximately a doubling of the scattered light intensity, which would be very obvious in the SPHERE data, but nothing is seen.}. We observe no such increase, the scattered light intensity distribution around the position of the \HCOp \ source is smooth.

The close coincidence between the location of the \HCOp \ source and the radial minimum in the surface density profile as derived from the SPHERE data supports interpretation involving an embedded planet. The location of the planet would coincide with that of the \HCOp \ source. The planet can be of only moderate mass ($\lesssim$~several~10\du\Mearth, see \sek\seksep\ref{sec:discussion:planet_masses}) and its presence is not directly observable at the disk surface in scattered light. The disk does show an approximately 60\degree \ wide azimuthal brightness enhancement in ring~\#2 that is roughly centered on the azimuthal direction of the \HCOp \ soure ($\approx$\ds233\degree \ E of N) but it is unclear whether the two are related. Whether local heating of disk material near a forming (accreting) planet is a plausible mechanism for producing the \HCOp \ emission, in particular to account for the line flux of the compact source detected by \cite{2015ApJ...799..204C}, remains to be studied.

\vspace{0.5cm}

\subsection{The spiral feature in Ring \#1}
\label{sec:discussion:spiral}

The dark spiral feature in ring~\#1 is intruiging. It is clearly seen in our H-band image; in our \Rp \ and \Ip \ images if falls outside the field of view and it is buried in the noise of our J-band observations that were optimized for the inner disk and are readout noise limited at the location of ring~\#1. The spiral is clearly seen in the STIS image of \cite{2013ApJ...771...45D} and is definitively a real feature of the \TW \ disk. The opening angle is very small: the dashed line in \fig\fsep\ref{fig:legend} is a logarithmic spiral with a pitch angle of 1.5\degree. This shape follows the dark spiral well. We did not attempt to comprehensively measure the shape of the feature; pitch angles larger than $\approx$\ds2\degree \ are too steep.

The structure is most easily perceived as a dark spiral and hence we refer to it as such. Physically this would require either a local decrease in scale height and hence incidence angle of stellar radiation, or a local reduction of the scattering efficiency. The latter could, for example, be caused by a local over-abundance of small $a<0.1$\du\mum \ grains in the disk surface, that absorb radiation much more effectively than they scatter it, hence making the disk appear dark. The dark spiral could also be intermediate between two brighter regions, due to e.g. local enhancements of the scale height, as may be caused by spiral shock waves that locally heat the disk material \citep[e.g.,][]{2015MNRAS.453.1768P}. We note, though that the small pitch angle of $\lesssim$\ds2\degree \ is not consistent with spirals induced by planet-disk interaction \citep[][]{2002ApJ...569..997R}.

Whether the spiral feature is leading or trailing depends on the sense of rotation of the disk. This can be derived from the mm line data, that yield the orientation of the major axis and the line of sight velocity field, \emph{if} it is known which side of the disk is the near side. The position angle of the major axis is $\approx$\ds151\degree E of N \citep{2012ApJ...757..129R}, which means the near side is at either $\approx$\ds61\degree \ (NEE) or $\approx$\ds241\degree \ (SWW). The SSE side of the disk is red-shifted and the NNW side is blue-shifted relative to the systemic velocity.

If the dust is somewhat forward scattering, then the near side is expected to be brighter than the far side. Because of the flared shape of the disk, the surface that we see at a given angular separation from the star on the near side is physically somewhat further from the star than the corresponding location on the far side. This causes the surface on the near side to receive fewer photons and be somewhat fainter. For \TW \ this geometrical effect is small, $\lesssim$\ds7\% in brightness. The combined effect of scattering phase function and the flared disk shape is that, for silicate particles and MIE calculations and a disk inclination of 7\degree, the far/near-side brightness ratio is approximately 0.95, 0.88, and 0.69 for compact spherical grains with radii of $a$\ds$=$\ds0.1, 0.2, and 0.3\du\mum, respectively\footnote{This illustrates the transition from close to isotropic to substantially forward scattering. It also shows how critically the brightness contrast depends on the dust properties that govern the scattering phase function, and hence why an accurate determination of the disk inclination angle from an observed scattered light profile is not possible.}.

In our data the Southern half of the disk is overall brighter than the Northern half. There is significant azimuthal sub-structure (see \fig\fsep\ref{fig:azimuthal_profiles}) and the division between a brighter and a fainter half does not quite line up with the disk position angle known from the mm data. Nonetheless, the disk hemisphere centered in azimuth around the SWW direction is overall clearly brighter than the hemisphere centered around the NEE direction. Hence, we consider the SWW direction to be the near side. This makes the disk of \TW \ rotate \emph{clockwise}, the spiral feature is then \emph{trailing}.

\subsection{Detecting potential embedded (proto-) planets}
\label{sec:discussion:planet_brightness}

\newcommand{\Reff}{$R_{\rm{eff}}$}
\newcommand{\Teff}{$T_{\rm{eff}}$}

\begin{figure}[t!]
\includegraphics[width=1.0\columnwidth,trim=0 0 0 0, clip]{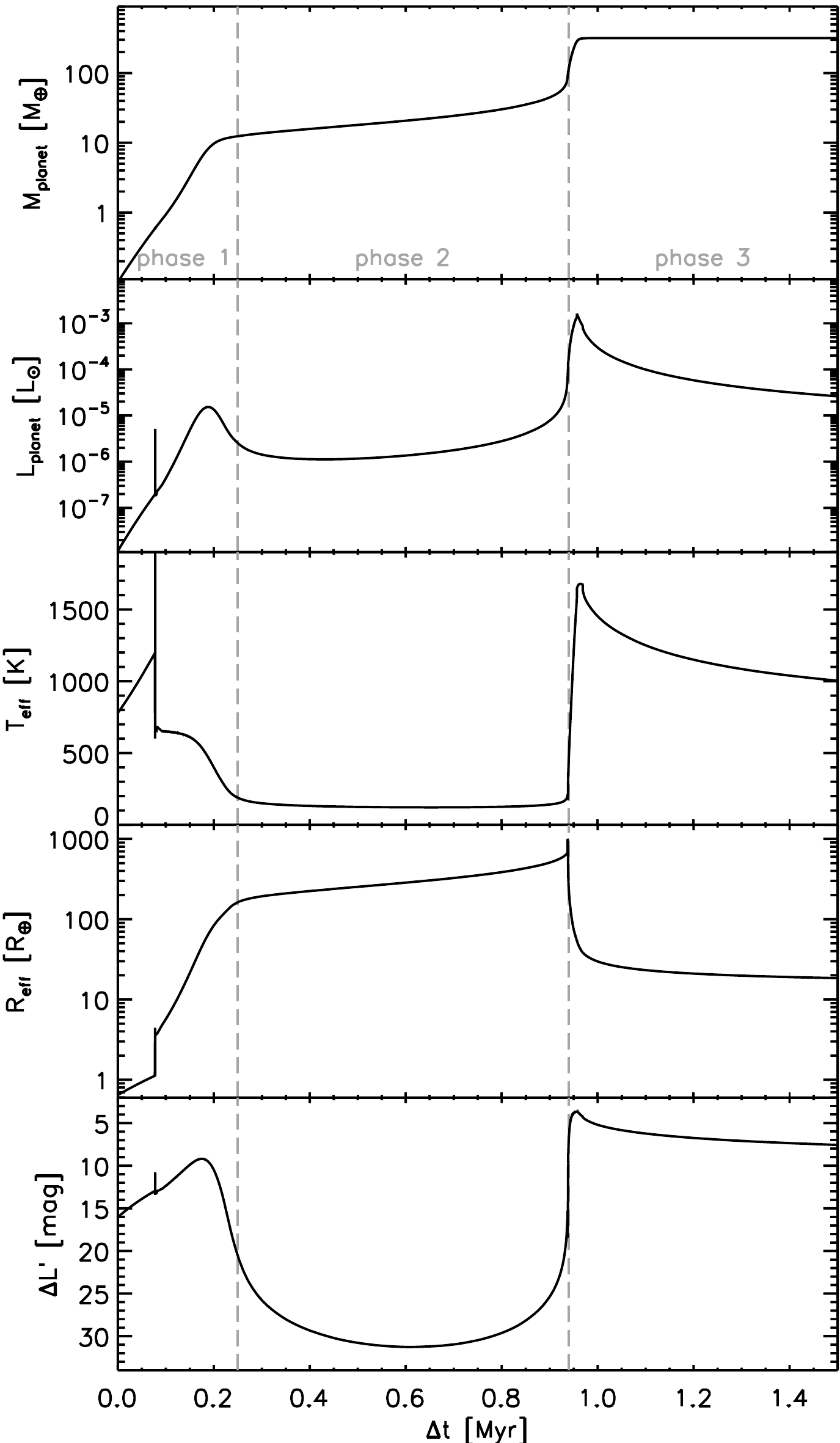}
\caption{\label{fig:mordasini_model} Example formation history of a planet with a final mass of 1\du\Mjup.}
\end{figure}

It is conceivable that embedded, still forming planets exist in the \TW \ disk, that are possibly accreting. In this section we explore the prospects of detecting them in the infrared. We will adopt the core-accretion scenario for giant planet formation 
\citep{1974Icar...22..416P,1978PThPh..60..699M,1986Icar...67..391B} and use the model of \cite{2012A&A...547A.111M} to estimate the brightness of the planets. We approximate the spectrum of the forming planet by blackbody emission from a sphere of radius \Reff, which is the radius where the gray optical depth $\tau = 2/3$ in the model of \cite{2012A&A...547A.111M}, and where the model temperature is \Teff. In \fig\fsep\ref{fig:mordasini_model} we show the formation history of an example planet with a final mass of 1\du\Mjup \ in order to illustrate the qualitative behavior and the three main observationally relevant formation phases. In the population synthesis calculations thousands of planetary formation tracks are calculated that are qualitatively similar in the sense that they follow the same phases, but the detailed evolution of a given planet depends on its formation location in the disk, whether the core formes early or later in the disk evolution, and on global parameters such as disk mass and metallicity. Later in this section we present predicted brightnesses of a full population from \cite{2012A&A...547A.111M}.

During the first phase a several Earth-mass core is built. Initially is has no substantial envelope but once the core reaches $\approx$\ds1\du\Mearth \ an envelope forms, which initially comprises only a minor fraction of the planet mass but is highly optically thick and determines the apparent size. Though it is much larger than the actual core, it is still relatively compact with a size of $R_{\rm{eff}} \lesssim$\ds$5$\du\Rjup. Through a combination of gravitational focusing and gas drag near the planet the accretion rate of planetesimals is high and the associated accretion luminosity makes the objects quite warm (\Teff\,$\approx$\ds500\du K) and luminous (of order $10^{-5}$\du\Lsun). During this phase the objects are potentially favorable for infrared observations despite their low mass, but not for very long ($\lesssim$\ds$10^5$ years).

The second phase starts when the planets reach a mass of order 10\du\Mearth, and acquire a massive gaseous envelope that fills the Hill sphere while remaining optically thick. The envelope forms a continuous structure with the surrounding disk which provides fresh material through the general disk accretion flow, and the objects are therefore said to be in the ``attached'' phase. The solution of the envelope's internal structure equations yields the rate at which the object can acquire further gas mass. Due to the envelope's large size the accreted material has relatively low energy when it is incorporated into the optically thick structure and the luminosity is correspondingly low (of order  $10^{-6}$\du\Lsun). Moreover, this small amount of emerging energy is emitted mostly at far-infrared wavelengths that are not accessible from the ground.

With increasing envelope mass the accretion rate allowed by the structure equations becomes higher, and at some point it surpasses the rate at which the surrounding disk can supply new gas. The envelope then contracts rapidly and the optically thick structure becomes much smaller than the Hill sphere. In this third, ``detached'' phase the objects are compact (a few \Rjup) and typically very luminous since the material that is now accreted reaches the optically thick structure with much more energy per unit mass. With effective temperatures around 1500\du K, and luminosities of up to $\approx$\ds10$^{-3}$\du\Lsun \ the objects are potentially very favorable to infrared observations.

\newcommand{\taut}{$\tau_{\rm{ex}}$}

\begin{table}[htb]
\begin{center}
\caption{\label{tab:tau_vertical} Vertical optical depth from the midplane for our model in the deepest parts of the gap regions. $\tau_{\rm{ex}}$ denotes the total extinction (absorption plus scattering).\vspace{0.0cm}}

all dust between midplane ($z=0$) and observer:\\
\begin{tabular}{ccccc}
$\lambda$ & \taut       & \taut      & \taut      & $\tau_{\rm{scat}}/$\taut \\
$\lbrack$\mum$\rbrack$    & (6.5\du\AU) & (21\du\AU) & (85\du\AU) &  \\
\tableline
    0.65 &  10.7 &   5.2 &   4.3 & 0.56 \\
    1.6  &   7.7 &   3.7 &   3.0 & 0.56 \\
    2.2  &   6.4 &   3.0 &   2.4 & 0.57 \\
    3.6  &   4.9 &   2.3 &   1.7 & 0.54 \\
    4.8  &   3.8 &   1.7 &   1.2 & 0.50 \\
    8.0  &   2.5 &   1.0 &   0.6 & 0.34 \\
   13    &   2.2 &   0.9 &   0.5 & 0.27 \\
   24    &   2.3 &   0.9 &   0.5 & 0.16 \\
  852    &   2.1 &   0.7 &   0.0 & 0.63 \\
 1300    &   2.3 &   0.8 &   0.0 & 0.67 \\
\end{tabular}

\vspace{0.3cm}
as above, but including only dust at $z > R_{\rm{Hill}}$:\\
\begin{tabular}{ccccc}
$\lambda$ & \taut       & \taut      & \taut      & $\tau_{\rm{scat}}/$\taut \\
$\lbrack$\mum$\rbrack$    & (6.5\du\AU) & (21\du\AU) & (85\du\AU) &  \\
\tableline
    0.65 &   3.2 &   2.5 &   2.8 & 0.57 \\
    1.6  &   2.1 &   1.6 &   1.9 & 0.58 \\
    2.2  &   1.6 &   1.3 &   1.5 & 0.59 \\
    3.6  &   1.1 &   0.9 &   1.0 & 0.58 \\
    4.8  &   0.8 &   0.6 &   0.7 & 0.55 \\
    8.0  &   0.4 &   0.3 &   0.4 & 0.38 \\
   13    &   0.3 &   0.2 &   0.3 & 0.27 \\
   24    &   0.3 &   0.3 &   0.3 & 0.08 \\
  852    &   0.0 &   0.0 &   0.0 & 0.00 \\
 1300    &   0.0 &   0.0 &   0.0 & 0.00 \\
\\
$r_{\rm{Hill}}$ [\AU]:  &    0.2 &   0.7 &   2.8 & \\ 
\end{tabular}
\end{center}
\end{table}

\begin{figure*}[t!]
\begin{center}
\hspace{-4mm}
\begin{tabular}{ccc}
\includegraphics[width=0.32\textwidth,trim=0 0 0 0, clip]{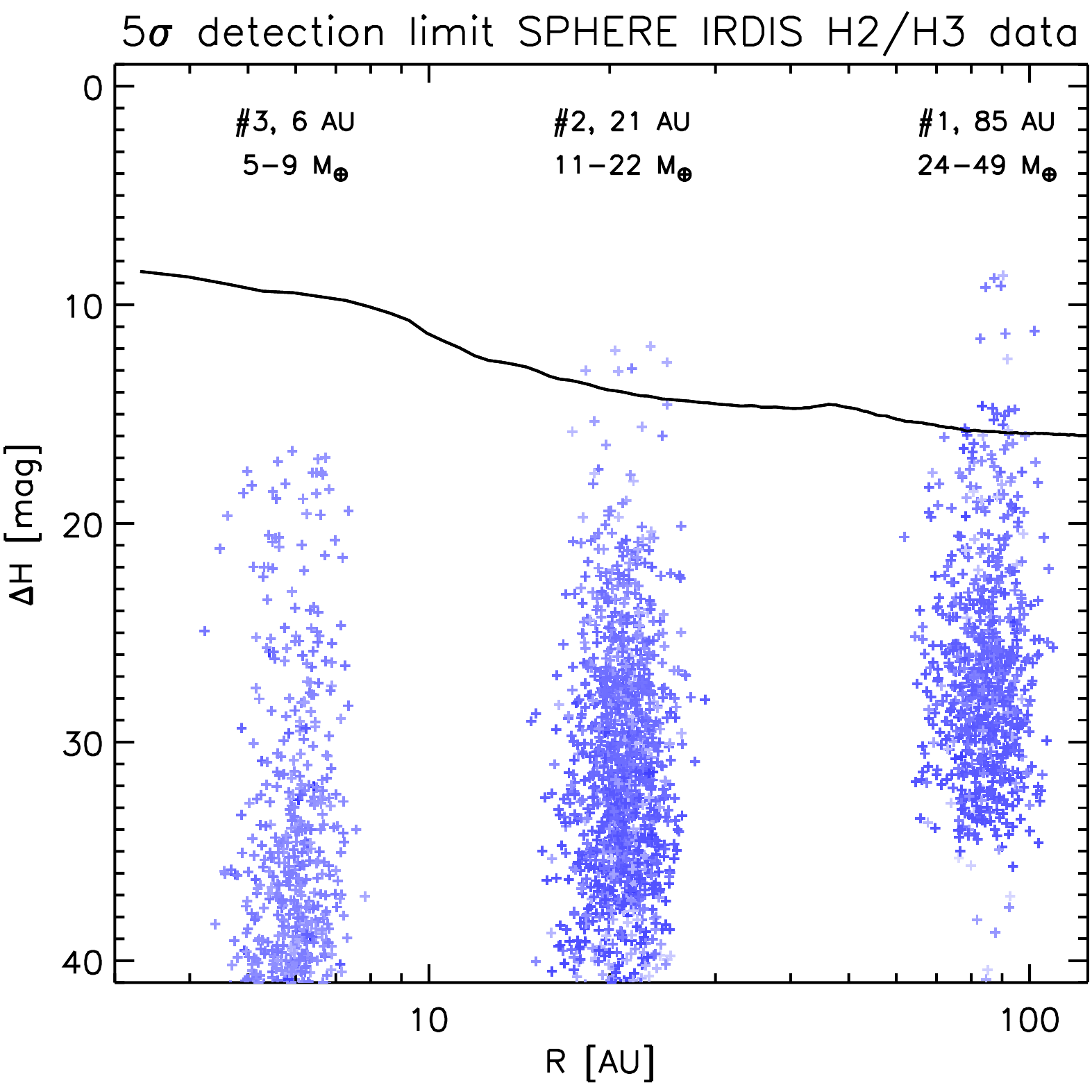} &
\includegraphics[width=0.32\textwidth,trim=0 0 0 0, clip]{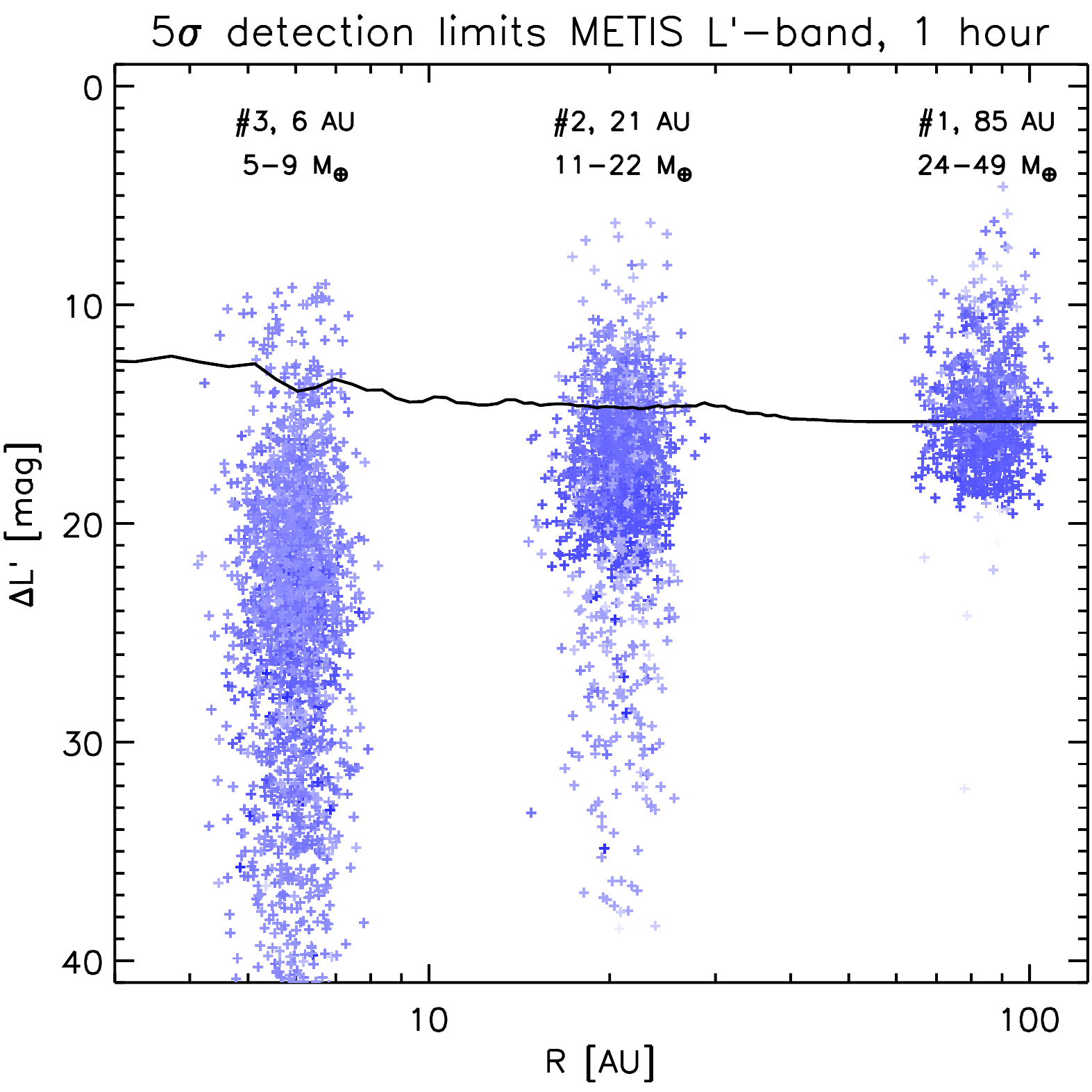} &
\includegraphics[width=0.32\textwidth,trim=0 0 0 0, clip]{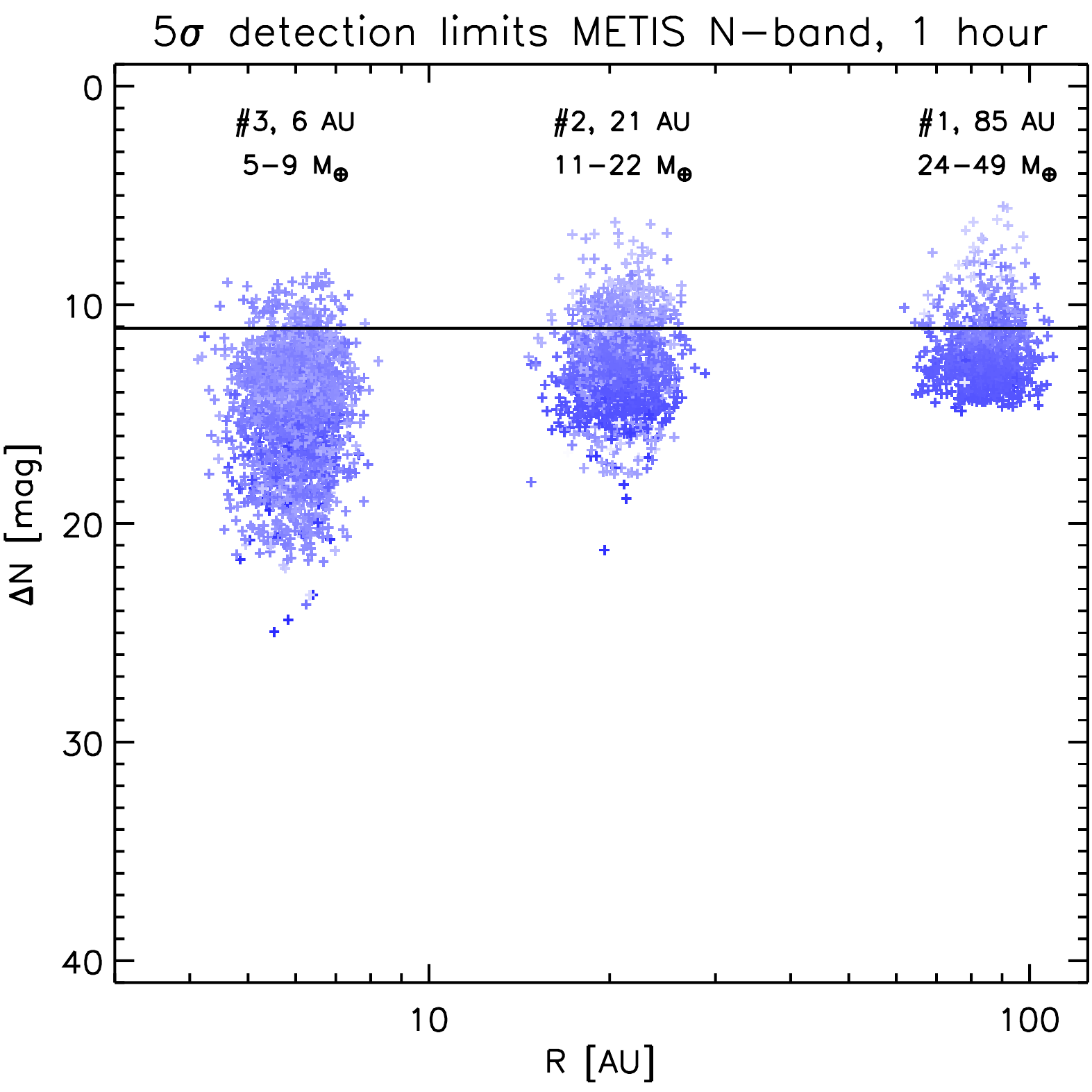}
\end{tabular}
\caption{\label{fig:CD753} Synthetic photometry of the CD753 planet population from \cite{2012A&A...547A.111M} at 2\,Myrs. The objects are color-coded according to physical size of the optically thick structure, with dark shades indicating compact and bright shades large objects. The black curves indicate the achieve detection limits of SPHERE (H-band, left panel) and the foreseen detection limits of METIS (\Lp \ and N-band, middle and right panel). A distance of \dpc\du pc has been assumed.}
\end{center}
\end{figure*}

\subsubsection{Dust extinction}
\label{sec:dust_extinction}

If forming planets are present in the \TW \ disk then they must still be embedded in disk material. We recall that there are no large local azimuthal variations in scattered light brightness anywhere in the gap regions. Thus, at least the higher disk layers appear largely unaffected by potential embedded planets, and we may assume that our modeling of the SPHERE data provides a reasonable approximation of the (vertical) dust distribution at the location of potential planets. Here we calculate the vertical optical depth of the disk material ``above'' the planet, i.e. the amount of opacity that radiation emitted by the planetary photosphere or the circumplanetary environment must pass through before it reaches the disk surface and can propagate to the observer.

In the upper half of \tab~\ref{tab:tau_vertical} we list the vertical optical depth from the midplane to the observer in our model, at various wavelengths accessible to observations, and at the radii of minimum column density in the three gap regions. The extinction values we derive are substantial at optical and infrared wavelengths, ranging from $\approx$\ds3 to 8~mag in the H-band and from $\approx$\ds2 to 5~mag in the L-band. However, if the objects are in the detached phase then we may assume that a region around the planet with approximately the size of the Hill sphere is comparatively empty (except for the plane of the circumplanetary disk). For the sake of the current argument we will assume the planets are in detached phase, recalling that prior to this phase the objects are in any case much too faint to be observed. We keep in mind that at the current time there is neither conclusive evidence for the presence of embedded planets, nor that they would be in the detached phase.

In the lower half of \tab~\ref{tab:tau_vertical} we show the vertical optical depths of the same model, but with removal of all dust at $z<r_{\rm{Hill}}$. In this case the extinction is substantially smaller, for example in the L-band it is roughly 1 magnitude. In the atmospheric N-band around 10\du\mum \ it is only a few tenths of a magnitude and at longer wavelengths it becomes negligible.

\subsubsection{The infrared brightness of a population}
\label{sec:population_brightness}

Assuming that the observed radial density depression in the \TW \ disk are caused by embedded planets in the mass ranges derived in \sek\seksep\ref{sec:discussion:planet_masses}, how bright may those objects be in the infrared, and could they be observed with current of future facilities? In order to assess this question, we analyzed the CD753 population from \cite{2012A&A...547A.111M}, which can be interactively inspected at \url{https://dace.unige.ch/evolution/index#}. We assumed mass ranges of 5-9\du\Mearth, 11-22\du\Mearth, and 24-49\du\Mearth \ for the objects in gaps \#3, \#2, and \#1, respectively. We then selected all objects in the CD753 that were in the given mass ranges after 2\du Myrs of disk evolution (the planets themselves have a range of ages $\leq$\,2\du Myr, depending on when in the population synthesis calculation their core formed). We then estimated the  brightness of the planets using a blackbody approximation, and investigated whether they would be observable in the infrared. In reality, the objects in the CD753 population are mostly much closer to the star than gaps \#1 and \#2, but for a first assessment of whether objects in this mass range are observable we artificially put them at the radii of the SPHERE gaps. We assume that the Hill spheres of the compact objects are optically thin but the disk region above and below the Hill sphere contains dust, and apply the extinction values listed in the lower half of \tab~\ref{tab:tau_vertical} to the synthetic photometry.

In \fig\fsep\ref{fig:CD753} we show the expected brightnesses in the H, \Lp, and N bands, where each point represents a specific planet of the CD753 population at 2\du Myr that fits the mass range for a given gap. In the radial direction small random offsets have been applied for better visibility. The points are color coded according to planet radius, with light shades indicating large objects (attached, in phase 2) and dark shades indicating compact objects (in gap \#3 these are typically still in phase 1, whereas in gap \#1 they are in phase 3, detached).

In the left panel we show the H-band brightnesses, together with the $5\sigma$ detection limits of our IRDIS ADI data (see \sek\seksep\ref{sec:methods:ADI_non_detection}). Our non-detection of any point sources in the IRDIS H-band ADI experiment is in agreement with the fact that the vast majority of objects in the simulation is much fainter than our detection limit.

In the middle panel we compare the brightness in the \Lp \ band with the expected sensitivity and contrast performance of the Mid-Infrared E-ELT Imager and Spectrograph \citep[METIS,][]{2014SPIE.9147E..21B}, which is the thermal infrared first generation instrument for the European ELT. The detection curve was taken from \cite{2010SPIE.7735E..2GB}. A reasonable fraction of the synthetic population would be observable with this facility, though many possible objects remain too faint even for a 40\du m class facility, particularly for gaps \#2 and \#3. In the right panel we show the brightness in the N-band, here taken to be a 1.5\du\mum \ wide passband centered on 11.2\du\mum. We assume that a METIS observation will be background-limited with a $5\sigma$ detection limit of 25\du $\mu$Jy in 1 hour \citep{2015IJAsB..14..279Q}. Also at this wavelength, a reasonable fraction of possible objects would be observable, though most remain too faint. The color coding in \fig\fsep\ref{fig:CD753} illustrates how the larger, cooler objects become more observable towards longer wavelengths.

Our analysis indicates that the non-detection of point sources in the SPHERE ADI data is all but surprising, and that potential embedded objects in the \TW \ disk may be observable with METIS. However, potential objects could quite plausibly be too faint even for the ELT. Note that our blackbody approximation will underestimate the true brightnesses of objects that are hot and whose opacity is dominated by molecules instead of dust. These will be brighter in the photometric bands of ground-based astronomy than the simple blackbody value, because similarly to Earth the atmospheric opacity of these objects is dominated by water and they are relatively transparent in our photometric bands so we effectively observe deeper, hotter layers at these wavelengths.

\subsubsection{The new GAIA distance to \TW}
\label{sec:gaia_distance}

During the acceptance stage of this paper the GAIA team announced their first data release \citep{2016arXiv160904172G} and the new parallax for \TW \ corresponds to a revised distance of $d$\eqsep$=$\eqsep$59.5^{+0.96}_{-0.93}$\du pc, i.e. $\approx$\ds10\% higher than the value of \dpc\du pc that we have assumed throughout this work. Here we briefly discuss the implications of this new distance. 

All physical length scales of the system trivally scale with $d$ and become $\approx$\ds10\%  larger. The inferred luminosity of the star scales with $d^2$ and increases by $\approx$\ds21\%. Because the pre-main sequence evolutionary tracks run approximately vertical in \TW's region of the Herzsprung-Russel diagram, this higher luminosity does not change the estimated stellar mass appreciably but will decrease the estimated age somewhat, it remains in the range of 5 to 10\,Myrs, however. The disk mass scales like $d^2$ and increases by $\approx$\ds21\%, where we note that the absolute disk mass was an input parameter of the model adopted from \cite{2013Natur.493..644B}; our model constrains how this matarial is distributed spatially but does deduce the absolute mass from the observations.

To assess the effect on the mass estimates $M_{\rm{p}}$ of potential embedded planets we use the relationship $(M_{\rm{p}}/M_*)^2$\eqsep$\propto$\eqsep$\mathcal{M}^{-5}$ (\eq~\ref{eq:planet_masses}). The temperature in the model grid cells scales approximately as $T$\eqsep$\propto$\eqsep$L^{1/4} R^{-1/2}$ and is therefore unaltered. The Mach number scales like $\mathcal{M}$\eqsep$=$\eqsep$R/H_{\rm{p}}$\eqsep$\propto$\eqsep$T^{-1/2}$ and hence remains unaltered. With the aproximately unaltered stellar mass this implies that the $M_{\rm{p}}$ estimates are not affected by the 10\% increase of the physical dimensions of the system. We recall that the actual uncertainties in the $M_{\rm{p}}$ estimates are substantial and are dominated by the uncertainty in the disk viscosity parameter $\alpha$ (see \sek\seksep\ref{sec:discussion:planet_masses}).

The inferred brightness of potentially embedded planetary objects (see \sek\seksep\ref{sec:population_brightness}) decreases because of the increasing extinction and the increased distance to Earth (the latter yields $\Delta m$\eqsep$\approx$\eqsep$+0.21$\du mag). Since the inferred planetary mass estimates are unaffected their intrinsic brightness will also remain the same. The amount of dust ``above'' the embedded planets causing extinction scales linearly with the disk mass and hence becomes $\approx$\ds21\% larger. The objects are more extincted by $\Delta m$\eqsep$\approx$\eqsep$0.21$\eqsep$\times$\eqsep$1.086$\eqsep$\times$\eqsep$\tau_0$ magnitudes, where $\tau_0$ denotes the extinction in the ``unscaled'' model (see \tab\tabsep\ref{tab:tau_vertical}). The combined effect yields fainter apparent magnitudes by $\Delta H$\eqsep$\approx$\eqsep$+0.65$\du mag ($\tau_{\rm{0,H}}$\eqsep$\approx$\eqsep$1.9$), $\Delta L$\eqsep$\approx$\eqsep$+0.45$\du mag ($\tau_{\rm{0,L}}$\eqsep$\approx$\eqsep$1.0$), and $\Delta N$\eqsep$\approx$\eqsep$+0.3$\du mag ($\tau_{\rm{0,N}}$\eqsep$\approx$\eqsep$0.3$); we have adopted the values corresponding to dust-free Hill spheres around the proto-planets (i.e. the lower half of \tab\tabsep\ref{tab:tau_vertical}). When compared to the very large brightness range in \fig\fsep\ref{fig:CD753}, where 1 tick mark represents 1\du mag, it is clear that the 10\% distance increase makes no qualitative difference to the discussion of the detectability of potential embedded sources.

\section{Summary}
\label{sec:summary}
We have presented new multi-band polarimetric imaging of the \TW \ disk obtained with the ZIMPOL and IRDIS sub-instruments of SPHERE at the VLT, detecting the disk down to $\approx$\ds5\du\AU \ in all four bands and possibly down to $\approx$\ds2\du\AU \ in the ZIMPOL \Rp and \Ip-band data. We detected the known ``gaps'' around 80 and 20\du\AU \ and discovered a third gap inward of 10\du\AU. The overall intensity distribution is highly symmetric but there are low-amplitude azimuthal brightness variations, which are similar over a large range of radii. There is a substantial correspondence in the substructure of the scattered light profile and that of the mm continuum data by \cite{2016ApJ...820L..40A}. We detected a spiral feature in the outer disk. We also observed the system in ADI mode to look for point sources in or near the disk, but did not detect any, with $5\sigma$ detection contrasts of $\Delta H$\,$=$\,9.5, 14.0, and 15.5\du mag at \Rthree, \Rtwo, and \Rone\du\AU, respectively. 

Because the scattered light signal is dominated by sub-micron sized grains, the scattering dust is very well coupled to the gas. We therefore explored a scenario in which radial variations in the bulk gas surface density are responsible for the observed scattered light brightness distribution. Using 2D radiative transfer modeling including self-consistent vertical structure calculation, grain size-dependent dust settling, and full non-isotropic scattering, we built a disk model that matches the observed scattered light profile. The main free parameter of the model is the radial surface density distribution of the gas. Starting from the continuous disk model of \cite{2014A&A...564A..93M}, we introduced radial depressions in the surface density which we varied in depth and shape until the resulting scattered light profiles matched the observations. The value of the turbulence parameter $\alpha$ and the smallest grain size in the dust population \amin \ are also important parameters. We include a separate population of large grains that is designed to match the ALMA continuum data.

We find that radial depressions of moderate depth with shallow outer flanks yield a good fit to the observations. The gap around 80\du\AU \ is reproduced by a reduction in the surface density $\Sigma(R)$ to $\approx$\ds1/2 of the value in the unperturbed model $\Sigma_0(R)$. The gap around 20\du\AU \ and the newly found gap at $\lesssim$\ds10\du\AU \ are reproduced by a wide depression in the surface density where the surface density drops to $\approx$\ds1/5 of that of the unperturbed model. A relatively small increase in the surface density by $\approx$\ds0.1\,$\Sigma_0$ is responsible for the bright ring separating the gaps at $\approx$\ds20 and $\lesssim$\ds10\du\AU. The defining characteristics of the inferred surface density depressions are their broad radial extent and their local surface density that is still a substantial fraction of that of a gap-less model, i.e. the gaps are far from ``empty''.

We explored a scenario in which planet-disk interaction is responsible for the radial inferred surface density variations and estimated the masses of possible embedded planets. Applying the model of \cite{2015ApJ...807L..11D} we find that relatively low mass objects, in the range of $\approx$\ds5 to several tens of Earth masses, would result in ``gaps'' of the right depth but of too narrow radial extent. The relatively high inferred surface density in the gap regions and the high degree of azimuthal symmetry argue against planets that are significantly more massive than the values we derive; these would yield nearly empty gap regions and may induce spiral structures that are not observed. 

The compact source of \HCOp(3$-$2) \ emission that \cite{2015ApJ...799..204C} detected to the South-West of the star lies in a gap region and its radial location coincides with a minimum in the surface density at 22\du\AU \ inferred from the SPHERE data. We exclude a local scale height enhancement leading to a local increase of irradiation by the central star as a cause for the compact \HCOp \ source, which is one of several possibilities suggested by \cite{2015ApJ...799..204C}. Such a structure would lead to an increase in scattered light brightness that is much larger than our images allow, we do not see any feature at the position of the \HCOp \ source in the SPHERE data. Another proposed possibility is that local heating of the midplane by an embedded planet causes an enhancement of the CO abundance, producing additional \HCOp. The close spatial coincidence between the \HCOp \ source and the radial minimum in surface brightness at 22\du\AU \ is intriguing in this respect. However, the \HCOp(3$-$2) line is likely optically thick in the higher disk layers and the large line flux of the \HCOp \ source requires an implausibly large luminosity of the potential planet. The origin of the compact \HCOp \ emission remains an open issue, which new ALMA observations with higher spatial resolution may help clarify.

Based on the overall scattered light brightness profiles in combination with literature mm data, we consider the SWW side of disk to be the near side. This means the \TW \ disk rotates clockwise and the detected spiral feature is trailing.

Should forming planets exist embedded in the disk of \TW \ it would be a prime future goal to directly detect them in the thermal infrared and measure their SEDs. We used the population synthesis simulations of \cite{2012A&A...547A.111M} to estimate the infrared brightness of embedded objects in the mass range consistent with the gap depths inferred from the SPHERE observations. We find that the SPHERE H-band non-detections are in line with expectations, and that the objects may be observable with an infrared instrument at an ELT such as METIS, though even with a 40\du m class telescope, detections are not guaranteed.

\acknowledgments
We gratefully acknowledge the many colleagues who conceived, designed, built, integrated and commissioned SPHERE, as well as the Instrument Operations team at ESO. We thank T.~Birnstiel, C.P.~Dullemond, and Ch.~Mordasini for clarifying discussions. We also thank the referee for a thorough and constructively critical evaluation of the manuscript. 

SPHERE is an instrument designed and built by a consortium consisting of IPAG (Grenoble, France), MPIA (Heidelberg, Germany), LAM (Marseille, France), LESIA (Paris, France), Laboratoire Lagrange (Nice, France), INAF - Osservatorio di Padova (Italy), Observatoire de Genève (Switzerland), ETH Zurich (Switzerland), NOVA (Netherlands), ONERA (France) and ASTRON (Netherlands) in collaboration with ESO. SPHERE was funded by ESO, with additional contributions from CNRS (France), MPIA (Germany), INAF (Italy), FINES (Switzerland) and NOVA (Netherlands). SPHERE also received funding from the European Commission Sixth and Seventh Framework Programmes as part of the Optical Infrared Coordination Network for Astronomy (OPTICON) under grant number RII3-Ct-2004-001566 for FP6 (2004-2008), grant number 226604 for FP7 (2009-2012) and grant number 312430 for FP7 (2013-2016). 

We acknowledge financial support from the Programme National de Planétologie (PNP) and the Programme National de Physique Stellaire (PNPS) of CNRS-INSU. This work has also been supported by a grant from the French Labex OSUG@2020 (Investissements d’avenir – ANR10 LABX56). 

This work has made use of the SPHERE Data Centre, jointly operated by OSUG/IPAG (Grenoble), PYTHEAS/LAM (Marseille), OCA/Lagrange (Nice) and Observatoire de Paris/LESIA (Paris).

{\it Facilities:} \facility{VLT (SPHERE)}

\bibliographystyle{apj}
\bibliography{references}

\end{document}